\documentclass[11pt]{article}

\usepackage[skip=14pt plus2pt, indent=0pt]{parskip}
\usepackage{verbatim}
\usepackage{epsfig,graphics}
\usepackage {graphicx}
\usepackage {epsfig}
\usepackage {subfigure}
\usepackage {tabularx} 
\usepackage{rotate}	
\usepackage{slashed}
\usepackage{enumerate}
\usepackage{comment}
\usepackage{fancybox}

\usepackage{amsmath}

\usepackage{amsfonts}
\usepackage{amssymb}
\usepackage{graphicx}
\usepackage{xcolor}
\usepackage{cite}
\usepackage{tcolorbox}
\tcbuselibrary{breakable}

\usepackage{fancyhdr}
\usepackage[hidelinks]{hyperref}

\usepackage{braket}
\usepackage{cancel}

\def\beq{\begin{equation}}
	\def\eeq{\end{equation}}
\def\beqa{\begin{eqnarray}}
	\def\eeqa{\end{eqnarray}}

\def\Mpd{M_{\text{pl,}\, d}}

\catcode`\@=11   
\newdimen\@rotdimen
\newbox\@rotbox  

\def\@vspec#1{\special{ps:#1}}%  passes #1 verbatim to the output
\def\@rotstart#1{\@vspec{gsave currentpoint currentpoint translate
		#1 neg exch neg exch translate}}% #1 can be any origin-fixing transformation
\def\@rotfinish{\@vspec{currentpoint grestore moveto}}% gets back in synch 
\def\@rotr#1{\@rotdimen=\ht#1\advance\@rotdimen by\dp#1%
	\hbox to\@rotdimen{\hskip\ht#1\vbox to\wd#1{\@rotstart{90 rotate}%
			\box#1\vss}\hss}\@rotfinish}
\def\@rotl#1{\@rotdimen=\ht#1\advance\@rotdimen by\dp#1%
	\hbox to\@rotdimen{\vbox to\wd#1{\vskip\wd#1\@rotstart{270 rotate}%
			\box#1\vss}\hss}\@rotfinish}%

\def\@rotu#1{\@rotdimen=\ht#1\advance\@rotdimen by\dp#1%
	\hbox to\wd#1{\hskip\wd#1\vbox to\@rotdimen{\vskip\@rotdimen
			\@rotstart{-1 dup scale}\box#1\vss}\hss}\@rotfinish}%
\def\@rotf#1{\hbox to\wd#1{\hskip\wd#1\@rotstart{-1 1 scale}%
		\box#1\hss}\@rotfinish}%
\def\rotate{\@ifnextchar[{\@rotate}{\@rotate[l]}}
\def\@rotate[#1]#2{\setbox\@rotbox=\hbox{#2}\@nameuse{@rot#1}\@rotbox}

\catcode`\@=12
\topmargin
-1.5cm
\textwidth
15.5cm
\textheight
23.5cm
\oddsidemargin
0.7cm
\evensidemargin
0.7cm

\begin{document}
\makeatletter
\@addtoreset{equation}{section}
\makeatother
\renewcommand{\theequation}{\thesection.\arabic{equation}}
%----------------------------------------------------------------------%
%  title page
%----------------------------------------------------------------------%
\pagestyle{empty}
\vspace{-0.2cm}
\rightline{IFT-UAM/CSIC-23-70}
\rightline{CERN-TH-2023-113}
\vspace{1.2cm}
\begin{center}
		
\LARGE{Entropy Bounds and the Species Scale Distance Conjecture \\[13mm]}

           \large{J. Calder\'on-Infante$^\clubsuit$, A. Castellano$^\diamondsuit$, A. Herr\'aez$^\spadesuit$ and L.E. Ib\'a\~nez$^\diamondsuit$ \\[12mm]}

		\small{
                $^\clubsuit$ {Theoretical Physics Department, CERN, CH-1211 Geneva 23, Switzerland}  \\[5pt]
			$^\diamondsuit$ {Departamento de F\'{\i}sica Te\'orica
				and Instituto de F\'{\i}sica Te\'orica UAM/CSIC,\\
				Universidad Aut\'onoma de Madrid,
				Cantoblanco, 28049 Madrid, Spain}  \\[5pt]
			$^\spadesuit$ {Institut de Physique Th\'eorique, Universit\'e Paris Saclay, CEA, CNRS\\
				Orme des Merisiers, 91191 Gif-sur-Yvette CEDEX, France} 
			\\[8mm]
		}
		\small{\bf Abstract} \\[6mm]
	\end{center}
	\begin{center}
		\begin{minipage}[h]{15.60cm}
            The Swampland Distance Conjecture (SDC) states that, as we move towards an infinite distance point in moduli space, a tower of states becomes exponentially light with the geodesic distance in any consistent theory of Quantum Gravity. Although this fact has been tested in large sets of examples, it is fair to say that a bottom-up justification based on fundamental Quantum Gravity principles that explains both the \emph{geodesic} requirement and the \emph{exponential behavior} has been missing so far. In the present paper we address this issue by making use of the Covariant Entropy Bound as applied to the EFT. When applied to backgrounds of the Dynamical Cobordism type in theories with a moduli space, we are able to recover these main features of the SDC. Moreover, this naturally leads to universal lower and upper bounds on the `decay rate' parameter $\lambda_{\text{sp}}$ of the species scale, that we propose as a convex hull condition under the name of Species Scale Distance Conjecture (SSDC). This is in contrast to already proposed universal bounds, that apply to the SDC parameter of the lightest tower. We also extend the analysis to the case in which asymptotically exponential potentials are present, finding a nice interplay with the asymptotic de Sitter conjecture. To test the SSDC, we study the convex hull that encodes the large-moduli dependence of the species scale. In this way, we show that the SSDC is the strongest bound on the species scale exponential rate which is preserved under dimensional reduction and we verify it in M-theory toroidal compactifications. 
		\end{minipage}
	\end{center}
\newpage
%----------------------------------------------------------------------%
%  Resetting of counters
%----------------------------------------------------------------------%
%\setcounter{page}{1}
\pagestyle{empty}
\renewcommand{\thefootnote}{\arabic{footnote}}
\setcounter{footnote}{0}
%---------------------------------------------------------------%
%  Paper begins
%---------------------------------------------------------------%
	
	\tableofcontents

\pagestyle{empty}
\newpage
\setcounter{page}{1}
\pagestyle{plain}
\section{Introduction}
\label{s:intro}

The goal of the Swampland Program (see \cite{Brennan:2017rbf,Palti:2019pca,vanBeest:2021lhn,Grana:2021zvf,Harlow:2022gzl,Agmon:2022thq,VanRiet:2023pnx} for reviews) is to identify which effective field theories (EFTs) may be completed in the ultraviolet (UV) in a way which is consistent with Quantum Gravity (QG). One of the best supported outputs of the program is the \emph{Swampland Distance Conjecture} (SDC)
\cite{Ooguri:2006in,Ooguri:2018wrx,Grimm:2018ohb,Font:2019cxq}. It states that in a theory consistent with QG, when traversing the moduli space, moving from a point $P$ to a point $Q$ located at a \emph{geodesic distance} $\Delta_{\phi}=d(P,Q)$ away, an infinite  tower of states becomes exponentially light (in Planck units) at a rate
\beq
m_{\text{tower}}(Q)\, \sim\, m_{\text{tower}}(P)\, e^{-\lambda \, \Delta_{\phi}}\, ,
\label{SDC}
\eeq
as $\Delta_{\phi}\rightarrow \infty$, with $\lambda$ some order one constant. A further refinement of the SDC, known as the \emph{Emergent String Conjecture} \cite{Lee:2019wij}, claims that along any such infinite distance limit the only possible physics that one can encounter corresponds (in some duality frame) either to a decompactification process, where some internal manifold grows large, or rather emergent string limits, in which some critical weakly coupled and asymptotically tensionless string emerges. Somewhat related to the SDC is the \emph{AdS Distance Conjecture} (ADC) \cite{Lust:2019zwm}, which asserts that for any family of scalar potentials with AdS minima, the limit $V\rightarrow 0$ is always accompanied by an infinite tower of states with characteristic mass scale (in Planck units) given by
\beq
m_{\text{tower}}\, \lesssim\, |V|^\alpha\, ,
\label{ADC}
\eeq
with $\alpha$ again some $\mathcal{O}(1)$ coefficient. A stronger version of the ADC posits  that $\alpha \geq 1/2$ with $\alpha=1/2$ in the supersymmetric case. It has also been argued that a similar condition could also be valid in dS space \cite{Lust:2019zwm}. Currently, there is compelling evidence for the SDC coming from a large variety of string theory vacua (see e.g. \cite{Ooguri:2006in, Blumenhagen:2017cxt, Grimm:2018ohb,Lee:2018urn, Lee:2018spm, Grimm:2018cpv, Buratti:2018xjt,Corvilain:2018lgw,Lee:2019tst, Joshi:2019nzi, Marchesano:2019ifh, Font:2019cxq, Lee:2019xtm,Lee:2019wij, Baume:2019sry, Cecotti:2020rjq, Gendler:2020dfp, Lee:2020gvu,Lanza:2020qmt,Klaewer:2020lfg, Lanza:2021udy, Palti:2021ubp,Etheredge:2022opl}) and in AdS/CFT \cite{Baume:2020dqd,Perlmutter:2020buo,Baume:2023msm}, and in some cases it may be directly connected with the (sub-)Lattice Weak Gravity Conjecture (sLWGC) 
\cite{Heidenreich:2015nta,Heidenreich:2016aqi,Harlow:2022gzl}. Additionally, a good amount of evidence has been gathered in favor of the (weaker version of the) ADC, whilst the stronger one (with $\alpha=1/2$) has been challenged in specific examples, which are still under discussion (see the recent review \cite{VanRiet:2023pnx} and references therein).
 
In spite of this vast evidence in favour of both conjectures, many important questions have not been fully answered yet. In particular, one could raise the following points: ($i$) what is the fundamental QG principle underlying these conjectures, ($ii$) what is the reason for the exponential and power-law functional behaviors and ($iii$) what are the allowed values for the constants $\lambda$ and $\alpha$, and why. Upon closely examining large classes of models, certain ranges of values for $\lambda$ have been proposed, see e.g.
\cite{Grimm:2018ohb,Gendler:2020dfp,Andriot:2020lea,Etheredge:2022opl}. However they are mostly empirical guesses, not based on general QG principles whatsoever. 

A fundamental explanation for the origin of the ADC based on the Covariant Entropy Bound (CEB) 
\cite{Bousso:1999xy,Bousso:1999cb} was proposed in \cite{Castellano:2021mmx}. By exploiting the holographic principle, which implies the existence of a maximum information content in a given spacetime region
\cite{Bekenstein:1980jp,Bekenstein:1993dz,Susskind:1994vu}, one is able to provide a bottom-up derivation for the ADC in equation \eqref{ADC}. This is done by imposing the CEB condition for an EFT in AdS spacetime with an associated UV cut-off, $\Lambda_{\text{UV}}$, which is  naturally identified with the species scale. The entropy considered here is the one associated to the EFT of the massless fields with cut-off $\Lambda_{\text{UV}}$, and in a region with typical volume-to-area ratio given by $L$ (which could be understood as a kind of IR cut-off), which in this case is identified with the AdS length $\ell_{\text{AdS}}$. In this way one is able to derive a lower bound for the constant $\alpha$ depending on the number of dimensions $d$ and the density parameter $p$  of the relevant tower of states.\footnote{The density parameter $p$ is defined  by $m_n=n^{1/p} m_{\text{tower}}$, with $m_{\text{tower}}$ the lightest mass in the tower. Thus a standard  KK tower has $p=1$ and  a string tower is better described by $p \to \infty$, see \cite{Castellano:2021mmx}. In the main text we present explicit examples with $p=2,3$ as well, which play an important  role in M-theory vacua (see section \ref{s:SSDC}).} This lower bound is consistent with all string examples analyzed so far. The explanation  so obtained for the ADC is intuitively simple  and is based on perhaps the  most characteristic property of Quantum Gravity, namely Holography. Note that supersymmetry is not needed for the derivation, contrary to other schemes proposed in the literature (see e.g. \cite{Montero:2022ghl, Cribiori:2022trc}).

One could expect somewhat similar entropic arguments to apply for the SDC. However, in flat space there is no characteristic length scale like in the AdS case for the IR cut-off. To apply such an entropic argument based on the degrees of freedom of the EFT,  a link between the field-space distance, $\Delta_{\phi}$, and the characteristic spacetime distance associated to the region under consideration, $L$, is needed. We argue in this work that the idea of {\it Dynamical Cobordism} \cite{Buratti:2021yia,Buratti:2021fiv,Angius:2022aeq,Blumenhagen:2022mqw,Angius:2022mgh,Blumenhagen:2023abk,Angius:2023xtu,Huertas:2023syg} nicely completes this scheme providing such a link between field-space and spacetime lengths. The Cobordism Conjecture \cite{McNamara:2019rup} states that any QG vacuum admits, at the topological level, a boundary ending spacetime. The dynamical cobordism idea tries to explore the dynamics as one approaches these boundaries. In the presence of scalar fields, these are described as codimension-one running solutions in which the scalars explore infinite field distance as one approaches the end-of-the-world (ETW) singularity. In string theory examples, a remarkable exponential dependence between the spatial distance to the ETW-singularity and the field space distance was found \cite{Buratti:2021fiv,Angius:2022aeq}. We analyze this kind of codimension-one running solutions for theories with exactly massless scalars, paying special attention to the region very far away from the ETW-singularity, as it is the region in which our CEB argument is applied. Remarkably, such a background exists for any geodesic in moduli space, and there is a universal relation between spacetime and (geodesic) field space distance: $L\sim e^{\sqrt{\frac{d-1}{d-2}} \, \Delta_\phi}$.

In this paper, we show how combining the CEB with the dynamical cobordism input one can provide a bottom-up argument for the SDC and its known properties.  In particular, we are able to recover the  asymptotic exponential behavior of the UV cut-off (and hence the mass scale of the tower) with the \emph{geodesic} distance in moduli space, and to give specific lower and upper limits for the exponential decay rates depending on the dimension (and the density of states  in the tower, $p$). Moreover, we also extend the  dynamical cobordism analysis to a situation in which  (both positive or negative) exponential potentials are present. The results obtained reproduce the previous analysis for the AdS conjecture and give rise to new bounds 
on the decay rate of the exponential potentials both in AdS and dS. In particular we can reproduce the TCC lower bound  for the scalar potential rate.

In the presence of multiple moduli and towers, the SDC becomes more involved and it is useful to define a Convex Hull of scalar charges \cite{Calderon-Infante:2020dhm}, in analogy with the case of multiple $U(1)$’s in the WGC \cite{Cheung:2014vva}. The points that define this hull are the analogues of the charge-to-mass ratios in the WGC, namely the log derivatives of the masses of the towers of states.  Based on the scalar WGC  (sWGC) \cite{Palti:2017elp}, it was argued that this convex hull should contain the ball in which the sWGC would be violated. Recently, a universal lower bound $1/\sqrt{d-2}$ for these decay rates (i.e. the radius of this ball) has been proposed in \cite{Etheredge:2022opl}.

We revisit the issue of the convex hull related to the SDC  from the point of view of the CEB. The first point to remark is that, unlike other analysis in the literature, the very existence of this convex hull constraint appears automatically and does not rely on the sWGC. Furthermore, we put the main focus on the species scale \cite{Dvali:2007hz,Dvali:2007wp}, instead of the mass of the lightest tower. In this regard, we point out that it is particularly useful to define the convex hull for the species scale itself, rather than for the masses of the towers, because it is well-suited for situations where multiple towers lie below the UV cut-off of the theory and effectively give rise to combined towers \cite{Castellano:2021mmx}. Furthermore, it efficiently encodes the relevant UV physics of the different infinite distance limits. Thus, we define a {\it Species Scale Distance Conjecture} (SSDC) in terms of the convex hull obtained from the species scale for any given theory. We find that the decay rate for the species scale has an absolute lower bound 
\beq
 \lambda_{\text{sp}}\, \geq\, 
\frac {1}{\sqrt{(d-1)(d-2)}}\ ,
\eeq
which is based on the CEB and the dynamical cobordism idea, with no reference to the sWGC. We also test that the existence of this lower bound is stable under dimensional reduction, although to obtain consistency with the convex hull conditions requires  in general the existence of additional towers, including e.g. `effective towers’ with higher density ($p>1$). When working out specific string vacua, the required towers to complete the convex hulls do indeed appear in a non-trivial way.

We test and illustrate the validity of our convex hull bounds in a number of M-theory/string theory vacua. The case of M-theory on $T^2$ is illustrative in that it contains Kaluza-Klein (KK), string and winding types of towers, and combined effective towers already play a crucial role in the fulfillment of the convex hull condition. The lower bound is saturated by KK-like towers associated to a single dimension decompactifying (i.e. $p=1$) whereas the vertices of the convex hull correspond to string towers and an effective KK-like tower associated to full decompactification of the internal space (i.e. $p=2$). We also work out more elaborated cases, such as M-theory on $T^3$, in which similar patterns arise. Namely, the string oscillator modes and the effective highest density towers corresponding to full decompactification ($p=3$) stay at the vertices,  the single KK-like towers ($p=1$) lie at the faces (saturating the bound) and the edges joining vertices in the convex hull contain the intermediate density KK-like towers ($p=2$).

It is remarkable that all this rich structure, including specific numerical bounds and exponential behavior at large moduli may be described in simple terms on the basis of two fundamental principles of Quantum Gravity: Holography as formulated by the CEB and the Cobordism Conjecture, in its dynamical implementation. Interestingly, this analysis does not require supersymmetry. We believe that the present work gives support to the idea that all distance conjectures in QG are based, one way or another, on general holographic principles. In particular, the Cobordism Conjecture is intrinsically motivated by the absence of global symmetries and the latter may be justified in the context of AdS/CFT Holography \cite{Harlow:2018tng}. So all in all, it may be argued that the Distance Conjectures are deeply rooted in Holography.

The structure of the rest of this paper is as follows. In the next section we first revisit the results from \cite{Castellano:2021mmx}, including some new insights, and then show how an analogous reasoning based on the CEB combined with the dynamical cobordism setup allows us to motivate the SDC from the bottom-up, providing also for non-trivial constraints on the decay rates $\lambda_{\text{sp}}$. The analysis is extended to the case in which an exponentially damped potential is present in section \ref{s:potential}, obtaining the known bounds on $\alpha$ for the ADC but also additional ones which apply to dS runaway potentials, making contact with the asymptotic dS conjecture. In section \ref{s:SSDC} we define the Species Scale Distance Conjecture as a convex hull condition on the asymptotic decay rates of the species-scale. We illustrate the validity of this constraint in certain toroidal M-theory compactification, in which all the new crucial ingredients are exemplified. In sections \ref{s:dimensionalreduction}
we study the behavior of the SSDC under dimensional reduction from different perspectives. Finally, in section \ref{s:Mthycompactifications} we verify the SSDC in further toroidal M-theory compactifications. We leave section \ref{s:conclusions} for general comments and conclusions. In the appendix we provide a proof of principle for the UV uplift of the running solutions described in section \ref{s:SDCbottomup}. We do so by means of a specific string theory background interpreted as a non-compact orbifold. This may suggest certain degree of generality as captured by the dynamical cobordism approach, which seems to provide a \emph{universal way} of approaching the boundaries of our QG EFTs.

\section{The Covariant Entropy Bound and Distance Conjectures}
\label{s:SDCbottomup}	

In this section we present our entropic argument for the Distance Conjecture. For this, we first review the Covariant Entropy Bound (CEB) as applied to effective field theories, and how it recovers the AdS Distance Conjecture (ADC) when considered for a family of AdS vacua. With this motivation in mind, we study dynamical cobordism solutions in theories with massless scalars, which give the link between the spacetime and moduli space distances required to recover the SDC. Finally, applying the CEB to these backgrounds, we show how to recover all these features of the SDC from the bottom-up.

\subsection{Holography and UV/IR Mixing}
\label{ss:ADCreview}

Let us start by reviewing how the holographic principle, formulated in the form of a maximal information content associated to a given spacetime region, can in principle lead to interesting constraints for the effective field theories (EFTs) that we use to describe the low energy physics taking place within the said region. These ideas were in fact exploited in \cite{Castellano:2021mmx} so as to provide a bottom-up rationale for the Anti-de Sitter Distance Conjecture (ADC) \cite{Lust:2019zwm}. In the following, we will briefly summarize the line of reasoning followed there, and our aim in this section is to extend such considerations so as to motivate the Swampland Distance Conjecture from this perspective.

Suppose that we restrict ourselves to some codimension-one spacelike region $\Sigma$ of a $d$-dimensional spacetime satisfying the constraint that its future-directed light-sheets are complete. This means, in turn, that the causal past associated to these light rays contains the full spatial region we are interested in. For concreteness, one could think of $(d-1)$-dimensional spheres of proper radius $L$. Then, upon applying the Covariant Entropy Bound (CEB) \cite{Bousso:1999xy}, one thus finds that the information content in that region presents a maximal upper bound determined by the area of its boundary, $A(\partial \Sigma)$, as follows
\beq \label{eq:BHentropy}
	S_{\text{max}} = \frac{A(\partial \Sigma)}{4 G_d}\, ,
\eeq
where $G_d$ is the $d$-dimensional gravitational constant. For the canonical example of spatial balls of proper radius $L$ in flat space, the holographic entropy becomes  essentially proportional to $(L/\ell_d)^{d-2}$, with $\ell_d$ the $d$-dimensional Planck length.

The idea is then to use this Quantum Gravity (QG) restriction to constrain the range of validity of any EFT that we use to describe the low energy physics taking place in $\Sigma$. Therefore, if we assume to have such a \emph{local} effective description of the massless/light modes of our theory, the \emph{maximum} field theory entropy associated to $\Sigma$ would scale extensively as follows
\beq \label{eq:EFTentropy}
	S_{\text{EFT}} \sim N_0\, \Lambda_{\text{UV}}^{d-1}\, \text{vol}(\Sigma) \, , %\sim N_0 \left(\Lambda_{\text{UV}}\, L\right)^{d-1}\, ,
\eeq
where we have introduced some UV cut-off, $\Lambda_{\text{UV}}$, beyond which our effective description breaks down, whilst $N_0$ denotes the number of massless/light fields described by the EFT, which we will henceforth consider to be of order one for simplicity. Hence, upon imposing the holographic upper bound on the entropy for the region $\Sigma$, \eqref{eq:BHentropy}, one finds the following QG upper bound for the UV cut-off in terms of the area-to-volume ratio of the region
\beq\label{eq:Boussobound}
\left(\Lambda_{\text{UV}}\, \ell_d\right)^{d-1} \, \lesssim \, \frac{A(\partial \Sigma)\  }{\text{vol}(\Sigma)}\,  \ell_d \, .
\eeq
Note that particularizing to the case of a $(d-1)$ spatial ball of proper radius $L$ in flat space, and identifying the IR cut-off simply as the inverse of the maximum length scale set by the system under consideration (namely $\Lambda_{\text{IR}} = L^{-1}$), we get the UV/IR relation $\left(\Lambda_{\text{UV}}\ \right)^{d-1} \lesssim \Lambda_{\text{IR}}$ (in Planck units). The moral is that as the area-to-volume ratio decreases (in Planck units), the UV cut-off, $\Lambda_{\text{UV}}$, has to decrease in order not to surpass the maximal holographic entropy \eqref{eq:BHentropy}. Let us stress that, even though having an UV cutoff that depends on the size of the experiment might seem a bit shocking from the Wilsonian perspective, it is to be interpreted as a form of UV/IR mixing, something pervasive in Quantum Gravity.

An alternative strategy followed in \cite{Cohen:1998zx, Banks:2019arz,Cohen:2021zzr} was to consider the collapse of thermodynamic configurations into black holes (BHs) and requiring them not to be part of the EFT. This yields similar (although asymptotically stronger) bounds for the ADC \cite{Castellano:2021mmx}.  We will not contemplate this possibility any further in this work, since our strategy is to confront the extensive entropy characteristic of a local field theory, which would be obtained in the absence of gravity, with the CEB.  Moreover, the results obtained using this approach give strong support (\emph{a posteriori}) that it is a sensible thing to do.

Let us also mention an important point regarding the entropy that we are considering throughout this work. One of our main goals is to use entropy bounds in order to obtain constraints on EFTs, by confronting the (extensive) EFT entropy with the Quantum Gravity bounds. In particular, the state counting that accounts for this entropy is due to the phase space available upon considering potential finite temperature configurations in the presence of a UV cut-off, as opposed to the zero-temperature entropy associated to e.g. the counting of the number of species or the microstate counting of extremal black holes studied in \cite{Cribiori:2022nke, vandeHeisteeg:2023ubh,vandeHeisteeg:2023uxj,Cribiori:2023ffn}. Of course, both contributions come into play whenever one wants to account for all the entropy of a given configuration, but in our case the former is the one that becomes crucial to obtain the constraints.\footnote{In fact, one can recover some zero-temperature entropy results from our analysis if the UV cut-off is set to $L^{-1}$, with $L$ the typical size of the region under consideration, since in such configurations the different species present are \emph{frozen} and all their entropy comes just from the counting of states.}  

Given that equation \eqref{eq:Boussobound} is a quantum-gravitational constraint, it is natural to think that $\Lambda_{\text{UV}}$ must be some intrinsically quantum-gravitational cut-off. Indeed, from our experience in string theory, we know that
the appearance of an infinite tower of states becoming light  seems to be the natural way in which the UV cut-off of our description may be lowered. This is indeed the usual mechanism by which QG censors pathological limits in familiar EFTs coupled to gravity in a consistent manner. Therefore, it is natural to identify $\Lambda_{\text{UV}}$ with the highest cut-off scale for a Quantum Gravity theory, namely the species scale, which is defined as follows \cite{Dvali:2007hz,Dvali:2007wp}
\beq 
\Lambda_{\text{sp}}\ \simeq \ \frac {M_{\text{pl}, d}}{N_{\text{sp}}^{\frac{1}{d-2}}} \ ,
\label{eq:speciesscale}
\eeq
with $N_{\text{sp}}$ the number of light species and $M_{\text{pl},d} \propto \ell_d^{-1}$ the $d$-dimensional Planck mass. Thus, let us consider a system with a tower of states parametrized by $m_{n}^p = n\, m_{\text{tower}}^{p}$, such that the species scale reads
\beq
\Lambda_{\text{sp}}^p\, \simeq\,  N_{\text{sp}}\, m_{\text{tower}}^p\, ,
\eeq 
where $m_{\text{tower}}$ denotes the mass scale of the tower. Notice that one indeed recovers the case of a single Kaluza-Klein tower for $p=1$, whereas $p \to \infty$ would instead correspond to a string tower (modulo log corrections, see \cite{Castellano:2022bvr}).\footnote{As explained in \cite{Castellano:2021mmx}, even though naively a critical string tower would seem to have $p=2$, namely $m_n= \sqrt{n} m_s$, taking into account the exponential degeneracy of states amounts, for our pruposes here, to regard it as an effective tower of density parameter $p\to \infty$. For example, its associated species scale becomes then identical to the string mass scale (c.f. eq. \eqref{eq:LambdaQGmtower}), which is in fact the correct result.} In the case of multiple towers, several of them can contribute to the species scale, and the computation becomes more involved. However, it turns out that if several different towers contribute to the species scale one can construct an equivalent \emph{effective tower}, with an effective mass scale and an effective density parameter that captures the right asymptotic behavior of both the number of total species and the species scale itself \cite{Castellano:2021mmx}. In particular, for a set of towers with mass scales $m_i$, and density parameters $p_i$, the effective mass $m_{\text{tower}}$ is in fact a geometric average of the masses of the towers involved (see section \ref{ss:convexhullspecies}) and  the effective $p_{\text{eff}}=\sum_i p_i>1$.

These expressions, when combined together, allow one to write $\Lambda_{\text{sp}}$ solely in terms of the mass scale of the tower
\beq
\Lambda_{\text{sp}}\, \sim\, m_{\text{tower}}^{\frac{p}{d-2+p}}\, \Mpd^{\frac{d-2}{d-2+p}}\, .
\label{eq:LambdaQGmtower}
\eeq
Inserting the above relation into the holographic bound \eqref{eq:Boussobound}  one finds \cite{Castellano:2021mmx}
\beq \label{eq:alpha_d}
m_{\text{tower}}\, \lesssim\, \left( \frac{A(\partial \Sigma)}{\text{vol}(\Sigma)} \right)^{2\alpha_d}\, , \qquad \alpha_d\, =\, \frac {(d-2+p)}{2p(d-1)} \ .
\eeq
Upon identifying the IR cut-off as $\Lambda_{\text{IR}}= A(\partial \Sigma)/\text{vol}(\Sigma)\, $, this bound can be rewritten (in Planck units) as
\beq
\Lambda_{\text{UV}}^{d-1} \, \lesssim \, \Lambda_{\text{IR}}\, ,
\eeq
which shows how the tower scale decreases as the IR cut-off goes down in a very precise way. With this we can now turn to two interesting applications of these ideas in the context of the Swampland Program.

\subsubsection{A Bottom-up Argument for the ADC}
\label{sec:ADC}

Let us first review how the application of the holographic ideas from the previous section may provide a bottom-up rationale for the AdS Distance Conjecture, as explained in \cite{Castellano:2021mmx}. We take $d$-dimensional Anti-de Sitter as our background spacetime, and parametrize it in terms of so-called Susskind-Witten\cite{Susskind:1998dq} coordinates, by which the spacetime can be viewed locally as a product of a $(d-1)$-dimensional spatial ball and an infinite time axis\footnote{\label{fnote:globalAdS}These coordinates can be seen to be related to the usual AdS$_d$ global coordinates, in which the metric reads as $ds_d^2= \ell_{\text{AdS}}^2 \left( -\cosh{\rho}^2 dt^2 + d \rho^2+\sinh{\rho}^2 d\Omega_{d-2}^2\right)$, via the change of radial coordinate $r=\tanh{\rho/2}$.}
\beq \label{eq:SWcoords}
	ds_d^2= \ell_{\text{AdS}}^2 \left[ -\left( \frac{1+r^2}{1-r^2}\right)^2 dt^2 + \frac{4}{\left( 1-r^2\right)^2} \left( dr^2+r^2 d\Omega_{d-2}^2\right)\right]\, ,
\eeq
where $\ell_{\text{AdS}}$ denotes the AdS length, $d\Omega_{d-2}^2$ is the distance element of the unit $(d-2)$ sphere and  $0 \leq r<1$. The conformal boundary lives at $r=1$.

%%%%%%%%%%%
\begin{figure}[tb]
\begin{center}
\includegraphics[width=0.6\textwidth]{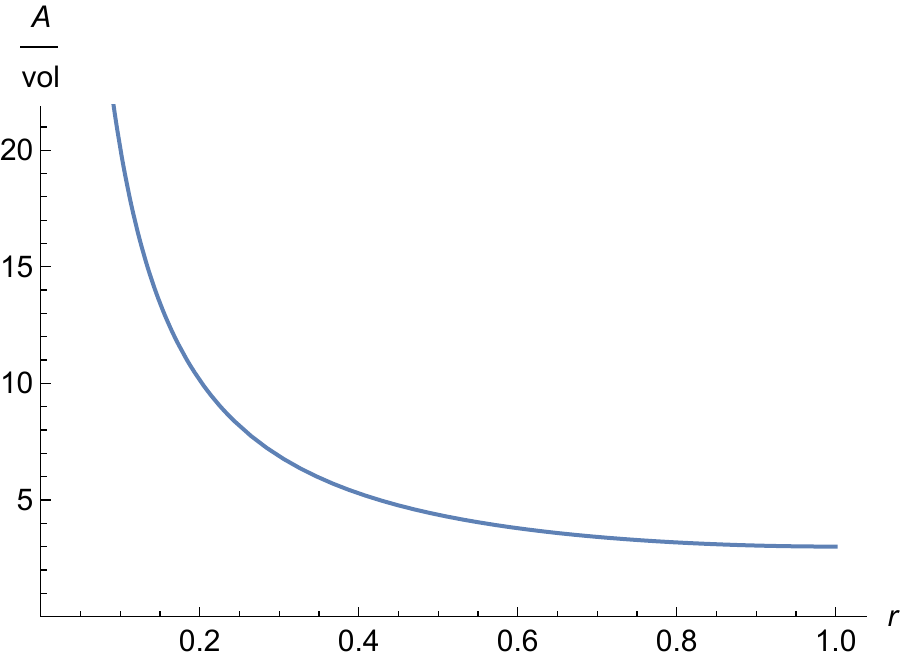}
\caption{\small Area-to-Volume ratio (with $\ell_{\text{AdS}}=1$) as a function of the distance from the boundary of AdS, located at $r=1$. For concreteness, we show here the dependence in five-dimensional Anti-de Sitter space. Notice that the minimum happens precisely when we take our ball to occupy all space.} 
\label{fig:Area/VolAdS}
\end{center}
\end{figure}
%%%%%%%%%%%

We now assume that the low energy physics taking place in AdS$_d$ admits a description in terms of a $d$-dimensional EFT weakly coupled to Einstein gravity up to some UV cut-off, $\Lambda_{\text{UV}}$, signalling the energy scale at which the effective local description starts to fail due to inherently quantum-gravitational effects. We consider a $(d-1)$ spatial ball (described by constant $r$ in Susskind-Witten coordinates, c.f. equation \eqref{eq:SWcoords}), with proper length $L$. Upon imposing the CEB --- in its spacelike projected version\cite{Bousso:2002ju} --- on the extensive EFT entropy given in \eqref{eq:EFTentropy}  one arrives at the following QG requirement
\beq \label{eq:CEB}
	\Lambda_{\text{UV}}^{d-1} \ell_d^{d-2} \lesssim \frac{A(L)}{\text{vol}(L)}\, .
\eeq
One can show that indeed such condition gives rise to constraints of the form
\beq \label{eq:CEB2}
	\left(\Lambda_{\text{UV}}\, \ell_d\right)^{d-1} \lesssim \left( \frac{\ell_d}{\ell_{\text{AdS}}}\right)\, ,
\eeq
for any $L \geq \ell_{\text{AdS}}\, $, since the quotient on the RHS of equation \eqref{eq:CEB} reduces to $1/\ell_{\text{AdS}}\, $, up to $\mathcal{O} (1)$ factors (see figure \ref{fig:Area/VolAdS}). Notice that this relation morally captures the content of the AdS Distance Conjecture, since the limit of large AdS radius (in Planck units) implies a vanishing UV cut-off. Indeed, upon taking the flat space limit $\ell_{\text{AdS}} \to \infty$, equation \eqref{eq:CEB2} tells us that the UV cut-off of the theory must go to zero, effectively invalidating our original description if we nevertheless insist on keeping the UV cut-off of our EFT fixed. 

Notice that this does not require from extra ingredients such as supersymmetry to work (as opposed to other arguments in the literature, see e.g. \cite{Montero:2022ghl, Cribiori:2022trc}), and directly spells what goes wrong if the EFT does not break upon taking the flat space limit: The absolute upper bound for the QG entropy is violated. It is important to remark, though, that the argument above only applies when one can somehow approach the flat space limit within the family of AdS vacua under consideration (perhaps in a discrete fashion via tuning some flux quanta), as oppossed to starting with a Minkowski vacuum directly. This is precisely the content of the ADC, since it does not prevent Minkowski vacua to exist in the first place, but rather censors a `smooth' interpolation between AdS and Minkowski within the same EFT. The reason is quantum-gravitational at its core, namely holographic entropies in flat space and AdS --- even though presumably infinite --- are very much distinct from each other due to the different asymptotics exhibited by these spacetimes.

Hence, upon identifying the QG cut-off with the species scale associated to a tower of light states,\footnote{One can indeed check that equation \eqref{eq:CEB2} with $\Lambda_{\text{UV}}$ taken to be the species scale is satisfied in $AdS_p \times S^q$ compactifications of type IIB string theory and M-theory\cite{Castellano:2021mmx}.}  eqs. \eqref{eq:LambdaQGmtower} and \eqref{eq:CEB2} yield the following result for the mass scale of such tower
\beq \label{eq:towerscale}
	m_{\text{tower}} \sim |\Lambda_{\text{c.c.}}|^{\alpha}\, ,
\eeq
with $\alpha\geq \alpha_d\, =\, \frac {(d-2+p)}{2p(d-1)}$. In general, this lower bound for the exponent varies in the range $1/2(d-1)\leq \alpha_d\leq 1/2$, which is consistent with all known examples in string theory. The case $p=1$, of an ordinary KK-like tower, yields $\alpha_d=1/2$ and does not allow scale separation. However, the constraint leaves room for scale separation for $p>1$, and the allowed range for $\alpha$ is maximized when the tower is of stringy type ($p \to \infty$).

Finally, the upper bound $\alpha\leq (d+p-2)/2p \leq (d-1)/2$ obtained in \cite{Castellano:2021mmx} can be understood as the requirement that the EFT as such makes sense, namely $\Lambda_{\text{UV}}\geq \Lambda_{\text{IR}}$. Notice that identifying this UV cut-off with the species scale, this requirement simply means that we start with an EFT weakly coupled to Einstein Gravity, which is precisely the kind of EFT we are interested in constraining.\footnote{In certain contexts, one can use AdS/CFT techniques to constrain even more general QG EFTs, such as Vasiliev-type (see e.g. \cite{Baume:2020dqd,Perlmutter:2020buo, Baume:2023msm}), but this is beyond the scope of this work.} In particular, note that this includes AdS vacua without scale separation (with $\alpha=1/2$), such as $\text{AdS}_5 \times S^5$, since the UV cut-off given by the species scale is always above that of the KK-scale.

\subsection{A Bottom-up Argument for the SDC}
\label{ss:bottom-upSDC}

In the case of an AdS background, there is a natural length scale, given by the AdS radius, which can be related to the UV cut-off upon imposing the CEB within a box larger than the AdS scale. A reasonable question is whether a similar logic may be applied in the absence of such non-vanishing cosmological constant, since this is the relevant scenario to investigate the Swampland Distance Conjecture \cite{Ooguri:2006in}. To do so, the key point is then to find some generic field configuration that links the spacetime distance with the moduli space distance, so that we can consequently relate the latter with the CEB.

\subsubsection{Running Solutions in Moduli Space}
\label{sss:runningsol}

In this section we introduce the EFT running solutions that precisely realize this dependence of the spacetime distance on the moduli space distance. In particular, the main result we find is the following exponential dependence of the  spacetime distance 
\beq
L\, \sim\, e^{\frac{\delta}{2}\,\Delta_{\phi}}\, , \qquad \text{with} \qquad \delta=2 \sqrt{\dfrac{d-1}{d-2}}\, .
\label{eq:mapdistancefields}
\eeq
These running solutions are asymptotically flat and explore geodesics in moduli space, as required by the Distance Conjecture, with the exponential coefficient depending just on the number of spacetime dimensions, $d$. In fact, there is one such running solution for each locally geodesic trajectory in moduli space. They are completely generic at the level of the EFT, and the only assumption is that the associated spacetime singularity is resolved in QG, as suggested by the Cobordism Distance Conjecture \cite{Buratti:2021yia,Buratti:2021fiv,Angius:2022aeq}. 
In the remainder of this section we explain how to obtain such solutions and discuss the physics behind them. Even though several important insights can be extracted from this discussion, the reader interested solely in the relation between the Distance Conjecture and the CEB can safely take equation \eqref{eq:mapdistancefields} and jump directly to section \ref{sss:entropyargument}.

Let us consider $d$-dimensional Einstein gravity with a moduli space $\mathcal{M}$ parametrized by scalars ${\phi^{i}}$. The action is given by
\begin{equation}\label{eq:EFT}
	S = \frac{1}{\kappa_d^2} \int d^{d}x\, \sqrt{-g}\,  \left[ \frac{1}{2}R - \frac{1}{2} G_{ij} (\phi) \partial_{\mu} \phi^{i} \partial^{\mu} \phi^{j}\right] \, . 
\end{equation}
Notice that this action is tailored to include the minimal ingredients required to formulate the SDC, i.e., gravity coupled to exactly massless scalars with no potential (we leave the inclusion of a non-vanishing potential for section \ref{s:potential}). In principle, the EFT could contain extra matter or $p$-form gauge fields, as typically happens in string theory constructions, whose gauge couplings are usually determined precisely by the scalar field vevs $\braket{\phi^{i}}$. In any event, by restricting to \emph{uncharged} solutions, namely those with field strength $F_p^2=0$ everywhere in spacetime, one can effectively use the truncated action \eqref{eq:EFT}. Furthermore, for this bottom-up approach we remain agnostic about the mechanism protecting the scalars from getting a mass through quantum corrections, and we just assume the existence of a moduli space. Typically, this would require the presence of some degree of supersymmetry and, again, \eqref{eq:EFT} would be a consistent truncation to the relevant sector of the theory.

Our goal is to obtain running solutions, in which the moduli-space and spacetime distances are linked dynamically. For that, it is natural to consider $(d-1)$-Poincar\'{e} preserving solutions, i.e., a domain-wall ansatz of the form
\begin{align} \label{eq:metricprofile}
	ds^{2} &= e^{2\sigma(r)} \eta_{\alpha \beta} dx^{\alpha} dx^{\beta} + dr^{2} \, ,\\
    \phi^{i} &= \phi^{i}(r) \, .
\end{align}
Notice that the second equation is that of a trajectory in moduli space, parametrized by the spacetime distance in the $r$-direction. This will be of relevance to our analysis, since it will allow us to determine dynamically not only the relation between the moduli space and spacetime distances, but also the trajectory that the solution is exploring in moduli space.

Plugging this ansatz into the equations of motion of \eqref{eq:EFT} yields: 
\begin{align}
	& \phi^{\prime\prime\, k} + \Gamma^{k}_{ij} \phi^{\prime\, i} \phi^{\prime\, j} + (d-1) \sigma^{\prime} \phi^{\prime\, k} = 0 \label{eq:scalars} \, , \\
    & (d-2)(d-1) \sigma^{\prime \, 2} = G_{ij} \phi^{\prime\, i} \phi^{\prime\, j} \, , \label{eq:Einstein-1}  \\
    & \sigma^{\prime \prime} + (d-1) \sigma^{\prime \, 2} = 0 \label{eq:Einstein-2} \, .
\end{align}
Here we have used primes to denote derivation with respect to $r$ and $\Gamma^{k}_{ij}$ is the connection associated to the moduli space metric. Equations \eqref{eq:Einstein-1} and \eqref{eq:Einstein-2} come from the Einstein's equations, while \eqref{eq:scalars} are the equations of motion for the scalars.

These equations have been written in a suggestive way. The last one, equation \eqref{eq:Einstein-2}, does not contain the scalars and completely determines the spacetime metric via the warp factor $\sigma(r)$. Recalling the definition of the line-element used to compute the moduli space distance in terms of the spacetime one,
\begin{equation} \label{eq:line-element}
    \Delta_{\phi}^{\prime\,2}=G_{ij} \phi^{\prime\, i} \phi^{\prime\, j} \, ,
\end{equation}
equation \eqref{eq:Einstein-1} then relates this to the spacetime metric. This is, having $\sigma(r)$, equation \eqref{eq:Einstein-1} allows us to solve for $\Delta_{\phi}(r)$. Finally, equation \eqref{eq:scalars} is then a constraint on the trajectory that is explored as we move along the $r$-direction.

Following this reasoning, let us first solve \eqref{eq:Einstein-2}. The solution takes the form
\begin{equation} \label{eq:warp-factor}
    \sigma(r) = \frac{1}{d-1} \log \left( \frac{r - r_\star}{r_0} \right) \, ,
\end{equation}
where $r_\star$ and $r_0$ are the two integration constants. We can set $r_\star =0$ without loss of generality by a shift in the $r$-coordinate, which does not change the ansatz \eqref{eq:metricprofile}. On the other hand, fixing $r_0$ amounts to choosing some units to measure spacetime distances. We will keep it explicitly here, although it has no effect whatsoever in our conclusions (see footnote \ref{fnote:r0}). From now on we work in Planck units and take
\begin{equation} \label{solution-sigma}
    \sigma(r) = \frac{1}{d-1} \log \left( \frac{r}{r_0} \right) \, .
\end{equation}

As discussed before, this completely fixes the spacetime metric. However, let us postpone the analysis of its properties and, for the moment, take this as an intermediate step needed to get $\Delta_{\phi}(r)$, which is our main focus.

Plugging \eqref{eq:warp-factor} and \eqref{eq:line-element} into \eqref{eq:Einstein-1} and solving the differential equation we get
\begin{equation} \label{eq:solution-Delta}
    \Delta_{\phi}(r) = \pm \sqrt{\frac{d-2}{d-1}} \log\left( r\right) \, ,
\end{equation}
where we have fixed an additive integration constant without loss of generality. In this case, it corresponds to setting the zero of the moduli space distance, this is, with respect to which point we measure it. In addition, the plus-minus sign indicates that we can choose to measure the distance as we move toward smaller or larger values of $r$. Let us remark that this equation will be key for our bottom-up rationale for the SDC. Just as we wanted, it relates dynamically the spacetime to the moduli space distances in an exponential fashion. Moreover, the exponential rate is fixed by the number of spacetime dimensions, as announced in equation \eqref{eq:mapdistancefields}.

We are then left with the constraint on the moduli space trajectory imposed by equation \eqref{eq:scalars}. One can see that it takes the form of a geodesic equation in moduli space in a non-affine parametrization $\phi^{i}(r)$. Therefore, the trajectory is constrained to be a local geodesic. To see this more clearly, let us write the trajectory in the proper-length parametrization, $\phi^k(\Delta_{\phi})$. Performing a reparametrization $\Delta_{\phi}(r)$ to equation \eqref{eq:scalars}, we then get to
\begin{equation}
    \Delta_{\phi}^{\prime \, 2} \left(\Ddot{\phi}^k + \Gamma^{k}_{ij} \dot\phi^{i} \dot\phi^{j}\right) + \left( \Delta_{\phi}^{\prime\prime} + (d-1) \Delta_{\phi}^{\prime} \sigma^{\prime} \right) \dot\phi^k = 0 \, ,
\end{equation}
where the dot denotes derivation with respect to $\Delta_{\phi}$. One can readily check that the last term vanishes for $\sigma(r)$ and $\Delta_{\phi}(r)$ given in equations \eqref{solution-sigma} and \eqref{eq:solution-Delta}, respectively. This is necessary for $\Delta_{\phi}(r)$ to be a consistent reparametrization to the proper-length, and can be interpreted as a compatibility condition between equations \eqref{eq:Einstein-1}, \eqref{eq:Einstein-2} and \eqref{eq:scalars}. After taking this and $\Delta_{\phi}^\prime \neq 0$ into account we end up with
\begin{equation}\label{eq:geodesic}
    \Ddot{\phi}^k + \Gamma^{k}_{ij} \dot\phi^{i} \dot\phi^{j} = 0 \, ,
\end{equation}
which is the geodesic equation for $\phi^k(\Delta_{\phi})$. We thus conclude that any local geodesic in moduli space can be explored by one of these running solutions. This is very suggestive from the SDC point of view, since it indeed applies to any geodesic reaching infinite distance in moduli space. Furthermore, the relation between spacetime and moduli space distances is universal to all such geodesic paths and given by \eqref{eq:solution-Delta}. This will later translate into having \emph{universal} upper and lower bounds on the exponential rate of the UV cut-off.

Finally, let us study the properties of the spacetime associated to these running solutions. For that, let us compute the Ricci scalar as
\begin{equation} \label{curvature}
	R = - (d-1) \left(2\sigma^{\prime \prime} + d \, \sigma^{\prime\,2} \right) =\frac{d-2}{d-1}\frac{1}{r^2} \, .
\end{equation}
First, we see that it goes to zero as $r\to \infty$, thus suggesting the fact that the spacetime is asymptotically flat in this limit (which can be checked by various methods). On the contrary, the scalar curvature blows up as $r\to0$, i.e., there is a singularity at that position in spacetime. Even though this may look dangerous for the validity of the solution, we now argue that this is an end-of-the-world (ETW) singularity as those appearing in the context of dynamical cobordisms to nothing \cite{Buratti:2021yia,Buratti:2021fiv,Angius:2022aeq,Blumenhagen:2022mqw,Angius:2022mgh,Blumenhagen:2023abk,Angius:2023xtu,Huertas:2023syg}. Indeed, given \eqref{eq:solution-Delta} and \eqref{curvature}, we get
\begin{equation}
\begin{split}
    |R| \sim e^{2\sqrt{\frac{d-1}{d-2}} \, \Delta_{\phi} } \, , \qquad r \sim e^{-\sqrt{\frac{d-1}{d-2}} \, \Delta_{\phi}} \, , \qquad \text{as } \Delta_{\phi}\to\infty \, ,
\end{split}
\end{equation}
where we have chosen the minus sign in \eqref{eq:solution-Delta} to measure the moduli space distance as the singularity is approached. These equations match the scaling relations put forward in \cite{Buratti:2021fiv,Angius:2022aeq}, with the critical exponent $\delta$ defined there being given by $\delta = \sqrt{(d-1)/(d-2)}$. These EFT solutions in which spacetime is capped off at a singularity of this type are expected to describe dynamical cobordisms to nothing, and the singularity should be resolved in the UV complete Quantum Gravity theory. We will thus assume these solutions to be physically meaningful. Furthermore, while the small $r$ region should receive strong corrections coming from higher-curvature terms in the EFT action that depend on its UV completion, we can trust the solution for large $r$, that is, if we work very far away from the ETW-singularity.

In fact, these solutions are nothing but the EFT realization of the Cobordism Distance Conjecture. Indeed, we found a set of solutions with ETW-branes exploring any infinite distance limit in moduli space. This is necessary for the conjecture to hold in theories exhibiting a moduli space. Of course, the most non-trivial part of the conjecture is the statement that such ETW-singularities are resolved in the UV complete theory of Quantum Gravity. This is the part that we will be assuming, and that remains to be addressed in generality so as to verify the Cobordism Distance Conjecture. 

As a proof of principle, we present in Appendix \ref{ap:runningKK} a particular example of UV uplift of this running solution in $d$ spacetime dimensions, and how the ETW-singularity gets resolved in the new frame. By assuming that the running scalar field corresponds to the radius of an extra dimension, we find that the solution gets uplifted to a $(d+1)$-dimensional theory with generic deficit angle, encoded by some of the integration constants in the running solution. This includes (non-compact) orbifold backgrounds which are known to be consistent within string theory\cite{Bagger:1986wa, Kachru:1995sj}, or even regular $(d+1)$-dimensional flat space. Interestingly, in this latter case the UV resolution of the ETW-singularity turns out to be purely geometrical and does not require from the presence of a localized object with the right properties to seal off the singularity. This mechanism is then complementary to that studied in \cite{Blumenhagen:2023abk}.

As a summary, we find running solutions exploring infinite distance in moduli space at asymptotic regions of spacetime. At the level of the EFT, these solutions exist for any geodesic and the relation between moduli space and spacetime distances, $\Delta_{\phi}(r)$, turns out to be universal. In the next section, we will use the CEB bound, as applied to EFTs, to provide a bottom-up rationale for the SDC.

\subsubsection{Entropy and Distance Conjectures}
\label{sss:entropyargument}

%%%%%%%%%%%
\begin{figure}[t]
\begin{center}
\includegraphics[width=0.8\textwidth]{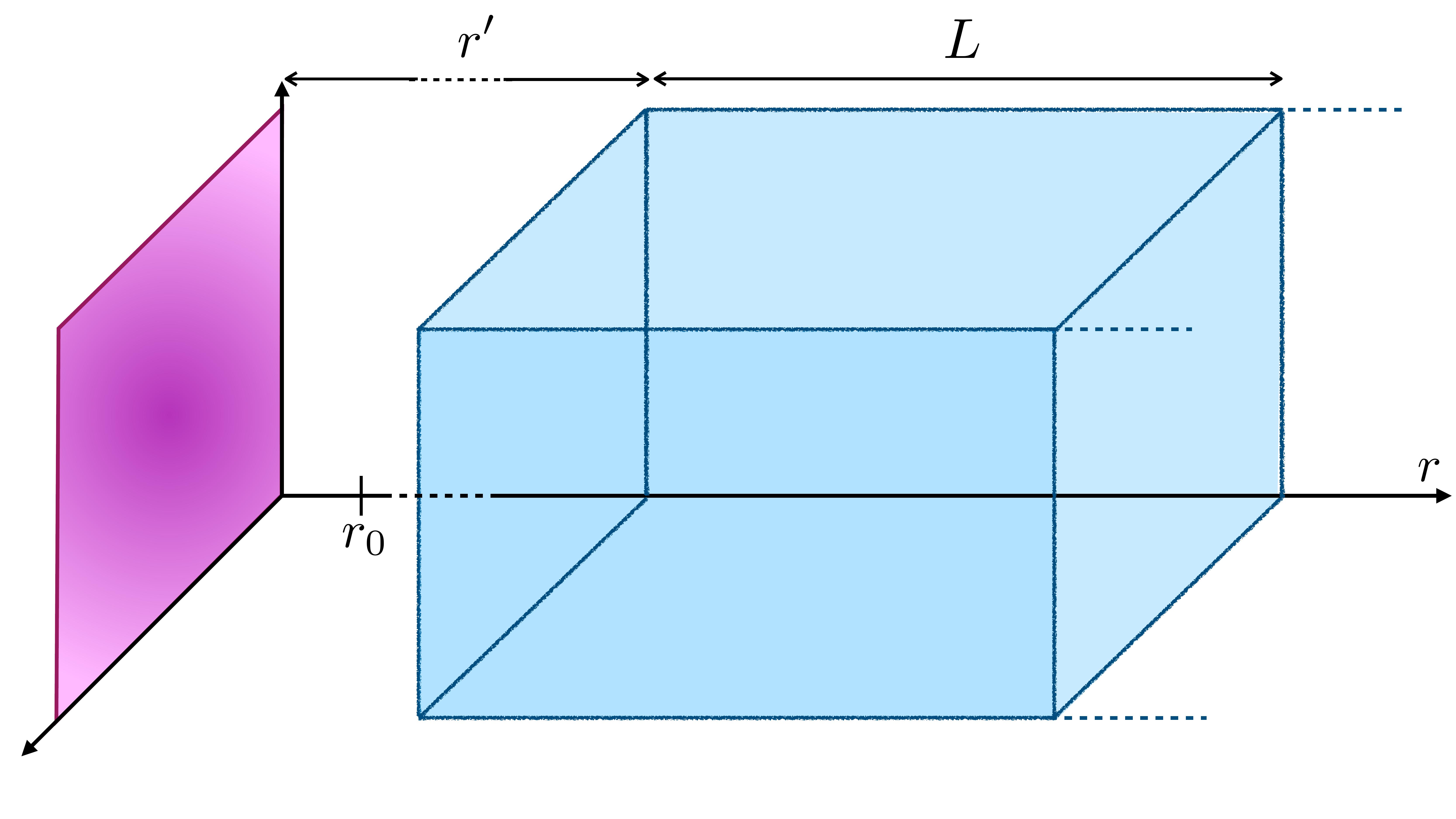}
\caption{\small Schematic representation of the region to which we apply the CEB and the singularity associated to the running solution at $r=0$, represented as the ETW-brane in the left. This region begins at a distance $r^\prime \gg r_0$, very far from the singularity and explores the direction perpendicular to the running solution along the length, $L$.  } 
\label{fig:CEBregion}
\end{center}
\end{figure}
%%%%%%%%%%%

In the following, we show how the application of the Covariant Entropy Bound to the previously found running solutions can actually give rise to the behavior predicted by the Distance Conjecture. In particular, we consider a region with boundary parallel to the ETW-brane and with length, $L$, exploring the $r$-direction. Moreover, we impose that the whole region is very far away from the singularity at $r=0$, so that we can trust the EFT solution (and thus our arguments are not sensitive to the details of the UV completion of the singularity). This region is schematically displayed in figure \ref{fig:CEBregion}. The ratio between the area and the volume of such region is precisely the quantity that bounds the UV cut-off upon imposition of the Covariant Entropy Bound (c.f. equation \eqref{eq:CEB2}), and in the aforementioned region it takes the form\footnote{\label{fnote:r0}Notice that the integration constant $r_0$ in eq. \eqref{eq:warp-factor} disappears when computing the area-to-volume ratio in \eqref{eq:area-volume} of the spacetime region we are interested in here (see figure \ref{fig:CEBregion}), thus having no net impact on our argumentation.}
\begin{equation} \label{eq:area-volume}
	\frac{A}{\text{vol}} \sim \frac{1}{L} \, .
\end{equation}
The connection with the Distance Conjecture is then transparent via the exponential mapping between spatial distance, $L$, and the distance in moduli space, $\Delta_{\phi}$, that characterizes our running solutions (c.f. eqs. \eqref{eq:mapdistancefields} and \eqref{eq:solution-Delta})
\beq
L\, \sim\, e^{\sqrt{\frac{d-1}{d-2}}\,\Delta_{\phi}}\, .
\label{eq:mapdistancefields2}
\eeq
Introducing these two equations into the entropy bound \eqref{eq:Boussobound}, yields the following relation between the Quantum Gravity cut-off and the field space distance
\beq
\label{eq:LambdaUVSDC}
\Lambda_{\text{UV}}\, \lesssim\, e^{-\frac{1}{\sqrt{(d-1)(d-2)}} \, \Delta_\phi} \, , \, \text{ as } \, \Delta_\phi \to \infty \, .
\eeq
This sets an exponential upper bound for such a UV cut-off, which is very reminiscent of the Distance Conjecture. In particular, upon identifying such an upper bound with the species scale we get:\footnote{Preliminary results, including this bound, were already presented by some of the authors in \cite{STRINGPHENO2022,SP22IR-UV}. More recently, it has also been observed in relation with the Emergent String Conjecture in \cite{vandeHeisteeg:2023uxj}.}
\beq
\label{eq:lambdaspeciesbound}
\boxed{
\Lambda_{\text{sp}}\, \sim\, e^{-\lambda_{\text{sp}}\, \Delta_{\phi}} \, , \, \text{ with } \, \lambda_{{\text{sp}}} \geq \lambda_{\text{sp}, \, \text{min}}\, = \, \frac {1}{\sqrt{(d-1)(d-2)}} \, , \, \text{ as } \, \Delta_\phi \to \infty\, .
}
\eeq
Moreover, this condition should be satisfied for \emph{any} locally geodesic trajectory in moduli space, with $\Delta_\phi$ being the distance along it. This comes from the fact that, from the EFT perspective, we can build a running solution of the kind considered here exploring the infinite distance limit of any such trajectory as $r\to\infty$. This will motivate our proposal of a Convex Hull condition for the species scale later on in section \ref{ss:convexhullspecies}.
	
Additionally, let us mention that it is also natural to impose that the UV cut-off should be above the area-to-volume ratio in our region, namely $\Lambda_{\text{sp}}\, \geq\, L^{-1}\, \sim \, e^{-\sqrt{\frac{d-1}{d-2}}\,\Delta_{\phi}}$. In this way we also obtain an upper bound on $\lambda_{\text{sp}}$, namely,
\beq \label{eq:upper-bound}
\lambda_{\text{sp}}\, \leq\,  \sqrt{\frac{d-1}{d-2}}\, .
\eeq
In fact, this bound is accounting for the fact that the background can be described by the EFT containing weakly coupled Einstein gravity, which includes all the Quantum Gravity vacua that we are interested in constraining from this perspective. In the AdS case discussed in \ref{sec:ADC}, this is satisfied as long as the species scale is above the cosmological constant scale, which includes any semiclassically consistent AdS background.\footnote{Notice that this does not exclude vacua without scale separation, such as $AdS_5 \times S^5$, since the fact that the Kaluza-Klein and the cosmological constant scales coincide automatically implies that the Quantum Gravity cut-off, namely the species scale, is above the latter.} For the case of the running solutions considered here, whether this constraint is a mandatory requirement for the EFT to be consistent is a bit more subtle, since it is not the vacuum of the theory. In any case, it is required for the rest of the analysis to be valid and it turns out that equation \eqref{eq:upper-bound} is satisfied in all string theory examples with room to spare, so it seems like a reasonable constraint. Actually, this can be seen as an asymptotic realization of the bound proposed in \cite{vandeHeisteeg:2023ubh}, which was claimed to be valid within the interior of the moduli space. It was further argued there that the asymptotic bound $\lambda_{\text{sp}}\leq 1/\sqrt{d-2}$ also holds, with the possibility of corrections that may increase its value.

Furthermore, we can always translate an upper bound for the species scale to an upper bound for the mass scale of the tower of states that lie below it (and hence are responsible for its decrease). This yields
\beq
m_{\text{tower}}\, \lesssim\,  e^{-2\alpha_d \sqrt{\frac{d-1}{d-2}}\, \Delta_{\phi}} \, , \, \text{ as } \, \Delta_\phi \to \infty \, ,
\eeq
with $\alpha_d$ as defined in equation \eqref{eq:alpha_d}. This again sets an exponential upper bound for the asymptotic behavior of $m_{\text{tower}}$, such that it can grow at most as
\beq\label{eq:SDCexponential}
m_{\text{tower}}\, \sim\,  e^{-\lambda \Delta_{\phi} }\, , \text{ with } \, \lambda\geq \lambda_{\text{min}}\, =\, \frac {d-2+p}{p\sqrt{(d-1)(d-2)}} \, , \, \text{ as } \, \Delta_\phi \to \infty \, , 
\eeq
which is precisely the exponential behavior predicted by the Distance Conjecture. Since this condition should again hold for any geodesic exploring infinite distance, it also recovers this feature of the Distance Conjecture. Moreover, this includes a precise lower bound for the exponential decay rate depending on the (effective) density parameter of the tower, $p$, as well as on the number of dimensions. The allowed values  of these lower bound exponents, $\lambda_{\text{min}}$, must thus be in the range
\beq
\label{eq:boundslambdatower}
\frac {1}{\sqrt{(d-2)(d-1)}}\, \leq\, \lambda_{\text{min}}\, \leq\, \sqrt{\frac {d-1}{d-2}}\, .
\eeq
as we vary $p$. Interestingly, all known string examples have exponents compatible with this range. In fact, the upper limit is saturated by a single KK-like tower, whereas the lower limit is saturated by e.g. the tower of the D0-D2 branes (with fixed D2 charge) in the large volume limit of type IIA compactified on a CY$_3$.

For completeness, let us comment on the alternative, more conservative approach of identifying the UV cut-off of our EFT with the mass of the first state of the tower. In this case, equation \eqref{eq:LambdaUVSDC} would directly give rise to the upper bound $m_{\text{tower}}\lesssim e^{-\frac{1}{\sqrt{(d-1)(d-2)}}\, \Delta_\phi}$, so that the exponential behavior of the SDC is again recovered and the lower bound on the exponential decay rate, $\lambda\geq \lambda_{\text{min}}= 1/\sqrt{(d-1)(d-2)}$, is relaxed with respect to the one given in \eqref{eq:SDCexponential}. Notice that this coincides with the lowest possible value for such $\lambda_{\text{min}}$ given in \eqref{eq:boundslambdatower}, but in that case it applied only to stringy towers  (i.e. with $p \to \infty$), in which both the species scale and the mass scale of the tower coincide, whereas here it is the lower bound for any tower. In fact, this value coincides with previous bounds discussed in the literature, see \cite{Grimm:2018ohb,Gendler:2020dfp,Andriot:2020lea}. Similarly, by requiring $\Lambda_{\text{sp}}\, \geq\, L^{-1}$ we also obtain $\lambda \leq \sqrt{(d-1)/(d-2)}$, which coincides with the upper bound proposed in \cite{Etheredge:2022opl}. As expected, the bounds obtained with this identification are enough to recover the exponential behavior predicted by the Distance Conjecture.  We will not consider this choice of cut-off any further in the paper, since we strongly believe that the identification of the species scale with the UV cut-off is legit and furthermore yields stricter bounds for the exponential coefficients. The main point of this brief detour is just to emphasize that the relation between the exponential behavior of the Distance Conjecture and the CEB goes beyond such precise identification of the UV cut-off, and therefore seems rather robust even if one wanted to remain agnostic about the choice of UV cut-off.

Finally, let us weigh in on the applicability of this strategy of identifying the UV cutoff appearing in the CEB with the species scale, which in turn relate it to towers of states. For instance, if it is to apply to flat space, it would be in a non-trivial way. A priori, there is no direct relation between taking a larger region and the tower of states becoming lighter. Even though a careful analysis is beyond the scope of this work, the fact that our approach recovers the ADC and the SDC when applied to certain backgrounds may suggest some universality in this link between CEB and towers of states.

\subsubsection*{Entropy variation rate}

It  is also interesting to consider the entropy variation rate in the aymptotic large $\Delta_{\phi}$ limits. The extensive EFT entropy may be written as
\beq 
S_{\text{EFT}}\, \simeq\, (\Lambda_{\text{sp}} L)^{d-1}\, =\, e^{\left( \sqrt{\frac{d-1}{d-2}} -\lambda_{\text{sp}}\right)(d-1)\Delta_{\phi}}\, ,
\eeq
and it grows in general exponentially with the moduli space distance, $\Delta_{\phi}$. Defining $S_{\text{EFT}}=e^{\lambda_S\, \Delta_{\phi}}$ the entropy rate is given by
\beq
\lambda_S\, =\, \frac{1}{S} \frac {\partial S}{\partial \Delta_{\phi}}\, =\,\left(\sqrt{\frac{d-1}{d-2}}-\lambda_{\text{sp}}\right)(d-1)\, ,
\eeq
and it is positive for $\lambda_{\text{sp}}\leq \sqrt{\frac{d-1}{d-2}}$, which as we said, corresponds to the consistency condition $\Lambda_{\text{IR}}\leq \Lambda_{\text{UV}}$.
From the bound $\lambda_{\text{sp}}  \geq 1/\sqrt{(d-1)(d-2)}$  we obtain
\beq
\frac{1}{S} \frac {\partial S}{\partial \Delta_{\phi}}\, \leq\,  \sqrt{(d-1)(d-2)}\, =\, \frac {1}{\lambda_{\text{sp},\, \text{min}}}\, .
\label{cajita}
\eeq
The variation rate of entropy is bounded above by $1/\lambda_{\text{sp},\, \text{min}}$ and the entropy cannot grow arbitrarily fast when a tower appears. Therefore, the minimum value for the species variation rate, $\lambda_{\text{sp},\, \text{min}}$, corresponds to an upper bound on the variation rate of the EFT entropy. This bound is saturated in specific examples by KK towers.
On the contrary, there is no minimum on the variation rate of entropy. It may vanish in the aforementioned limit $\lambda_{\text{sp}} \to \sqrt{\frac{d-1}{d-2}}$.

\section{Inclusion of a Potential and Relation to dS Conjectures}
\label{s:potential}

Following the discussion in the previous section, a natural next step is to introduce a potential in the discussion about CEB in running solutions. When applying the CEB, the relevant part of the running solution was its asymptotics, $r\to\infty$, and how infinite distance in moduli space is explored in that limit. Therefore, this will be our main focus in this section: by considering the asymptotic behavior of the potential in the infinite field distance limit, we will determine how this is explored in the running solution as $r\to\infty$. In particular, we remain agnostic about the form of the potential in the interior of the field-space, as well as the form of the running solution for small $r$, which in turn will depend on the former.

\subsection{Running Solutions with Exponential Potential}

Adding a potential to the previously considered effective action \eqref{eq:EFT}, this is,
\begin{equation}\label{eq:actionwithpotential}
  S = \frac{1}{\kappa_d^2} \int d^{d}x\, \sqrt{-g}\,  \left[ \frac{1}{2}R - \frac{1}{2} G_{ij} (\phi) \partial_{\mu} \phi^{i} \partial^{\mu} \phi^{j} - V(\phi^i) \right] \, . 
\end{equation}
and introducing the ansatz in equation \eqref{eq:metricprofile} for the $d$-dimensional line element, the equations of motion read
\begin{align}
	& \phi^{\prime\prime\, k} + \Gamma^{k}_{ij} \phi^{\prime\, i} \phi^{\prime\, j} + (d-1) \sigma^{\prime} \phi^{\prime\, k} = G^{ki} \partial_i V \, , \\
	& \frac{1}{2} (d-1)(d-2) \sigma^{\prime\,2} = \frac{1}{2} |\phi^{\prime}|^{2} - V \, , \\
	& (d-2) \sigma^{\prime\prime} = - |\phi^{\prime}|^{2} \, .
\end{align}

This set of equations is equivalent to the one obtained in cosmological scenarios with varying scalar potential up to flipping its sign, which can be understood as the change from space-like to time-like running coordinate. This will allow us to borrow some results from previous studies of these kind of solutions. For instance, as discussed in \cite{Calderon-Infante:2022nxb} (see also \cite{Achucarro:2018vey}), geodesic motion in this setup corresponds to gradient flow solutions and viceversa. In the following, we assume that the potential allows for this type of asymptotic solutions as we explore infinite distance in field space. This was indeed found to be the case in general flux compactifications of F-theory  on CY$_4$ near different asymptotic regimes in complex structure moduli space \cite{Calderon-Infante:2022nxb}. Under this assumption, we can reduce this multi-scalar setup to a single (canonically normalized) scalar that parametrizes the distance along the geodesic trajectory, which we denote (with a slight abuse of notation) by $\phi$. This scalar will be subject to the potential obtained by restricting $V(\phi^i)$ to the geodesic trajectory parametrized by $\phi$.\footnote{This also means that now $\phi$ coincides with the moduli space distance $\Delta_{\phi}$, as defined before (see equation \eqref{eq:line-element}).} The equations of motion for $\phi$ are then given by
\begin{align}
	& \phi^{\prime\prime} + (d-1) \sigma^{\prime} \phi^{\prime} = \partial_\phi V \, , \label{eq:with-potential-1}\\
	& \frac{1}{2} (d-1)(d-2) \sigma^{\prime\,2} = \frac{1}{2} \phi^{\prime\,2} - V \, , \label{eq:with-potential-2} \\ 
	& (d-2) \sigma^{\prime\prime} = - \phi^{\prime\,2} \label{eq:with-potential-3}\, .
\end{align}

As we learned in the previous section, the relevant quantity for the application of the CEB is the behavior of $\phi(r)$ as $r\to\infty$, which will thus be sensitive to the asymptotic form of the potential $V(\phi)$, only. With the intention of connecting with the asymptotic dS conjecture \cite{Ooguri:2018wrx}, let us consider an exponentially falling potential, i.e.,
\begin{equation} \label{exp-potential}
  V(\phi) = \pm V_0 \, e^{- c \,\phi} , \quad V_0 > 0 \, .
\end{equation}
For clarity we display the sign of the potential explicitly and, in what follows, we will treat both cases separately. Let us remark that to make direct contact with the asymptotic dS conjecture, it is crucial that geodesics correspond to gradient flow trajectories as mentioned above. Only if $\phi$ parametrizes a gradient flow trajectory it is actually true that
\begin{equation}
  c = - \frac{\partial_\phi V(\phi)}{V(\phi)} = \frac{|\vec{\nabla} V(\phi^i)|}{|V(\phi^i)|} \, .
\end{equation}
In the RHS we thus recognize the quantity bounded by the asymptotic dS conjecture, i.e. the dS coefficient. In other words, only in this case the scalar $\phi$ is sensitive to the whole slope of the multi-field potential $V(\phi^i)$, and not to just part of it, therefore guaranteeing that the quantity $c$ appearing in \eqref{exp-potential} is indeed the dS coefficient. On the contrary, if the scalars explore non-gradient-flow trajectories, the former would be strictly smaller than the latter, and the connection to the asymptotic dS conjecture would not be as straightforward.

Our goal is then to find the asymptotic profile for $\phi(r)$ as $r\to\infty$ as a function of the dS coefficient, $c$. For this, we again borrow techniques from the literature on cosmological setups. The late-time cosmology of positive exponential potentials in 4d was studied in \cite{Copeland:1997et} and generalized to arbitrary spacetime dimension in \cite{Rudelius:2022gbz}. Here we extend this analysis to account for potentials of both signs. Hence, we first define the dimensionless parameters
\begin{equation}
	x = \frac{\phi^{\prime}}{\sqrt{(d-1)(d-2)}\,\sigma^{\prime}} \, , \quad y = \frac{\sqrt{2|V|}}{\sqrt{(d-1)(d-2)}\,\sigma^{\prime}} \, .
\end{equation}
Given equation \eqref{eq:with-potential-2}, these parameters are subject to the constraint
\begin{equation} \label{constraint}
  x^2 \mp  y^{2} = 1 \, .
\end{equation}
Defining also the dimensionless coordinate $dN=\sigma^{\prime}\,dr$, and using equations \eqref{eq:with-potential-1} and \eqref{eq:with-potential-3}, we find the following evolution equations for $x$ and $y$:
\begin{equation} \label{dynamical-system}
\begin{split}	
    & \frac{dx}{dN} = \mp \frac{c y^2 \sqrt{(d-1)(d-2)}}{2} \pm (d-1) x y^2 \, , \\
	& \frac{dy}{dN} = - \frac{c x y \sqrt{(d-1)(d-2)}}{2} + (d-1) y (1 \pm y^2)  \, .  
\end{split}
\end{equation}
In this way, the equations of motion are reduced to a set of coupled differential (flow) equations subject to the constraint \eqref{constraint}, and one can analyze the asymptotic behavior of the solution by finding the attractors and repellers associated to this flow. Since the system is symmetric under $y\to -y$, in what follows we focus just on the $y\geq 0$ region. Physically, this amounts to choosing $r$ such that $\sigma^\prime \geq 0$. In this case, the sign of $x$ indicates the direction of field space that we explore as $r\to \infty$.

\subsubsection*{Negative exponential potentials}

Let us remind that the upper and lower signs in all these equations correspond to positive and negative potentials, respectively. Choosing the lower sign one recovers the set of equations in \cite{Rudelius:2022gbz}, which does not come as a surprise since our negative potentials precisely map to positive ones when considering cosmological solutions. One finds one attractor point for \eqref{dynamical-system} that satisfies the constraint \eqref{constraint}, and that is given by
\begin{equation}\label{eq:attractorsnegpot}
\begin{array}{l r}
  \left( x,y \right) = \left( \dfrac{c}{2} \sqrt{\dfrac{d-2}{d-1}} , \sqrt{1 - \dfrac{c^2(d-2)}{4(d-1)}}\right) \qquad\qquad\qquad\qquad &\text{if} \  c< 2 \sqrt{\dfrac{d-1}{d-2}} \, ,  \\
  \left( x,y \right) = \left( 1 , 0 \right)  & \text{if} \  c\geq 2 \sqrt{\dfrac{d-1}{d-2}} \, .
  \end{array}
\end{equation}
An example of each of these two cases is shown in figure \ref{fig:flow-negative} below. The first attractor corresponds to a solution in which the kinetic term  associated to the scalar and the potential energy compete asymptotically (i.e. $y/x\to\text{const.}$). Imposing this, one automatically gets the field profile $\phi(r)$. The second attractor provides for a running solution in which the kinetic energy dominates asymptotically (i.e. $y/x\to0$). Effectively, this corresponds to setting $V(\phi)=0$, thus recovering the same equations of motion and the profile for the scalar field $\phi(r)$ as in section \ref{sss:runningsol}. One can then check that the kinetic energy dominates over the potential contribution asymptotically precisely when the condition on $c$ for this attractor to exist is satisfied. The asymptotic scalar profile for both cases can be recasted as
\begin{equation}\label{eq:asymptoticprofileneg}
  \phi(r) \simeq \frac{2}{\delta}\, \log(r) \quad \text{ as } r\to\infty \, , 
\end{equation}
with
\begin{equation}
\begin{array}{l r}
  \delta = c  &\text{if} \  c< 2 \sqrt{\dfrac{d-1}{d-2}} \, ,  \\
  \delta = 2 \sqrt{\dfrac{d-1}{d-2}}  \qquad\qquad\qquad\qquad &\text{if} \  c\geq 2 \sqrt{\dfrac{d-1}{d-2}} \, .
  \end{array}
\end{equation}
%

%%%%%%%%%%%%
\begin{figure}[htb]
\begin{center}
	\subfigure[$d=4$ and $c=1.5<\sqrt{6}$]{\includegraphics[width=0.6\textwidth]{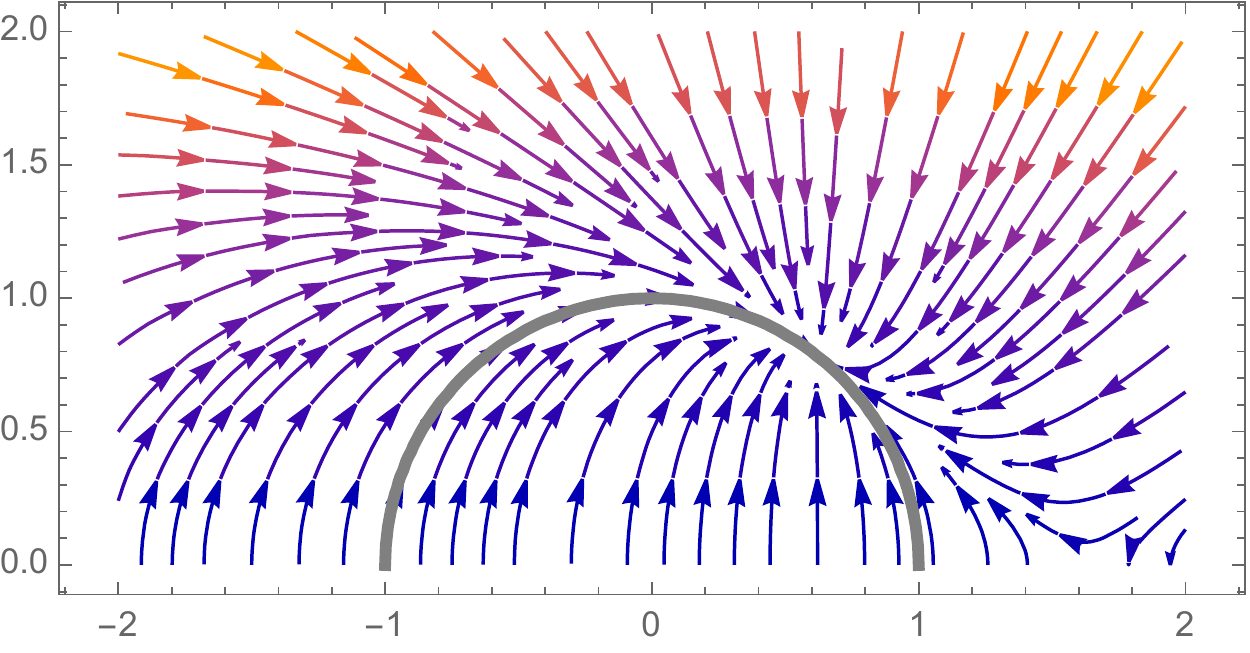}}
	\subfigure[$d=4$ and $c=5>\sqrt{6}$]{\includegraphics[width=0.6\textwidth]{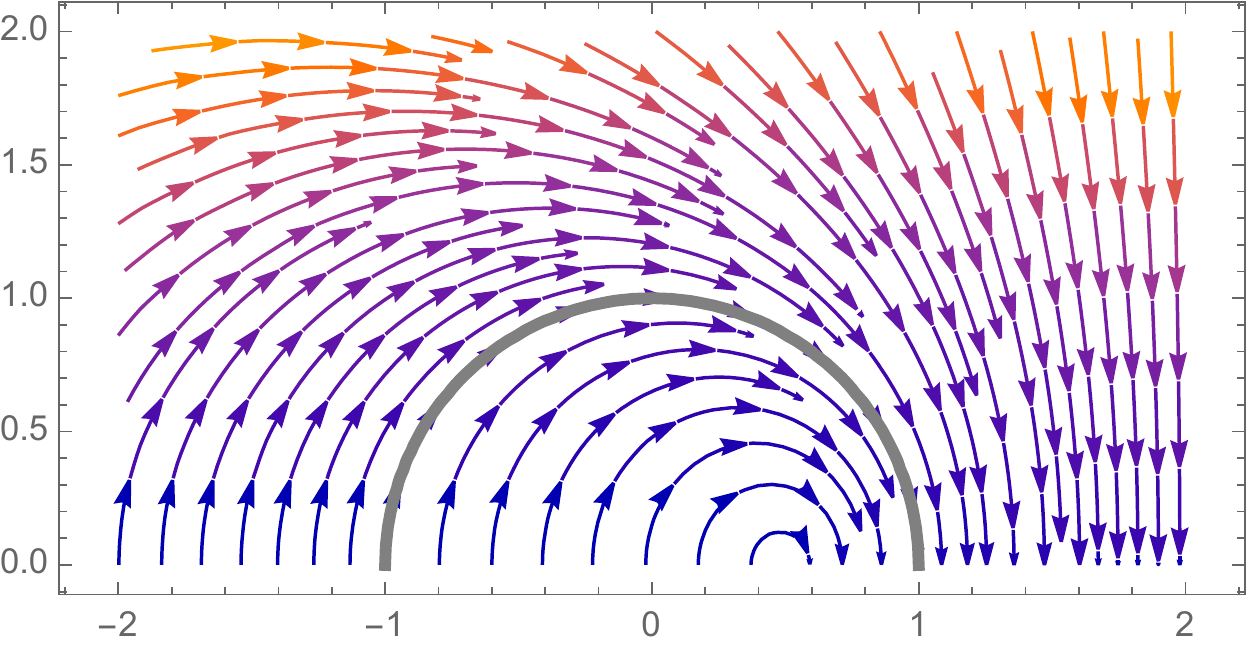}}
	\caption{Flow equations in \eqref{dynamical-system} and constraint in \eqref{constraint} (gray line) for the lower sign, corresponding to a negative potential of the form $V(\phi) = - V_0 \, e^{- c \,\phi}$.}
	\label{fig:flow-negative}	
\end{center}
\end{figure}
%%%%%%%%%%%%

\subsubsection*{Positive exponential potentials}

\begin{figure}[htb]
\begin{center}
	\subfigure[$d=4$ and $c=1.5<\sqrt{6}$]{\includegraphics[width=0.6\textwidth]{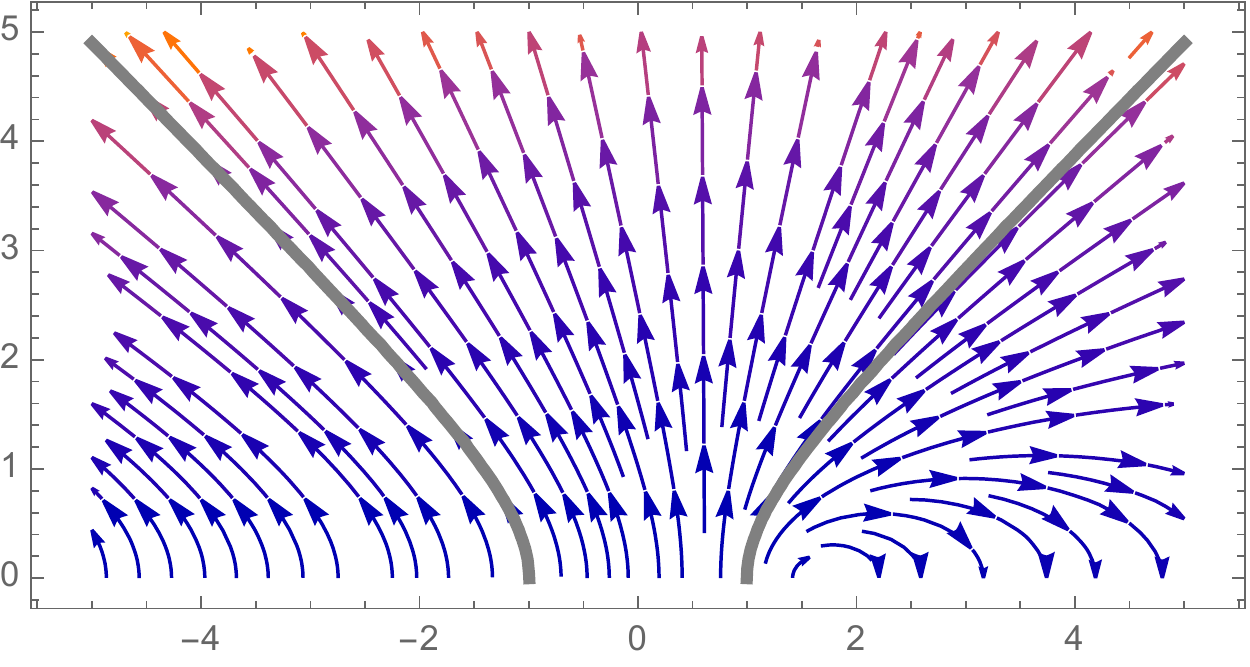}}
	\subfigure[$d=4$ and $c=5>\sqrt{6}$]{\includegraphics[width=0.6\textwidth]{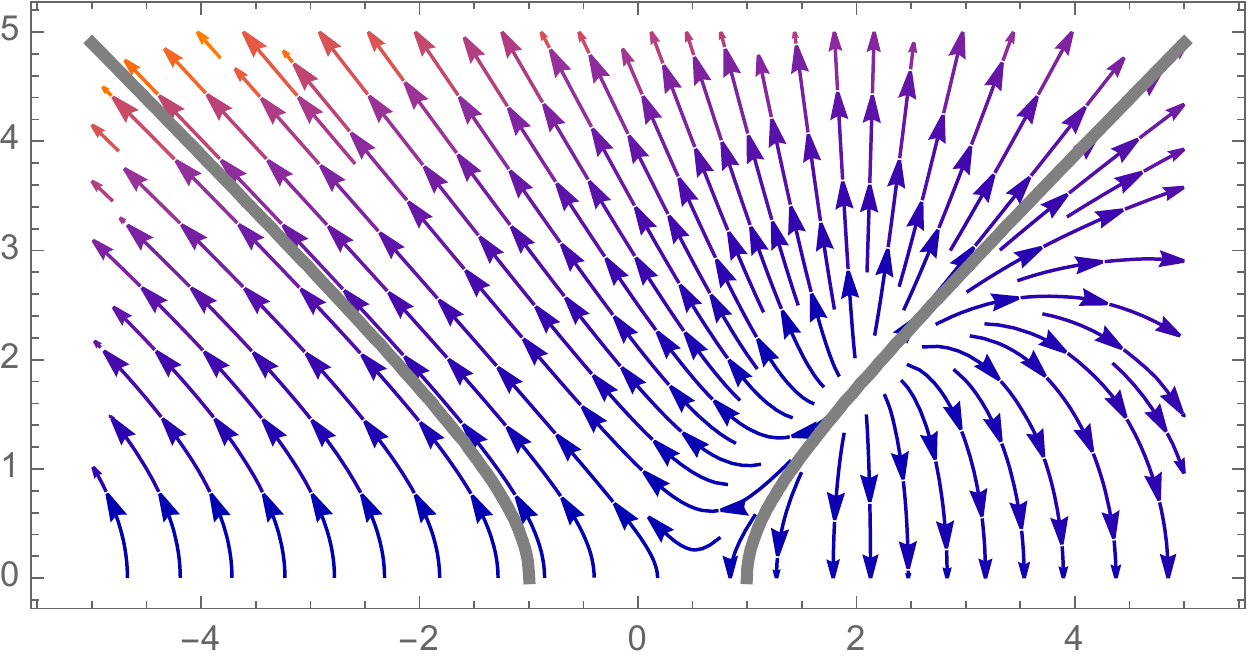}}
	\caption{Flow equations in \eqref{dynamical-system} and constraint in \eqref{constraint} (gray line) for the upper sign, corresponding to a positive potential of the form $V(\phi) = + V_0 \, e^{- c \,\phi}$.}
	\label{fig:flow-positive}	
\end{center}
\end{figure}

Let us now turn to the case of a positive potential, i.e., the upper sign in eqs. \eqref{constraint} and \eqref{dynamical-system}. As shown in figure \ref{fig:flow-positive}, one finds the following attractors
\begin{equation}
\begin{array}{l l}
  \left( x,y \right) \to \left( \infty , \infty \right)\, , \text{ with } \dfrac{y}{x}\to 1 \qquad\qquad\qquad\qquad &\forall \ c>0 \, , \\
  \left( x,y \right) = \left( 1 , 0 \right) & \text{if} \  c> 2 \sqrt{\dfrac{d-1}{d-2}}  \, ,
  \end{array}
\end{equation}
and the repeller
\begin{equation}
  \left( x,y \right) = \left( \frac{c}{2} \sqrt{\frac{d-2}{d-1}} , \sqrt{1 - \frac{c^2(d-2)}{4(d-1)}}\right) \qquad\quad \ \ \text{if} \  c> 2 \sqrt{\frac{d-1}{d-2}} \, ,
\end{equation}
all of them satisfying the constraint in \eqref{constraint}. Here we have ignored other attractors leading to $x<0$, since they can be seen to explore $\phi\to-\infty$ instead of $\phi\to\infty$ (see discussion after equation \eqref{dynamical-system}). The first attractor point corresponds to a solution in which the kinetic and potential energy coincide asymptotically, while the second one is again dominated by the kinetic energy alone. As for the case of negative potential, we can again recast the spacetime profile for the scalar field of both solutions as
\begin{equation}\label{eq:asymptoticprofilepos}
  \phi(r) \simeq \frac{2}{\delta}\, \log(r) \quad \text{ as } r\to\infty \, , 
\end{equation}
with
\begin{equation}
\begin{array}{l l}
  \delta = c  &\forall \ c>0 \, ,  \\
  \delta = 2 \sqrt{\dfrac{d-1}{d-2}} \qquad\qquad\qquad\qquad &\text{if} \  c > 2 \sqrt{\dfrac{d-1}{d-2}}  \, .
  \end{array}
\end{equation}
On the other hand, the solution corresponding to the repeller is expected to be unstable. In any event, its field profile coincides asymptotically with that of the first attractor, and therefore it gives the same results upon imposing the CEB, as we discuss in the next subsection. Notice that, unlike for negative potentials, there are now two different solutions for $c>2\sqrt{(d-1)/(d-2)}$. Therefore, the asymptotic behavior will be essentially determined by the initial conditions. This will make a difference when imposing the CEB, since having two different asymptotic behaviors leads to two different constraints for the \emph{same} theory.

%%%%%%%%%%%%
%%%%%%%%%%%%

\subsection{CEB Bounds on Exponential Potentials}

Now we turn to the constraints imposed by the CEB on these backgrounds. In order to do that, we first note that the spacetime curvature goes to zero asymptotically for all the solutions that we have described. We can then apply the CEB, similarly as we did in section \ref{sss:entropyargument} for the case of vanishing potential. As we know already, given the logarithmic profile that we found in eqs. \eqref{eq:asymptoticprofileneg} and \eqref{eq:asymptoticprofilepos}, this constrains the UV cut-off to fall exponentially as we explore infinite distance in field space, thus recovering the SDC behavior. Moreover, its exponential rate is bounded as follows
\begin{equation}
  \frac{\delta}{2(d-1)}\, \leq\, \lambda_{\text{sp}}\, \leq\, \frac{\delta}{2} \, .
\end{equation}
Let us stress that these two bounds are very different in nature. The lower bound captures the constraints arising from the CEB, whilst the upper bound corresponds to the asymptotics of the running solution admitting an effective description within the EFT, namely $\Lambda_{\text{UV}}>\Lambda_{\text{IR}}$.

Different running solutions as the ones explored above will give different values for $\delta$, sometimes related to the dS coefficient $c$. 

\subsubsection*{Bounds for negative exponential potentials}

Consider first the case of a negative potential of the form \eqref{exp-potential} with $c\geq 2\sqrt{(d-1)/(d-2)}\, $. Following the discussion around equation \eqref{eq:attractorsnegpot}, we thus recover the exact same bound on $\lambda$ as discussed previously in section \ref{sss:entropyargument}. On the other hand, for $c<2\sqrt{(d-1)/(d-2)}$ we find a bound that involves both $\lambda$ and $c$ at the same time. To sum up, in the presence of a negative exponential potential we find:
\begin{align}
  \frac{1}{2(d-1)}\, \leq\, &\frac{\lambda_{\text{sp}}}{c}\, \leq\, \frac{1}{2}  &\text{if} \  c< 2 \sqrt{\frac{d-1}{d-2}} \, , \\
  \frac{1}{\sqrt{(d-1)(d-2)}}\, \leq\, &\lambda_{\text{sp}}\, \leq\, \sqrt{\frac{d-1}{d-2}} &\text{if} \  c\geq 2 \sqrt{\frac{d-1}{d-2}} \, .
\end{align}
The excluded regions in the $(\lambda_{\text{sp}},c)$-plane are depicted in figure \ref{fig:CEB-potential-bounds} below.

%%%%%%%%%%%%
\begin{figure}[htb]
\begin{center}
	\subfigure[Negative potential]{\includegraphics[width=0.45\textwidth]{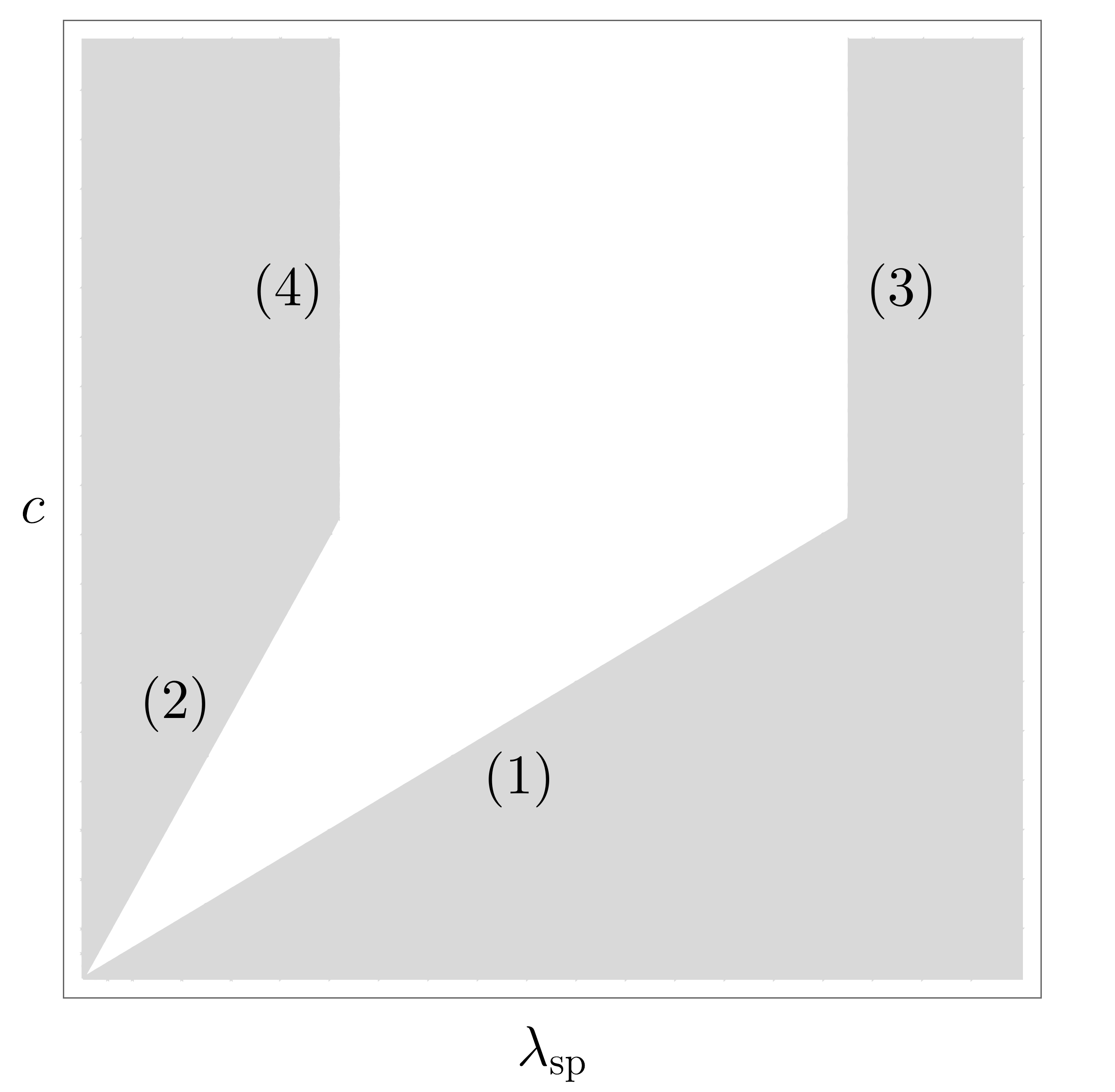}}
    \qquad
    \subfigure[Positive potential]{\includegraphics[width=0.45\textwidth]{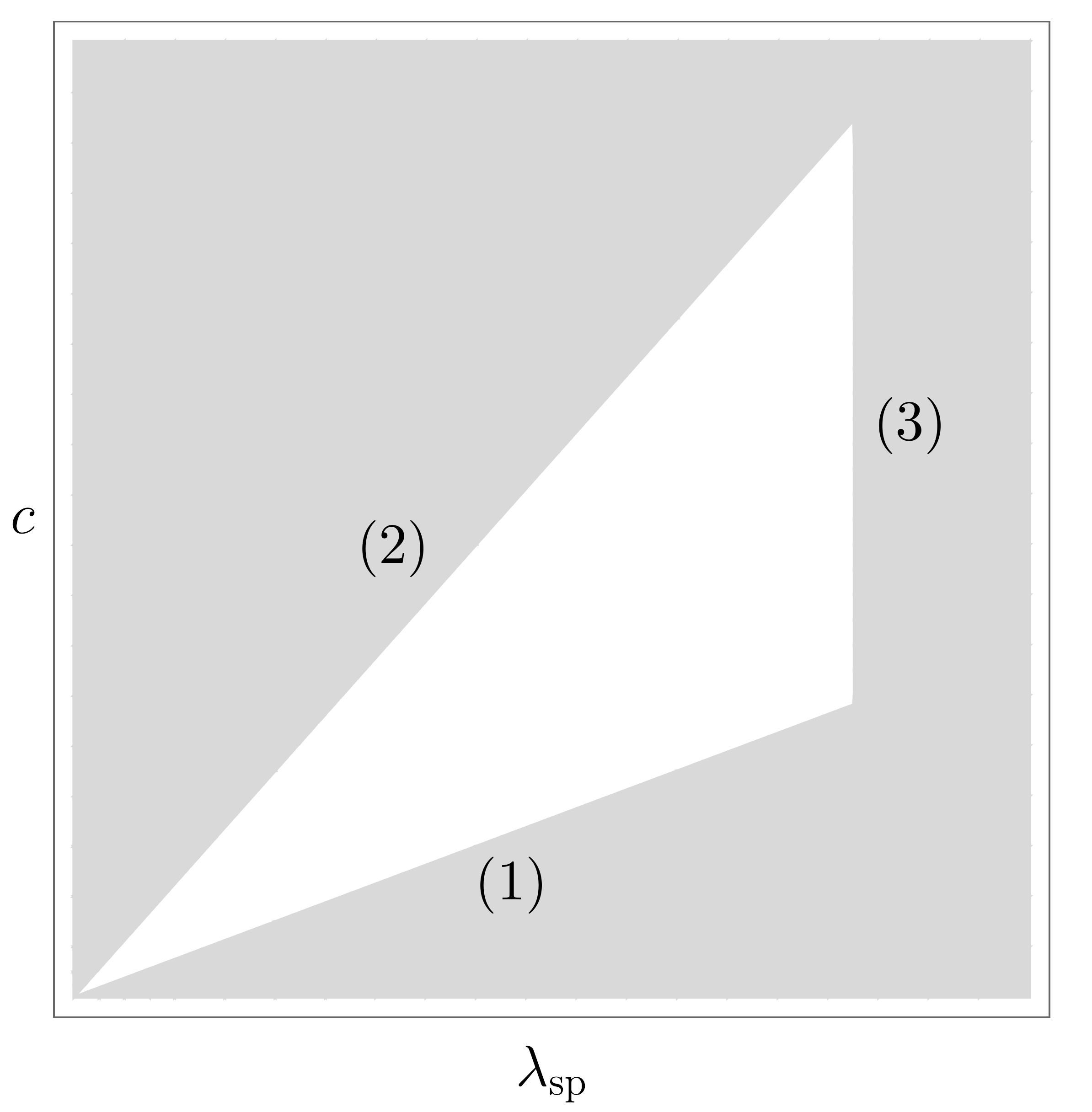}}
	\caption{Bounds imposed by applying the CEB to running solutions in theories with negative (a) and positive (b) exponential potential. The regions shaded in grey are excluded. The numbered lines correspond to: (1) $c = 2 \lambda_{\text{sp}}$, (2) $c = 2(d-1) \lambda_{\text{sp}}$, (3) $\lambda_{\text{sp}} = \sqrt{(d-1)/(d-2)}$ and (4) $\lambda_{\text{sp}} = 1/\sqrt{(d-1)(d-2)}$.}
	\label{fig:CEB-potential-bounds}	
\end{center}
\end{figure}
%%%%%%%%%%%%

\subsubsection*{Bounds for positive exponential potentials}

For a positive potential with $c<2\sqrt{(d-1)/(d-2)}$ the results are exactly the same as above. However, for $c\geq 2\sqrt{(d-1)/(d-2)}$ we found two backgrounds with different asymptotic behaviors, so both of them should be consistent with the CEB. Upon identifying CEB cut-off with a tower of states, this becomes a bound on the tower itself, which may be present in other backgrounds. Thus, various backgrounds may put several constraints on the same tower and should be mutually compatible. This is related to the idea of background independence in Quantum Gravity, in the sense that a well-defined theory should be consistent in any background. In particular, this is expected to be the case since its imposition is related to linking the cut-off in the CEB to a tower of states, which is something inherent to the theory and thus background independent. It turns out that the bound imposed by the $\delta=c$ solution is always stronger than the other. Therefore, for theories exhibiting a positive exponential potential we find:
\begin{equation} \label{eq:bounds-positive-potential}
    \frac{1}{2(d-1)}\, \leq\, \frac{\lambda_{\text{sp}}}{c}\, \leq\, \frac{1}{2} \, , \quad \lambda_{\text{sp}} \leq \sqrt{\frac{d-1}{d-2}} \, .
\end{equation}
Again, the excluded regions in the $(\lambda_{\text{sp}},c)$-plane are shown in figure \ref{fig:CEB-potential-bounds}. 

The bounds involving the ratio between $\lambda_{\text{sp}}$ and $c$ can be brought to a more suggestive form by writting them in terms of $\Lambda_{\text{sp}}$ and $V$ as follows
\begin{equation}
	\Lambda_{\text{sp}} \sim |V|^{\alpha_{\text{sp}}} \, , \qquad \frac{1}{2(d-1)}\, \leq\, \alpha_{\text{sp}}\, \leq\, \frac{1}{2} \, ,
\end{equation}
which take the same form as the ones found in \cite{Castellano:2021mmx} for the case of AdS and dS minima of the potential, and that we reviewed in section \ref{sec:ADC} above. The difference is that these bounds now apply to exponentially damped potentials. In particular, the upper bound on $\alpha_{\text{sp}}$ should always hold while the lower bound does not apply if the potential is negative and in addition $c> 2 \sqrt{(d-1)/(d-2)}$. In fact, the upper bound imposes the natural condition of the energy scale of the potential remaining below the UV cut-off, which was recently used in \cite{vandeHeisteeg:2023uxj} to find bounds on slowly-varying potentials, in relation to bounds on the species scale.

It is also interesting to discuss the presence of absolute upper and lower bounds on $\lambda_{\text{sp}}$ or $c$. For instance, we recover the absolute upper bound on $\lambda_{\text{sp}}$ found in the previous section, which now gets generalized to the case of a non-trivial (exponential) potential. Similarly, for positive potentials we find the absolute upper bound on $c$
\begin{equation} \label{eq:upper-bound-c}
  c\, \leq\, 2 (d-1) \sqrt{\frac{d-1}{d-2}} \, .
\end{equation}
Its physical meaning is that, for steeper exponential potentials, it is impossible to describe both running solutions with $\delta = c$ and $\delta = 2\sqrt{(d-1)/(d-2)}$ within the same EFT, since the UV cut-off would fall bellow the IR one for at least one of them. Nevertheless, it would be interesting to further test the plausibility of this bound against string theory examples.

Let us stress that these bounds are only valid for theories with an exponential potential as $\phi\to\infty$. For other types of potentials, our analysis of the running solutions stops being valid, and its extension to more classes of potentials falls beyond the scope of this work. In other words, these bounds should be understood as applicable to potentials as the one in \eqref{exp-potential} with $c$ being a constant, and not a function of $\phi$. For instance, potentials with an exponential of exponential behavior can be regarded as having $c(\phi)\to\infty$ as $\phi\to\infty$, but they are not excluded by the upper bound on $c$ discussed above. In fact, this kind of asymptotic behaviors for $V(\phi)$ can indeed occur in string theory, for example when the flux potential has some flat directions that only receive non-perturbative contributions (see e.g. \cite{Demirtas:2019sip,Demirtas:2021nlu,Demirtas:2021ote}).

In addition, note that these bounds should also be understood as applying to infinite distance limits of \emph{gradient flow} trajectories of the potential. For instance, if this gradient flow presents some attractor, then the bound only applies to this locus as it explores infinite distance. This makes difficult to test \eqref{eq:upper-bound-c} against previous results in the literature, in which moreover the focus was put on the lower instead of the upper bounds. However, the setup is very well suited for comparison with the recent results in \cite{Calderon-Infante:2022nxb}, in which an extensive study of the values of $c$ along gradient flows in various asymptotic limits in F-theory compactified on CY$_4$ with fluxes was carried out. The maximum value found there was $c=\sqrt{14}$, thus satisfying \eqref{eq:upper-bound-c} for $d=4$ by almost a factor of two.\footnote{Even though it may seem that this value only appears in table 3 in \cite{Calderon-Infante:2022nxb}, it also appears in tables 2 and 5 upon including the contribution from the K\"ahler sector.}

\begin{figure}[htb]
\begin{center}
	\subfigure[Negative potential]{\includegraphics[width=0.45\textwidth]{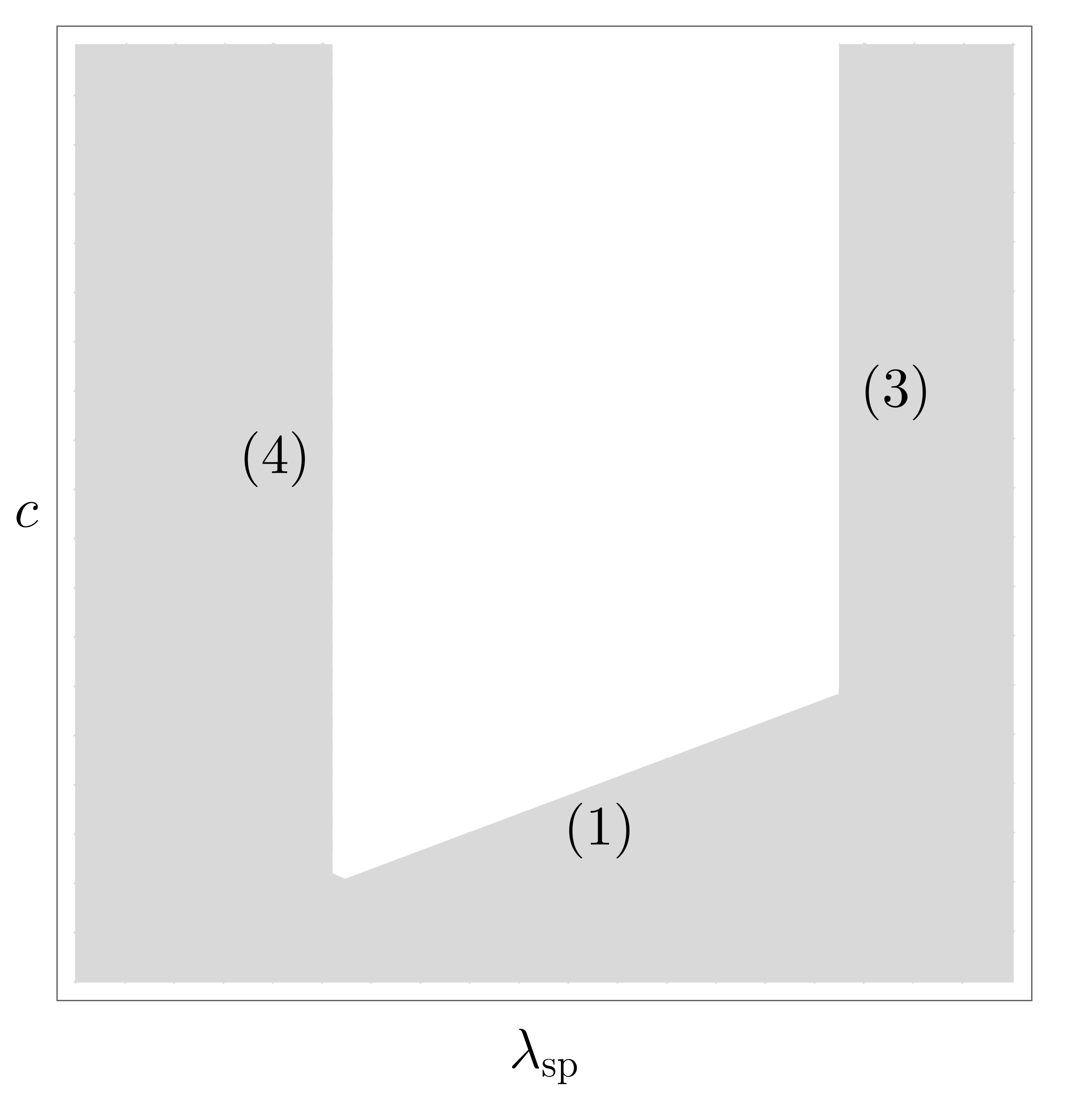}}
    \qquad
    \subfigure[Positive potential]{\includegraphics[width=0.45\textwidth]{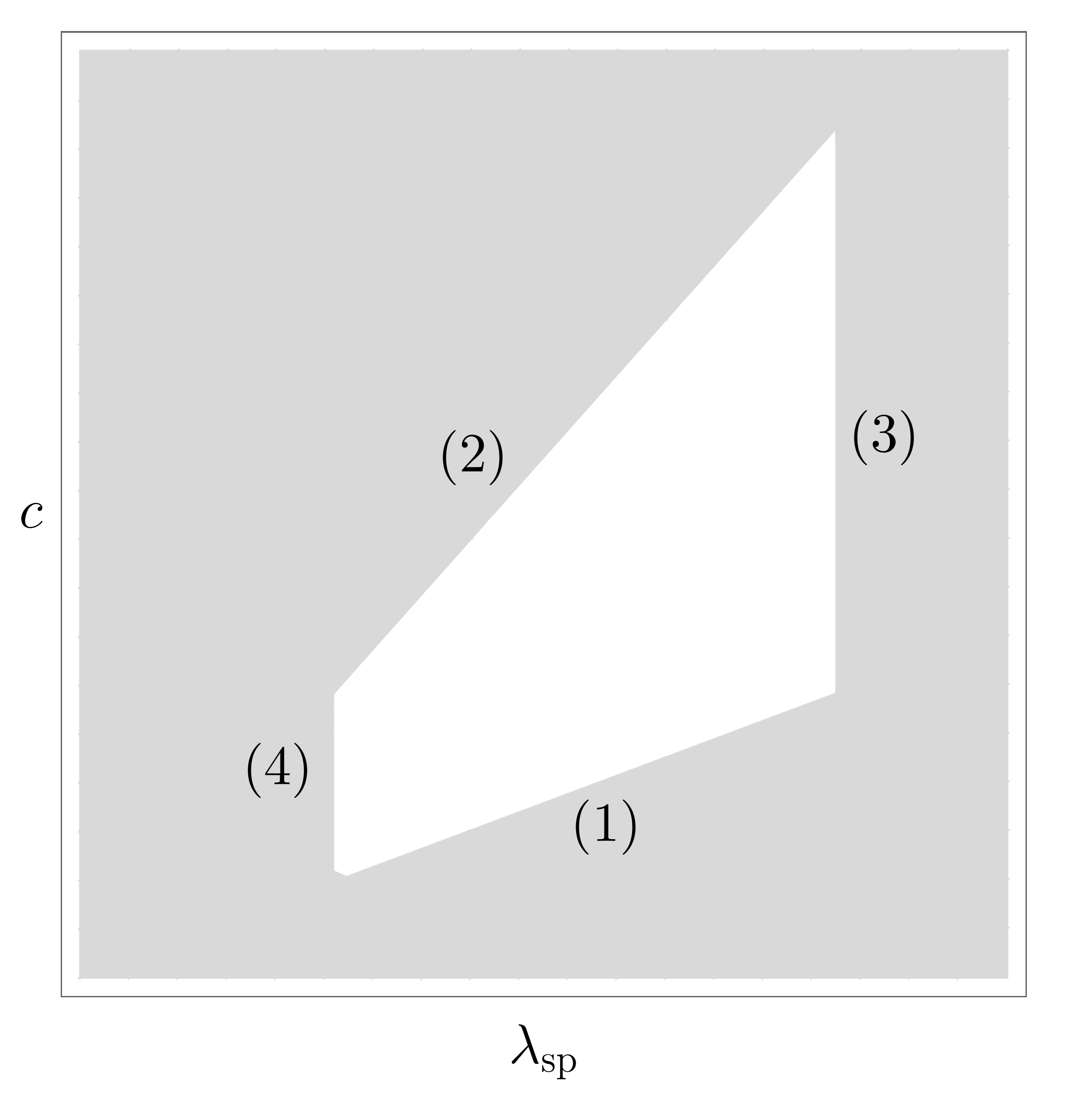}}
	\caption{Bounds imposed by applying the CEB to running solutions in theories with negative (a) and positive (b) exponential potential, in addition to the universal lower bound $\lambda_{\text{sp}}\geq 1/\sqrt{(d-1)(d-2)}$. The regions shaded in grey are excluded. The numbered lines correspond to: (1) $c = 2 \lambda_{\text{sp}}$, (2) $c = 2(d-1) \lambda_{\text{sp}}$, (3) $\lambda_{\text{sp}} = \sqrt{(d-1)/(d-2)}$ and (4) $\lambda_{\text{sp}} = 1/\sqrt{(d-1)(d-2)}$.}
	\label{fig:CEB-potential-extra}	
\end{center}
\end{figure}

Finally, let us note that we do not find any universal lower bounds on either $\lambda_{\text{sp}}$ or $c$ from these arguments. However, since we are assuming geodesic motion for the scalar fields, it seems reasonable to extend to the present analysis the lower bound on $\lambda_{\text{sp}}$ found in the previous section for the case of a vanishing potential. This is very natural when the potential comes from deforming a theory with a moduli space, in particular when this potential is small since one can then expect that this deformation does not destroy the UV structure of the towers. This happens e.g. in string theory flux potentials in the diluted flux limits. Upon imposing this lower bound on $\lambda_{\text{sp}}$, we obtain an analogous one on the dS coefficient as follows
\begin{equation}
  c\, \geq\, \frac{2}{\sqrt{(d-1)(d-2)}} \, ,
\end{equation}
which precisely coincides with the TCC bound \cite{Bedroya:2019snp}. This arises quite directly from our universal lower bound on $\lambda_{\text{sp}}$ and the lower bound on $c$ that guarantees that the mass scale of the potential is below the cut-off. From this latter perspective, this coincidence was recently noted in \cite{vandeHeisteeg:2023uxj}. The excluded regions in the $(\lambda_{\text{sp}},c)$-plane when adding the universal lower bound on $\lambda_{\text{sp}}$ are shown in figure \ref{fig:CEB-potential-extra} for the case of both negative and positive potentials. 

%%%%%%%%%%%%
%%%%%%%%%%%%

Finally, let us briefly come back to the equivalence between domain-wall and cosmological solutions upon flipping the sign of the potential. In fact, if one were to naively apply the CEB to these cosmological solutions, in which the role of the space distance $r$ is played by the cosmological time $t$ instead, then one would get the stronger bounds in \eqref{eq:bounds-positive-potential} also for negative potentials, and thus the absolute upper bound on $c$.
 
\section{The Species Scale Distance Conjecture}
\label{s:SSDC}

The purpose of this section is to take the lower bound \eqref{eq:lambdaspeciesbound} for the decay rate of the species scale seriously and investigate more closely what are its consequences in any EFT weakly coupled to Einstein gravity. Along the way, we will provide for additional evidence in its favour based on a simple nine-dimensional example arising from M-theory compactified on $T^2$.

\subsection{A Convex Hull for the Species Scale}
\label{ss:convexhullspecies}

Let us start by reviewing some generalities which will be useful both in this section and in the upcoming ones. We consider a gravitational EFT describing the low energy dynamics of a set of $d$-dimensional fields, including certain \emph{exactly} massless scalars $\phi^i$, parametrizing a moduli space $\mathcal{M}$. Upon exploring asymptotic directions within said moduli space, the SDC predicts the appearance of infinite towers of states becoming exponentially light with respect to the (traversed) moduli space distance, c.f. equation \eqref{eq:SDCexponential}. Hence, given a trajectory of this kind characterised by some unit tangent vector $\hat n$, one can readily evaluate the exponential decay rate of the tower of states as \cite{Calderon-Infante:2020dhm}
\begin{equation}\label{eq:decayrate}
  \lambda_{\text{t}} = \hat n \cdot \vec{z}_{\text{t}}\, ,
\end{equation}
where $\vec{z}_{\text{t}}$ denotes the scalar charge-to-mass vector of the tower, whose components in an orthonormal frame are defined by
\begin{equation} \label{eq:z-vectors}
  z_{\text{t}}^{a} = - \delta^{ab} e^{i}_{b} \, \frac{\partial_{i}m_{\text{tower}}}{m_{\text{tower}}}\, . 
\end{equation}
Here $\partial_i$ denotes differentiation with respect to the modulus $\phi^i$, $e^{i}_{b} (\phi)$ are nothing but the vielbein for the field space metric $G_{ij}$ (c.f. equation \eqref{eq:EFT}), whilst $m_{\text{tower}}$ is the mass scale associated to the tower of states. In the case in which there is more than one infinite tower of states becoming light upon taking some infinite distance limit, the SDC coefficient $\lambda$ is usually taken to be \emph{largest} one, namely 
\begin{equation}\label{eq:lambdaSDC}
  \lambda = \max_{\text{t}}\, \{ \lambda_{\text{t}} \} = \max_{\text{t}}\, \{ \hat n \cdot \vec{z}_{\text{t}} \} \, ,
\end{equation}
thus corresponding to the set of states which become light the fastest. The Convex Hull SDC\cite{Calderon-Infante:2020dhm} then comes as a natural way of imposing $\lambda \geq \lambda_{\text{min}}$ for any possible asymptotic direction within $\mathcal{M}$, see \cite{Etheredge:2022opl} for a recent proposal for the precise value of $\lambda_{\text{min}}$.

Our aim here will be to formulate an analogous condition for the QG cut-off in our EFTs, namely the species scale $\Lambda_{\text{sp}} (\phi)$. To do so, we first need to know how to properly compute such quantity in the presence of several towers of states, since a key point is that it may receive important contributions from towers other than the lightest one. The details of the calculation strongly depend on how the different sets of states relate to each other, i.e. whether they are additive or multiplicative. We will only review the latter case, since it is enough for our purposes in this work, and we refer the interested reader to \cite{Castellano:2021mmx,Castellano:2022bvr}. 

Let us start by considering a spectrum of mixed states with quantum numbers $n, k \in \mathbb{Z}\,$, associated to two different infinite towers. For concreteness, let us take their mass dependence to be of the form\cite{Castellano:2021mmx} 
\begin{equation} \label{eq:massmultiplicativetowers}
  m_{n,k}^{2} = n^{2/p_1} m_{\text{tower},\, 1}^2 + k^{2/p_2} m_{\text{tower},\, 2}^2 \, ,
\end{equation}
with $m_{\text{tower},\, 1}\leq m_{\text{tower},\, 2}$ without loss of generality. Notice that both mass scales typically depend on  where we sit at the moduli space $\mathcal{M}$, and they can do so in a very different manner. One useful way to think of this spectrum is as if it were coming from two distinct \emph{multiplicative} towers with mass scales $m_{\text{tower},\, 1},\, m_{\text{tower},\, 2}$ and density parameters $p_1$ and $p_2$, respectively. The canonical example being that of two Kaluza-Klein towers associated to two compact internal directions, verifying $m_{\text{tower},\, i}=1/R_i$ and $p_i=1$ for $i=1,2$, where $R_i$ denotes the radius of the corresponding 1-cycle.

As an intermediate step to compute the relevant QG cut-off, one can first associate to each separate tower a would-be species scale as follows (see section \ref{ss:ADCreview})
\begin{equation}
  \Lambda_{\text{sp},\, 1}\, \sim\, m_{\text{tower},\, 1}^{\frac{p_1}{d-2+p_1}} \, , \qquad \Lambda_{\text{sp},\, 2}\, \sim\, m_{\text{tower},\, 2}^{\frac{p_2}{d-2+p_2}} \, ,
\end{equation}
where each of them is computed by accounting just for the subset of states associated to the corresponding tower and ignoring those arising from the other (as well as mixed states thereof).

Next, we consider the combined effect of the two towers by essentially counting all states with mixed quantum numbers that fall below the species scale itself. This yields asymptotically (up to $\mathcal{O}(1)$ factors) \cite{Castellano:2021mmx}
\begin{equation}\label{eq:effspeciesscale}
  \Lambda_{\text{sp},\, \text{eff}}\, \sim\, m_{\text{tower},\, \text{eff}}^{\frac{p_{\text{eff}}}{d-2+p_{\text{eff}}}}\, ,
\end{equation}
where we have defined (geometric) `averaged' quantities as follows
\begin{equation}\label{eq:masseffectivetower}
  m_{\text{tower},\, \text{eff}}\, \sim\, \left( m_{\text{tower},\, 1}^{p_{1}} m_{\text{tower},\, 2}^{p_{2}} \right)^{1/p_{\text{eff}}}\, , \qquad p_{\text{eff}} = p_1 + p_2 \, .
\end{equation}
The main reason why it is useful to divide such computation into two steps is because depending on the asymptotic direction in moduli space $\hat n$ that is being explored, the states associated to either one of the two infinite towers can become arbitrarily lighter than those coming from the second one. Therefore, in certain circumstances it may be enough to just consider particle states arising from just one of the two towers so as to determine completely the QG cut-off. However, oftentimes one may still need to consider mixed states of the form \eqref{eq:massmultiplicativetowers} to fully determine $\Lambda_{\text{sp}}$, even if one of the two scales (say e.g. $m_{\text{tower},\, 1}$) becomes arbitrarily lighter than the other. Indeed, this happens when
\begin{equation}
  m_{\text{tower},\, 2}\, \lesssim\, \Lambda_{\text{sp},\, 1}\, ,
\end{equation}
and may be understood as the first tower not being able to saturate the species scale because of a (asymptotically) diverging number of states associated to the second one falling bellow $\Lambda_{\text{sp},\, 1}$. 

Rewriting $m_{\text{tower},\, 2}$ in terms of $\Lambda_{\text{sp},\, \text{eff}}$, the last condition turns out to be equivalent to
\begin{equation}
  \Lambda_{\text{sp},\, \text{eff}}\, \lesssim\, \Lambda_{\text{sp},\, 1}\, .
\end{equation}
This means, in particular, that the true species scale for each asymptotic direction that one may consider is simply given by the \emph{smallest} one out of the set $\{\Lambda_{\text{sp},\, 1},\, \Lambda_{\text{sp},\, 2},\, \Lambda_{\text{sp},\, \text{eff}}\}$. Hence, as also happens when studying the SDC (see discussion around equation \eqref{eq:lambdaSDC}), one must take the would-be species scale to be the one that is falling at the fastest rate. The crucial difference being that, in general, we do not only need to consider the different towers separately, but also possible `combinations' thereof. This is a direct reflection of the species scale being sensitive not only to the lightest tower, but in general to all the light towers of states.

With this in mind, we can now propose a Convex Hull condition for the species scale cut-off, which we call the \emph{Species Scale Distance Conjecture} (SSDC). Consider a spectrum of states formed by several towers with \emph{species scale vectors} defined, in analogy with the scalar charge-to-mass vectors of the towers (c.f. equation \eqref{eq:z-vectors}), by
\begin{equation}
  \mathcal{Z}_{\beta}^{a} = - \delta^{ab} e^{i}_{b} \,  \frac{\partial_{i}\Lambda_{\text{sp},\, \beta}}{\Lambda_{\text{sp},\, \beta}}\, .
\end{equation}
Here $\beta$ labels the species scales of all the towers, including the effective ones that combine several of them. The exponential decay rate of the species scale along a given trajectory with unit tangent vector $\hat n$ is given by
\begin{equation}
  \lambda_{\text{sp}} = \max_{\beta}\, \{ \hat n \cdot \vec{\mathcal{Z}}_{\beta} \} \, .
\end{equation}
Then, in analogy with the Convex Hull SDC discussed above, the SSDC can be formulated as:

\noindent\shadowbox{
\begin{minipage}{6in}
\textbf{Species Scale Distance Conjecture (SSDC):} \emph{the convex hull of species scale vectors $\{ \vec{\mathcal{Z}}_{\beta} \}$ should contain the ball of radius} $\lambda_{\text{sp}, \, \text{min}}= \dfrac{1}{\sqrt{(d-1)(d-2)}}$ .
\end{minipage}}

Notice that we have not only defined a convex hull for the species scale, but there is also a simple algorithm that translates the convex hull for the towers (with the additional information of their density parameter) to the convex hull for the species. Given the scalar charge-to-mass ratio associated to each (multiplicative) tower in the theory $\{\vec z_{\text{t}}\}$, as well as their density parameters $\{p_\text{t}\}$, the species scale vector of any effective tower is readily computable as follows:
\begin{equation} \label{eq:eff-vector}
    \vec{\mathcal{Z}}_{(\text{t}_1,\ldots ,\text{t}_n)} = \frac{1}{d-2+ \sum_\alpha p_{\text{t}_\alpha}} \, \sum_{\alpha=1}^{n} p_{\text{t}_\alpha} \, \vec{z}_{\text{t}_\alpha}\, .
\end{equation}
The subindex in the LHS indicates that this effective tower combines those from the set $\{\text{t}_\alpha\}$ with $\alpha=1,\ldots,n$.\footnote{One can analogously define the scalar charge-to-mass vector associated to the effective tower defined in equation \eqref{eq:masseffectivetower} above, by $\vec{z}_{(\text{t}_1,\ldots ,\text{t}_n)} = \frac{1}{\sum_\alpha p_{\text{t}_\alpha}} \, \sum_{\alpha} p_{\text{t}_\alpha} \, \vec{z}_{\text{t}_\alpha}$. From this one may rewrite \eqref{eq:eff-vector} as $\vec{\mathcal{Z}}_{(\text{t}_1,\ldots ,\text{t}_n)}= \frac{p_{\text{eff}}}{d-2+p_{\text{eff}}}\, \vec{z}_{(\text{t}_1,\ldots ,\text{t}_n)}$.} Therefore, with the extra information of the density of each tower and whether they are multiplicative or not, the convex hull of the towers can be algorithmically converted into the convex hull for the species scale.

As a summary, we have shown that a Species Scale Distance Conjecture of the form $\lambda_{\text{sp}} \geq \lambda_{\text{sp}, \, \text{min}}$ can be reformulated as a Convex Hull condition, encompassing all possible asymptotically geodesic directions that one can explore within the moduli space of the EFT under consideration. The main difference with the Convex Hull SDC is that the set of vectors $\{ \vec{\mathcal{Z}}_{\beta} \}$ not only includes information about single towers, but also combinations thereof in the form of effective towers.

In sections \ref{ss:MthyT2} and \ref{ss:MthyT3} below we will present a couple of explicit examples of species scale convex hulls associated to certain EFTs arising from Quantum Gravity. Apart from verifying the SSDC, one of the goals will be to remark how important it is to include the effective towers into the analysis. Without them, the convex hull would not capture the underlying physics correctly, and thus the SSDC will be violated. Furthermore, we will observe that the minimal ingredients to generate the convex hull (its vertices) always correspond to either maximal KK or stringy towers. In other words, we find that the maximum decompactification (to 11d M-theory) and the emergent string limits seem to contain all the relevant information required so as to build the convex hull for the species scale, effectively realizing the idea that in String/M-theory setups the species scale always ends up capturing the fundamental UV scale. In any event, we will point out that the remaining KK towers, appearing at the faces and edges of the convex hull, play an important role in determining the physics at the asymptotic regimes along partial decompactification limits.

\subsection{M-theory on $T^2$}
\label{ss:MthyT2}

We consider first a 9d $\mathcal{N}=2$ example arising from compactifying M-theory on a two-dimensional torus $T^2$. Thus, we start from the 11d supergravity bosonic action\cite{Cremmer:1978km}
\begin{align}\label{eq:Mthyactionbos}
			S^{\text{11d}}_{\text{M}} = \frac{1}{2\kappa_{11}^2} \int R \, \star 1-\frac{1}{2}  dC_3\wedge \star  dC_3 -\frac{1}{6} C_3 \wedge dC_3 \wedge dC_3 \, ,    
\end{align}
and we impose the following ansatz for the 11d metric\footnote{We henceforth ignore the Kaluza-Klein photon since we want to focus just on the scalar sector of the theory.}
\begin{equation}\label{eq:11dmetric}
	ds^2_{11} = e^{-2U/7} ds_9^2 + g_{mn} dz^m dz^n\, ,
\end{equation}
where the metric on the two-torus takes the usual form
\begin{equation}\label{eq:T2metric}
	g_{mn}= \frac{e^U}{\tau_2} \left(
			\begin{array}{cc}
				1 & \tau_1  \\
				\tau_1 & |\tau|^2  \\
			\end{array}
			\right) \, ,
\end{equation}
with $\tau=\tau_1+i\tau_2$ the complex structure of the $T^2$. This leads to a 9d supergravity action whose scalar and gravitational sectors read
\begin{equation}\label{eq:9d}
	S^{\text{9d}}_{\text{M}} \supset \frac{1}{2\kappa_9^2} \int d^{9}x\, \sqrt{-g}\,  \left( R - \frac{9}{14} \left( \partial U \right)^2 -\frac{\partial \tau \cdot \partial \bar \tau}{2 \left(\text{Im}\, \tau\right)^2} \right)\, .
\end{equation}
This theory enjoys an $SL(2, \mathbb{Z})$ U-duality symmetry whose origin can be seen to be clearly geometric from this perspective, since it is associated to the modular group of large diffeomorphisms of the internal torus\cite{Schwarz:1995dk,Aspinwall:1995fw}. 

The goal is to check our proposed convex hull condition for the species scale, namely the requirement
\begin{equation} \label{eq:bound9d}
  \lambda_{\text{sp}} \geq \frac{1}{\sqrt{(d-1)(d-2)}} \stackrel{\text{9d}}{=} \frac{1}{\sqrt{56}}\, .
\end{equation}
Since this condition should hold for any locally geodesic trajectory in moduli space as it explores infinite distance, one first needs to characterize them. For this we note that the 9d moduli space can be identified with $\mathcal{M}_{\text{9d}}=SL(2, \mathbb{Z})\backslash SL(2, \mathbb{R})/U(1) \times \mathbb{R}$. Notice that apart from the modular sector, it also possesses an additional non-compact direction parametrized by the overall volume field, $U$. As discussed in more detail in section \ref{sss:axions}, all geodesic trajectories are such that $\tau_1 \to \text{const.}$ asymptotically. This allows us to restrict the convex hull to the non-compact directions in the $U$--$\,\tau_2$ plane, which corresponds to the subspace of asymptotically geodesic tangent vectors introduced in reference \cite{Calderon-Infante:2020dhm}.

In a next step, one needs to account for the relevant towers of states, and then compute all the possible quantum gravity cut-offs that could arise depending on the infinite distance singularity that we choose to sample. Let us start by considering the $\frac{1}{2}$-BPS strings, which arise by wrapping the eleven-dimensional M2-branes on any $(p,q)$-cycle of the internal geometry. Their tension reads
\begin{equation}\label{eq:pqstrings9d}
	T_{p,q} = \frac{2\pi}{\ell_{9}^2} \frac{|p+q\tau|}{\sqrt{\tau_2}} e^{\frac{3}{14}U}\, ,
\end{equation}
where $\ell_{9}$ denotes the 9d Planck length. In the following, we will fix the axion vev $\tau_1=0$ since it can be seen to play no role in our present discussion, and only keep track of the saxionic dependence of the relevant mass scales. We will come back to the axions and justify this choice in detail in  section \ref{sss:axions} below. We moreover define canonically normalized fields $\hat U$ and $\hat \tau$ as follows
\begin{equation} \label{eq:canonicalnormalization}
  U =  \kappa_9 \sqrt{\frac{14}{9}}\, \hat U \, , \quad \tau_2 = \kappa_9\, e^{\sqrt{2} \, \hat\tau} \, ,
\end{equation}
in terms of which the mass parameter of the $(p,q)$-strings, i.e. equation  \eqref{eq:pqstrings9d}, reads
\begin{equation} \label{strings}
  m^{(\text{str})}_{p,q} = \sqrt{T_{p,q}} \sim M_{\text{pl},\, 9} \left( p^2 e^{-\sqrt{2} \,\hat\tau} + q^2 e^{\sqrt{2} \,\hat\tau}\right)^{1/4} e^{\frac{1}{2 \sqrt{14}} \, \hat U} \, .
\end{equation}
Equation \eqref{strings} has two different asymptotic behaviors depending on which infinite distance limit is probed and whether $p$ and $q$ are non-vanishing. As it is to be expected, any infinite distance limit is dominated by either $q=0$ or $p=0$ cases, which are associated to the fundamental and S-dual strings in the type IIB frame, respectively. These correspond to two relevant asymptotic scales for the Quantum Gravity cut-off
\begin{equation} \label{string-species}
  \Lambda_{\text{str, 1}} \sim e^{\frac{1}{2 \sqrt{14}} \, \hat U-\frac{1}{2\sqrt{2}} \,\hat\tau } \, , \quad \Lambda_{\text{str, 2}} \sim e^{\frac{1}{2 \sqrt{14}} \, \hat U + \frac{1}{2\sqrt{2}} \,\hat\tau } \, ,
\end{equation}
where we are taking into account that the species scale associated to a critical string is given at leading order by its own mass. The corresponding species scale vectors are thus given by
\begin{equation}
\label{eq:speciesscalevectorsstringsT2}
  \vec{\mathcal{Z}}_{\text{str, 1}} = \left( -\frac{1}{2 \sqrt{14}},\frac{1}{2\sqrt{2}} \right) \, , \quad 
  \vec{\mathcal{Z}}_{\text{str, 2}} = \left( -\frac{1}{2 \sqrt{14}},- \frac{1}{2\sqrt{2}} \right) \, ,
\end{equation}
where we use the notation $\vec{\mathcal{Z}} = \left(\mathcal{Z}_{\hat U}, \mathcal{Z}_{\hat \tau} \right)$.

On the other hand, the 9d theory also presents some particular spectrum of $\frac{1}{4}$-BPS particles, whose masses depend on where we sit on moduli space and are given by \cite{Obers:1998fb}
\begin{equation}\label{eq:pqparticles9d}
	m^{(\text{part})}_{p,q,w} = \frac{2\pi}{\ell_{9}} \frac{|p+q\tau|}{\sqrt{\tau_2}} e^{-\frac{9}{14}U} + |w| e^{\frac{6}{7}U}\, .
\end{equation}
Setting again the axion to zero and re-expressing it in terms of the canonically normalized fields \eqref{eq:canonicalnormalization}, we get 
\begin{equation} \label{particles}
  \frac{m^{(\text{part})}_{p,q,w}}{M_{\text{pl},\, 9}} \sim \left( p^2 e^{-\sqrt{2} \,\hat\tau} + q^2 e^{\sqrt{2} \,\hat\tau}\right)^{1/2} e^{-\frac{3}{\sqrt{14}} \, \hat U} + |w| e^{\sqrt{\frac{8}{7}} \, \hat U}\, .
\end{equation}
These particles arise as bound states of Kaluza-Klein modes along the compact directions (with charges $p, q \in \mathbb{Z}$) and non-perturbative states obtained by wrapping an M2-brane $w \in \mathbb{Z}$ times along the internal 2-cycle.\footnote{Alternatively, the M2-particles may be viewed as winding modes of the critical type IIA strings described in \eqref{strings}.} For us, it turns out to be enough to focus on towers comprised by $\frac{1}{2}$-BPS states, since only these become light and dense enough asymptotically so as to saturate the species scale at some infinite distance corner of the 9d moduli space. This may be understood heuristically by noticing that any other state not being half-BPS necessarily contains higher-spin fields, such that whenever they become nearly massless one expects some other critical string dominating the asymptotic physics. Therefore, we may divide the spectrum into two sectors, corresponding to either $w=0$ or $p=q=0$. 

Consider first the $w=0$ sector. It does behave as two multiplicative towers of Kaluza-Klein type --- indeed they are the KK modes corresponding to both 1-cycles of the torus --- with mass scales behaving in Planck units as
\begin{equation} \label{KK-mass}
  m_{\text{KK},\, 1} \sim e^{-\frac{3}{\sqrt{14}} \, \hat U - \frac{1}{\sqrt{2}} \,\hat\tau } \, , \quad 
  m_{\text{KK},\, 2} \sim e^{-\frac{3}{\sqrt{14}} \, \hat U + \frac{1}{\sqrt{2}} \,\hat\tau } \, ,
\end{equation}
and densities $p_1=p_2=1$. Their associated species scales can be easily computed:
\begin{equation}
  \Lambda_{\text{KK},\, 1} \sim e^{-\frac{3\sqrt{14}}{112} \, \hat U - \frac{\sqrt{2}}{16} \,\hat\tau } \, , \quad 
  \Lambda_{\text{KK},\, 2} \sim e^{-\frac{3\sqrt{14}}{112} \, \hat U + \frac{\sqrt{2}}{16} \,\hat\tau } \, , \quad
   \Lambda_{\text{KK},\, (12)} \sim e^{-\frac{\sqrt{14}}{21} \, \hat U} \, ,
\end{equation}
where the last one corresponds to the effective combination of the first two, see section \ref{ss:convexhullspecies}. Thus, their species scale vectors become
\begin{equation}
  \vec{\mathcal{Z}}_{\text{KK},\, 1} = \left( \frac{3\sqrt{14}}{112},\frac{\sqrt{2}}{16} \right)\, , \quad
  \vec{\mathcal{Z}}_{\text{KK},\, 2} = \left( \frac{3\sqrt{14}}{112},-\frac{\sqrt{2}}{16}\right) \, , \quad
  \vec{\mathcal{Z}}_{\text{KK},\, (12)} = \left( \frac{\sqrt{14}}{21}  , 0\right) \, .
\end{equation}

For the $p=q=0$ sector, \eqref{particles} tells us that the tower behaves essentially as some sort of KK spectrum. This is easy to understand, since they are nothing but the Kaluza-Klein replica of the 10d fields implementing the M-/F-theory duality\cite{Vafa:1996xn}. Their mass scale is thus given by
\begin{equation}\label{eq:Ftheorytower}
  \frac{m_{\text{M}2}}{M_{\text{pl},\, 9}} \sim e^{\sqrt{\frac{8}{7}} \, \hat U } \, ,
\end{equation}
and its associated species scale and charge-to-mass ratio are
\begin{equation}
  \Lambda_{\text{M}2} \sim e^{\frac{1}{2\sqrt{14}} \, \hat U }\, , \qquad \vec{\mathcal{Z}}_{\text{M}2} = \left( -\frac{1}{2\sqrt{14}} , 0\right) \, .
\end{equation}
A priori one should also include the vectors $\vec{\mathcal{Z}}^{\,(1)}_{\text{KK},\, \text{M}}$ and $\vec{\mathcal{Z}}^{\,(2)}_{\text{KK},\, \text{M}}$, which correspond to the effective towers formed by combining $m_{\text{KK},\, 1}$ and $m_{\text{KK},\, 2}$ with $m_{\text{M}2}$, respectively. However, they will not change the convex hull diagram since these scales turn out to be always above the mass scale of the strings in \eqref{string-species}. Similarly, if one were to include the species scale of an effective tower including all three KK-like towers, the result would be that it is always above the others. This is just another way of seeing that the states that mix the three towers (i.e. with $p,q,w\neq 0$) never become massless, so that they never help in lowering the species scale. In other words, this effective species scale is not physical in any asymptotic limit.

\begin{figure}[htb]
		\begin{center}
			\subfigure[]{
				\includegraphics[width=0.45\textwidth]{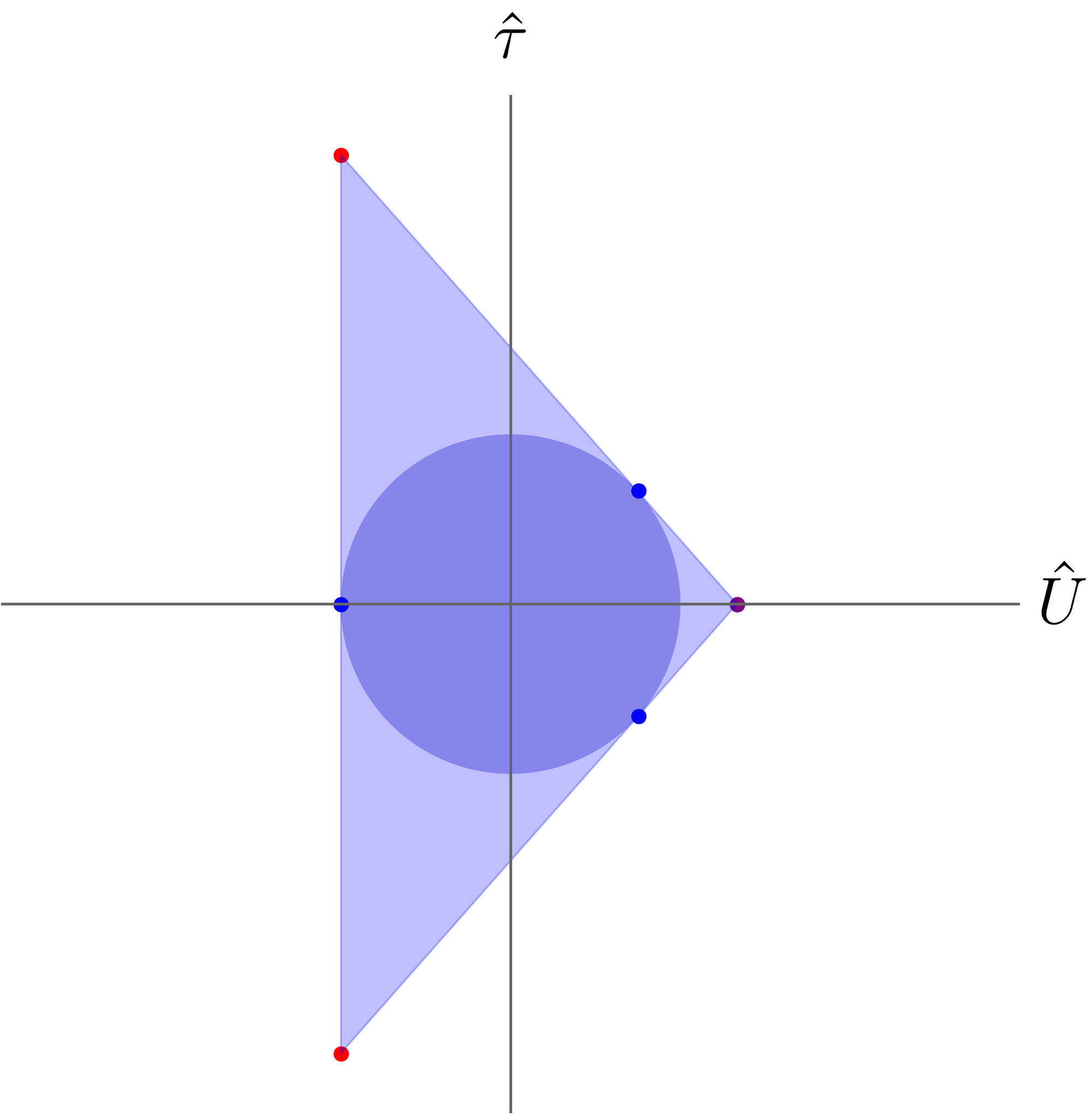}\label{sfig:9dCH}
			}
            \quad
			\subfigure[]{
				\includegraphics[width=0.45\textwidth]{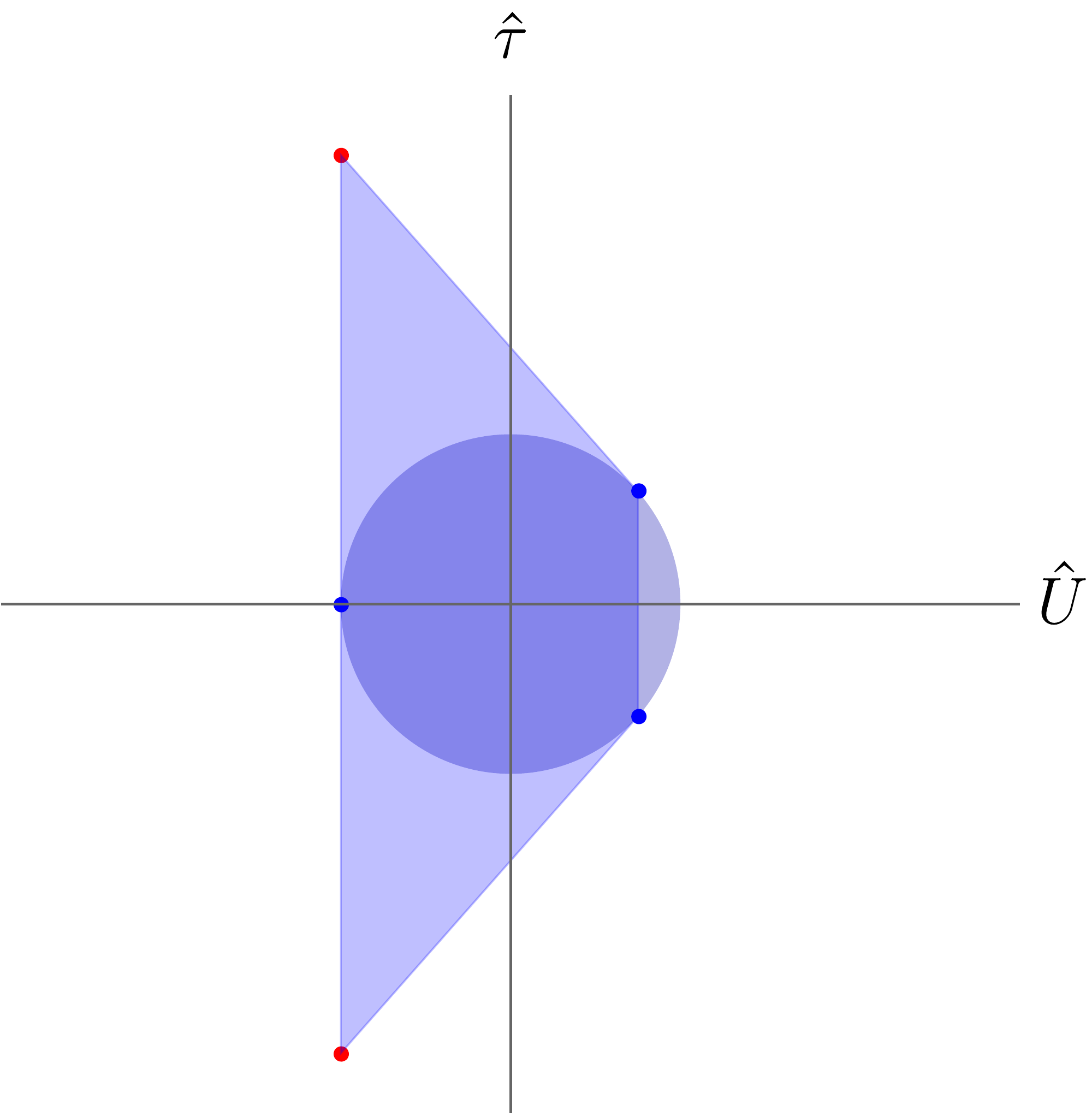}\label{sfig:9dCH-2}
			}
			\small \caption{a) Convex hull condition for the species scale in M-theory on $T^2$. b) Convex hull condition for the species scale in M-theory on $T^2$, if the effective double KK tower would not exist. The blue dots appearing in the faces of the convex hull are single KK towers, whereas the red and purple dots appearing at the vertices represent stringy and double KK towers.}
			\label{fig:ch1}
		\end{center}
\end{figure}

With this we already have all the necessary ingredients in order to draw the convex hull for the species scale. The result is shown in figure \ref{fig:ch1}, where we have also depicted the extremal radius corresponding to $\lambda_{\text{min}}$ in this 9d setup. For comparison, we also include how the convex hull would look in the absence of the effective tower. As advocated, the effective tower is crucial for capturing the underlying physics, and also for the SSDC to be satisfied. 

Notice also that there is a  $\mathbb{Z}_2$-symmetry relating the upper and lower halves of the convex hull in figure \ref{fig:ch1}, which is nothing but a manifestation of the $SL(2, \mathbb{Z})$ duality group of this theory (in particular of the S-transformation). On top of this, this convex hull for the species scale presents a lot of structure that nicely encodes the physics at the different asymptotic limits, corresponding to different directions in figure \ref{fig:ch1}.

Let us start with the vertices: Upon pointing towards one of the red dots, the species scale is dominated by a (critical) string tower. Therefore, the associated regime turns out to be an emergent string limit \cite{Lee:2019wij}. Similarly, upon pointing along the direction determined the purple dot, the species scale is dominated by the double KK tower, thus signalling towards decompactification of two extra dimensions, i.e., to 11d M-theory. In fact, this also holds upon exploring any intermediate asymptotic direction between the blue dots, since even though one KK tower becomes parametrically lighter than the other, the species scale is yet saturated by accounting for mixed states thereof (see discussion around equation \eqref{eq:masseffectivetower}).

Finally, let us discuss the directions associated to the blue dots. It turns out that these are always orthogonal to some face of the convex hull. In fact, if this were not the case, the convex hull condition for the SSDC would be violated since these single KK vectors precisely saturate the bound. Therefore, the three effective species scales contained in the face all fall at the same rate along the aforementioned asymptotic limit. Their (finite) ratios are not encoded in the convex hull, and depend on the non-divergent value of the moduli that are not sent to infinity. This has a nice interpretation as a decompactification to an extra dimension: the species scale of the single KK tower we are pointing to signals this decompactification, whereas the others correspond to towers that are already present in the higher-dimensional theory. In fact, we observe that the faces precisely reproduce the (one-dimensional) convex hull of the decompactified theory. Indeed, the vertical line on the LHS of the diagram reproduces the convex hull of 10d type IIB string theory, with the F1 and D1-strings becoming light at weak and strong coupling, respectively. Similarly, the other two faces correspond to the convex hull of 10d type IIA, with the fundamental string and the tower of D0-branes becoming light analogously at weak and strong coupling. 

In comparison to \cite{Etheredge:2022opl}, we also note that the roles of saturating/protecting the convex hull condition are exchanged between strings and KK towers for the case of the species scale. Indeed, it is instructive to superimpose both diagrams for the species scale and for the mass scales of the lightest towers. This is depicted in figure \ref{fig:ch2} below, where one can already appreciate some hidden symmetry relating both convex hulls. This symmetry will be further described and explored in \cite{Castellano:2023stg, Castellano:2023jjt}.

%%%%%%%%%%%
\begin{figure}[htb]
\begin{center}
\includegraphics[width=0.45\textwidth]{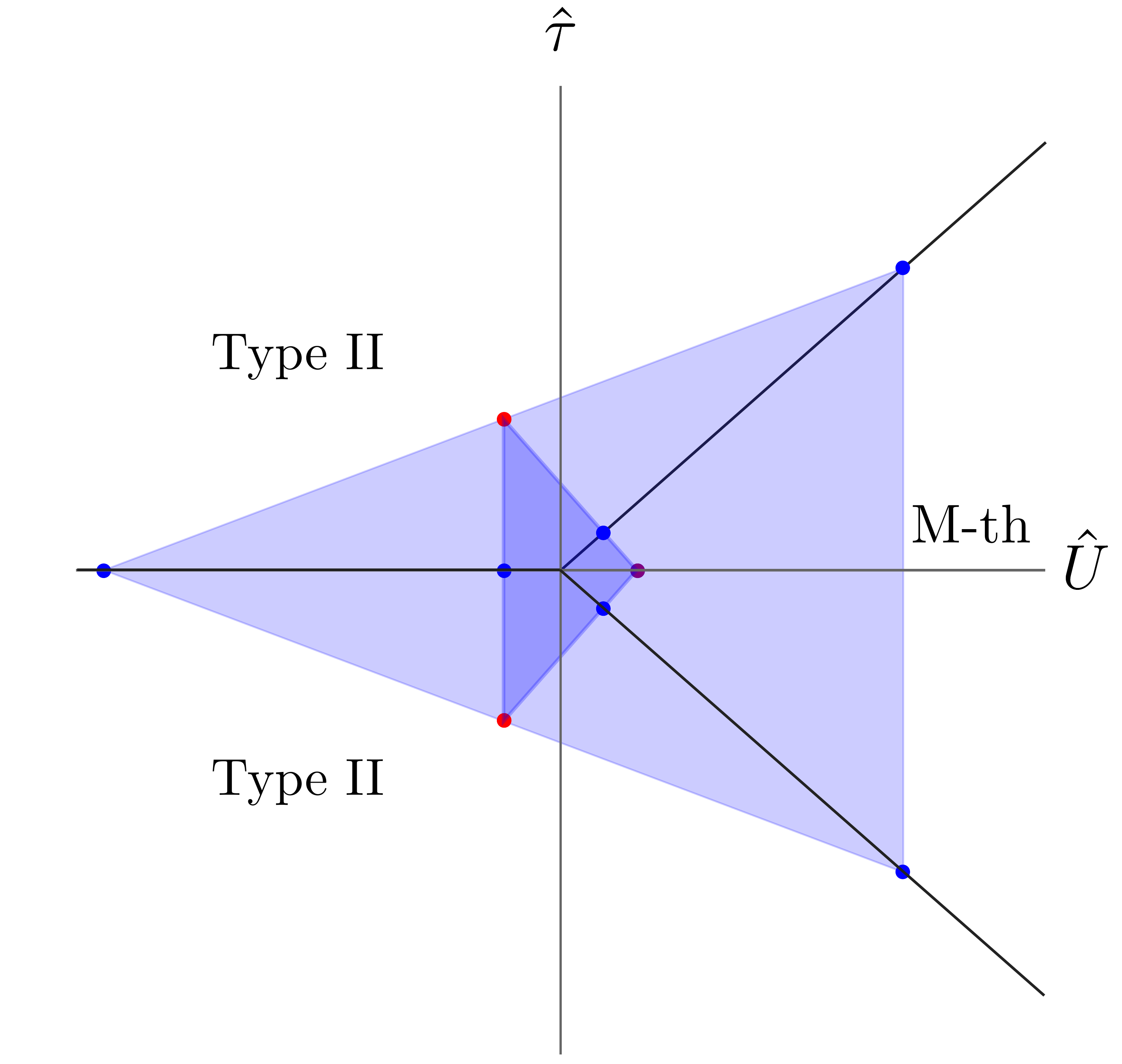}
\caption{\small Convex hulls superimposed for both the species scale and mass scales of the leading towers in M-theory on $T^2$. The black lines separate the different duality frames within the 9d setup, corresponding to type II string theory and 11d M-theory. The self-dual line, namely $\hat{\tau}=0$, is fixed under the $\mathbb{Z}_2$-symmetry.} 
\label{fig:ch2}
\end{center}
\end{figure}
%%%%%%%%%%%

\subsubsection{Revisiting the Axion}
\label{sss:axions}

Let us now go back and reconsider the role of the axion $\tau_1$ on our previous analysis. This exercise proves to be instructive and imparts a valuable lesson on the significance of including compact scalar fields in our analysis of the Species Scale Distance Conjecture.

Notice from equation \eqref{eq:9d} that the two scalar sectors of the moduli space are decoupled (at the two-derivative level) and only the modular one, which describes the complex structure of the internal $T^2$, contains the axion. Such complex-valued field parametrizes the manifold $SL(2, \mathbb{Z})\backslash SL(2, \mathbb{R})/U(1)$, with $SL(2, \mathbb{Z})$ being the U-duality group of the 9d theory.

Restricting ourselves to the fundamental domain, $\mathcal{F}$, of the moduli space (see figure \ref{sfig:funddom} below) it is transparent that there is just one infinite distance singularity, namely $\tau \to i \infty$. This corresponds to a weak coupling limit for the fundamental (F1) type IIA string, whose tension reads
\begin{equation}\label{eq:IIAtension}
	T_{\text{F1}} = \frac{2\pi}{\ell_{9}^2} \frac{1}{\sqrt{\tau_2}} e^{\frac{3}{14}U}\, ,
\end{equation}
leading to the species scale vector $\vec{\mathcal{Z}}_{\text{str, 1}} = \left(\mathcal{Z}_{U}, \mathcal{Z}_{\tau_2}, \mathcal{Z}_{\tau_1} \right)$ given by the components displayed in the first vector of equation \eqref{eq:speciesscalevectorsstringsT2} and the additional axionic component
\begin{equation} \label{eq:chargetomassF1}
 \mathcal{Z}^{\text{F1}}_{\tau_1} = -\sqrt{G^{\tau_1 \tau_1}}\, \partial_{\tau_1} \log \left(m_{\text{F1}}\right) = 0 \, .%\\
\end{equation}
The fact that this extra component vanishes (since the tension of the F1 does not depend on the axion) implies that the decay rate parameter $\lambda_{\text{sp}}$ of the species scale along directions fulfilling $\tau \to i \infty,\, U \to \infty$ is indeed controlled just by the saxionic dependence of the species scale vector, regardless of the particular geodesic we follow to approach the weak coupling point. Furthermore, geodesics reaching infinity within $\mathcal{F}$ are vertical straight lines --- i.e. they satisfy $\dot \tau_1=\frac{d\tau_1}{ds}=0$ with $s \in \mathbb{R}$ some affine parameter --- such that we can effectively forget about the axionic direction $\tau_1$.

\begin{figure}[htb]
		\begin{center}
			\subfigure[]{
				\includegraphics[width=0.5\textwidth]{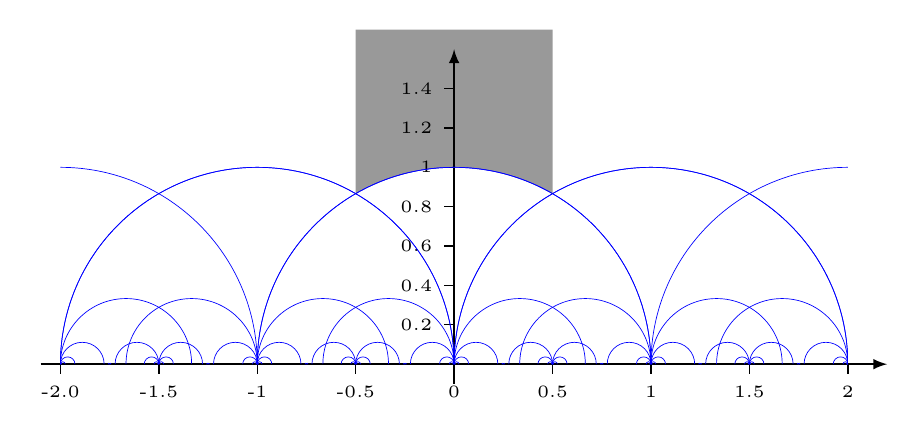}\label{sfig:funddom}
			}
            \quad
			\subfigure[]{
				\includegraphics[width=0.4\textwidth]{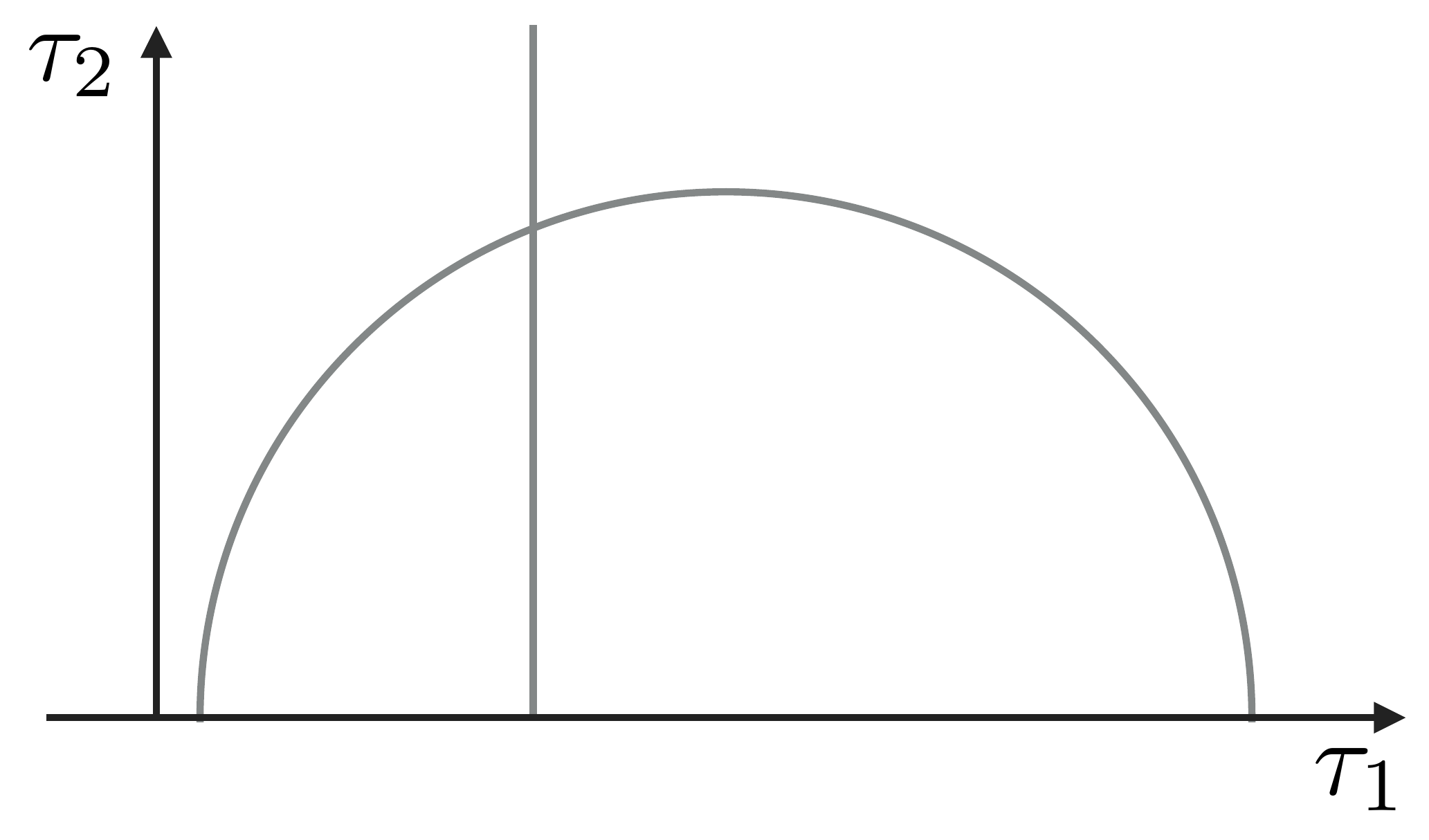}\label{sfig:geodesics}
			}
			\caption{\small a) The upper half-plane $\mathbb{H}$ and the different $SL(2,\mathbb{Z})$ domains one can consider, namely the `triangular' regions. The fundamental one $\mathcal{F}$ is shown shaded in grey. b) Geodesics in the hyperbolic plane are given by vertical straight lines as well as half-circles intersecting the real axis orthogonally.}
			
		\end{center}
\end{figure} 

At this point, one could argue that this might be an artifact of restricting ourselves to the fundamental domain, since the theory actually presents a whole plethora of $\frac{1}{2}$-BPS $(p,q)$-strings whose tension in Planck units is given by equation \eqref{eq:pqstrings9d} above. Furthermore, the latter may provide for the dominant tower of light states required by the SDC upon approaching different infinite distance boundaries of moduli space. In fact, such $(p,q)$ strings present the following species scale vectors
\begin{equation} \label{eq:chargetomasspq}
\mathcal{Z}_{U} = -\frac{1}{2 \sqrt{14}} \, , \qquad	
 \mathcal{Z}_{\tau_2} = \frac{(p+q\tau_1)^2-q^2\tau_2^2}{2\sqrt{2} |p+q\tau|^2}\, , \qquad \mathcal{Z}_{\tau_1} =  -\frac{q \tau_2 (p+q\tau_1)}{\sqrt{2} |p+q\tau|^2}\, ,
\end{equation}
which do not vanish in general. Moreover one can check that upon plotting these vectors taking also into account the $\tau_1$-direction they densely fill a circumference of radius $1/\sqrt{7}$\cite{Etheredge:2022opl}, regardless of the particular point in moduli space we are sitting at.\footnote{This point becomes subtle if one approaches infinite distance by taking e.g. $\tau_2 \to \infty$, since to actually fill the circumference one needs to consider $(p,q)$ bound states with increasingly bigger F1 string charge, namely $p \sim k \tau_2$ with $k\in \mathbb{Z}$, since otherwise their charge-to-mass vectors collapse to those of the F1 or D1 strings. Upon doing so, equation \eqref{eq:chargetomasspq} leads to $\vec{\mathcal{Z}}_{(p,q)} = \frac{1}{\sqrt{8}} \left( \cos 2\theta, -\sin 2\theta\right)$, with $\cos \theta =k/\sqrt{k^2+q^2}$.} This suggests that perhaps one might need to include the axionic component $\mathcal{Z}_{(p,q)}^{\tau_1}$ into our analysis, since it is non zero in general depending crucially on the value of $\braket{\tau_1}$. However, it turns out to be irrelevant whenever we reach an asymptotic boundary of $\mathcal{M}_{\text{9d}}$. The reason being that for geodesic paths reaching an infinite distance singularity other than $\tau_2 \to \infty$, namely any point with $\tau_2=0$ and $\tau_1 \in \mathbb{Q}$ in figure \ref{sfig:funddom}, one finds again both $\dot \tau_1 \to 0$ asymptotically and also $\mathcal{Z}_{\tau_1} \to 0$ for the \emph{lightest} $(p,q)$-string. Indeed,  the geodesics reaching such infinite distance points are given by half-circles orthogonal to the real axis and passing through $(\tau_1, \tau_2)=(-p/q,0)$, see figure \ref{sfig:geodesics}. Hence, upon substituting these asymptotic values for $\tau_2$ and $\tau_1$ we obtain $\vec{\mathcal{Z}}_{(p,q)} \to \left(-\frac{1}{2 \sqrt{14}}, \frac{1}{2\sqrt{2}}, 0 \right)$, matching those from \eqref{eq:speciesscalevectorsstringsT2} and \eqref{eq:chargetomassF1}. As expected, this is merely a manifestation of the fact that an $SL(2, \mathbb{Z})$ transformation relating F1 to any other $(p,q)$-string maps the vertical lines reaching $i\infty$ to the circles intersecting the real axis at $\tau_1=-p/q$, as well as the tension $T_{\text{F1}}$  to $T_{(p,q)}$.

Therefore, the take-home message is that whenever we find some modular sector within the moduli space of our theory and we want to check whether the SSDC is satisfied \emph{asymptotically}, it is enough to restrict ourselves to the fundamental domain and focus just on the saxionic components of the species scale vectors. This follows since any other path reaching an infinite distance boundary for which a different $(p,q)$ tower dominates the asymptotic physics can be effectively translated --- via some modular transformation --- to this simplified setup. Notice that restricting to the fundamental domain means, in turn, that when plotting the convex hull in e.g. figure \ref{sfig:9dCH} it is not necessary to consider directions exploring $\tau_2 \to 0$, since these are already accounted for once we sit in the appropriate duality frame.

\section{The SSDC under Dimensional Reduction}
\label{s:dimensionalreduction}

In this section we discuss the behavior of our bound on the exponential decay rate for the species scale under dimensional reduction. The goal in section \ref{ss:field-theory} is to show that the SSDC is stable under dimensional reduction of a field theory. In fact, we find that it is the strongest bound on the exponential rate of the species scale $\lambda_{\text{sp}}$ yet compatible with this procedure from the bottom-up perspective. For this, we restrict ourselves to those directions in moduli space in which this purely field-theoretical approach suffices to determine the spectrum of towers, i.e., without the need of including the presence of \emph{additional} extended objects such as strings. In section \ref{ss:preservation&saturation} we consider relations between this exponential rate of the species scale and the density parameter of the tower $p$ coming from imposing preservation and saturation of the SSDC under dimensional reduction. Finally, in section \ref{ss:compactificationstring} we go beyond this field theory analysis and we also include the presence of strings (together with their winding modes). We find that, for theories in less than ten dimensions, the SSDC convex hull condition requires from the existence of extra non-perturbative objects for it to be satisfied after the compactification process.

\subsection{Dimensional Reduction of Field Theory} 
\label{ss:field-theory}

Let us consider a field theory in $D=d+1$ dimensions with a single (canonically normalized) modulus that we denote by $\hat \phi$. Even though we restrict to only one modulus, this can be seen as parametrizing any geodesic trajectory in a multi-moduli theory without loss of generality.  Our starting point is to assume that the SSDC is satisfied along this trajectory, so we introduce a tower of states with density parameter $p$ and exponential rate $\lambda_{\text{t}}$ verifying
\begin{equation}
    \lambda_{\text{sp}} = \frac{p}{D-2+p} \lambda_{\text{t}} \geq \frac{1}{\sqrt{(D-1)(D-2)}} \, .
\end{equation}
The idea would be then to dimensionally reduce this theory on a circle and see under which conditions the SSDC is still satisfied in $d$-dimensions. In order to keep the discussion as general as possible, we will consider $\{\lambda_{\text{t}},p\}$, or equivalently $\{\lambda_{\text{sp}},p\}$, as independent and free parameters. This also fits well the interpretation of $\hat \phi$ as parametrizing any geodesic trajectory, since depending on the latter the same tower may have different exponential rates. Let us mention that, even though for the purpose of being general and exhaustive here we are taking these two parameters to be independent, it is reasonable to consider some correlation between them, particularly in relation with saturation of the bound. We will elaborate more on this in the next subsection, where a more UV-inspired guess for the correlation between $p$, $\lambda_{\text{sp}}$, and saturation of the bound will be examined, but for now we remain completely neutral about this correlation and see how far we can go by applying this logic.

After compactifying this field theory on a circle, we end up with a $d$-dimensional one with an extra modulusas well as an extra tower. These are the (canonically normalized) radion $\hat \sigma$ and the KK tower. The relevant scalar charge-to-mass ratios (in $d$-dimensional Planck units) can be found in \cite{Etheredge:2022opl}:
\begin{equation}\label{eq:zvectorafterdimreduction}
	\vec z_{\text{KK}} = \left( 0 \ , \ \sqrt{\frac{d-1}{d-2}} \right) \, , \quad \vec z_{\text{t}} = \left( \lambda_{\text{t}} \ ,\ \frac{1}{\sqrt{(d-1)(d-2)}} \right) \, .
\end{equation}
Using equation \eqref{eq:eff-vector}, we can easily translate this into the species scales vectors:
\begin{equation} \label{vectors}
\begin{split} 
	&\vec{\mathcal{Z}}_{\text{KK}} = \frac{1}{d-1} \, \vec z_{\text{KK}} = \left( 0 \ ,\ \frac{1}{\sqrt{(d-1)(d-2)}} \right) \, ,\\
	&\vec{\mathcal{Z}}_{\text{t}} = \frac{p}{d-2+p} \, \vec z_{\text{t}} = \left( \frac{d-1+p}{d-2+p} \ \lambda_\text{sp} \ ,\ \frac{p}{(d-2+p)\sqrt{(d-1)(d-2)}} \right) \, ,\\
	&\vec{\mathcal{Z}}_{(\text{KK, t})} = \frac{1}{d-1+p} \left( \vec z_{\text{KK}} + p\, \vec z_{\text{t}} \right) = \left( \lambda_\text{sp} \ ,\ \frac{1}{\sqrt{(d-1)(d-2)}} \right) \, ,
\end{split}
\end{equation}
where we have used that the KK tower of the circle has density parameter $p=1$. Additionally, we have expressed everything in terms of $\lambda_{\text{sp}}$ instead of $\lambda_{\text{t}}$, since that is the relevant parameter for the SSDC in $D$-dimensions.

\begin{figure}[htb]
\begin{center}
\includegraphics[width=0.45\textwidth]{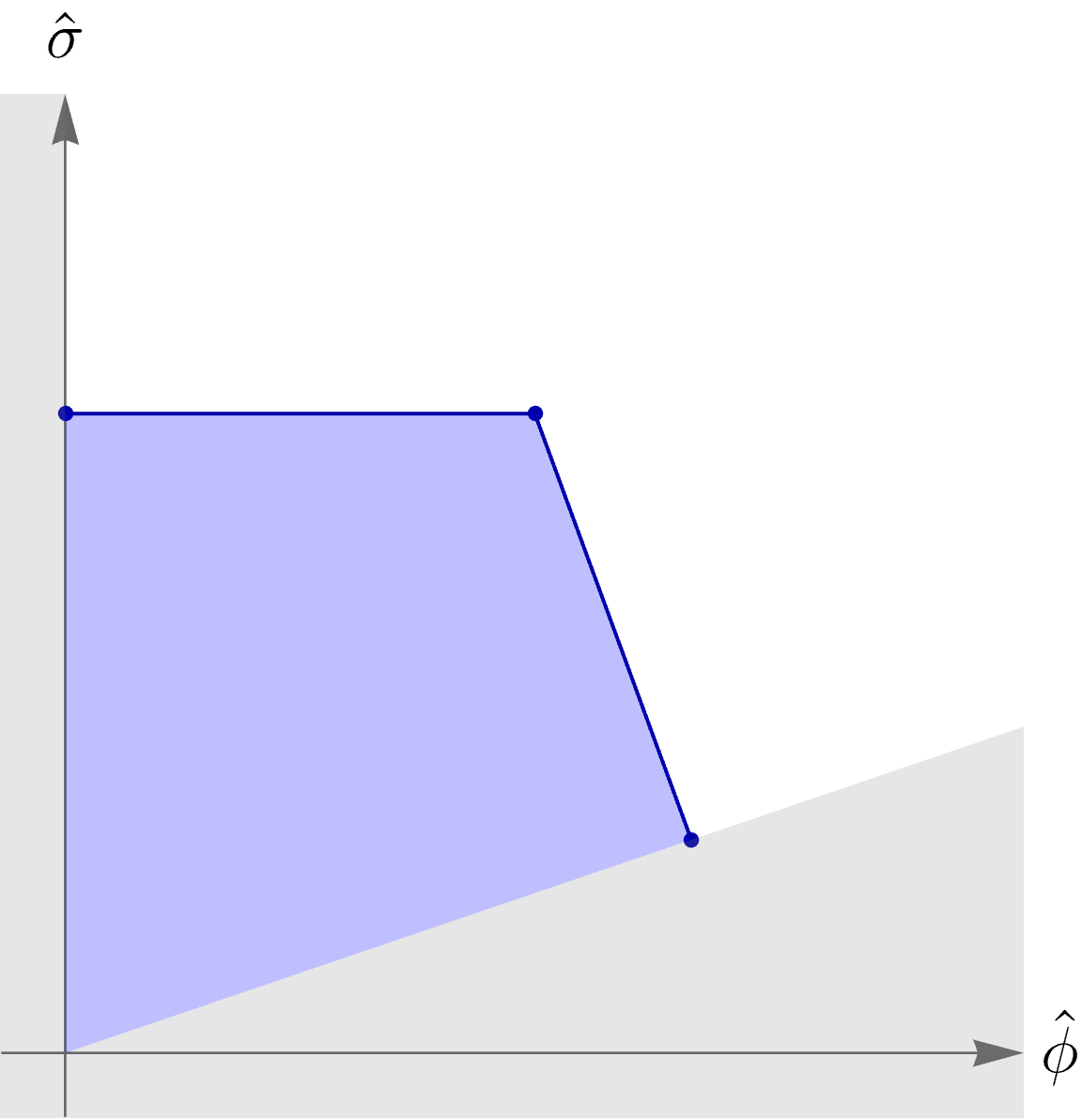}
\caption{\small Example of the convex hull generated by the vectors in \eqref{vectors}, restricted to the region in which these towers of states are enough to build it. In the gray region, new towers coming from extended objects in $D$-dimensions are expected to become relevant and complete the rest of the convex hull.} 
\label{fig:ch-example}
\end{center}
\end{figure}

With this information, we are ready to test the SSDC in the $d$-dimensional theory. The first thing to be noticed is that these towers are not enough to satisfy the conjecture in all geodesic directions in moduli space. As it is well-known, the SDC requires elements beyond field theory for it to be satisfied after dimensionally reducing on a circle, for instance in the limit in which the $S^1$ becomes very small. Indeed, the SDC requires the presence of $D$-dimensional extended objects that can wrap the $S^1$, e.g., a string along with its winding modes. In order to remain agnostic about these ingredients beyond field theory, in this section we focus on the directions in which the vectors in \eqref{vectors} are enough to build the convex hull (i.e. those between the vectors $\vec{\mathcal{Z}}_{\text{KK}}$ and $\vec{\mathcal{Z}}_{\text{t}}$), and study the conjecture in that regime, leaving the inclusion of extended objects in this context for section \ref{ss:compactificationstring}. An example of this restriction and the part of the convex hull generated by the species vectors in \eqref{vectors} is shown in figure \ref{fig:ch-example}. There we see that the boundary of the convex hull is given by the two lines joining $\vec{\mathcal{Z}}_{\text{KK}}$ with $\vec{\mathcal{Z}}_{(\text{KK, t})}$ and $\vec{\mathcal{Z}}_{(\text{KK, t})}$ with $\vec{\mathcal{Z}}_{\text{t}}$. In the following, we will discuss the implications of these two lines for the SSDC under dimensional reduction separately.

%%%%%%%%%%%%
%%%%%%%%%%%%

Let us start with the first line, namely that going from $\vec{\mathcal{Z}}_{\text{KK}}$ to $\vec{\mathcal{Z}}_{(\text{KK, t})}$. As one can see from the form of these vectors in \eqref{vectors}, this line is horizontal for any value of $\lambda_{\text{sp}}$ and $p$. Furthermore, the distance from this line to the origin is always given by the SSDC minimum value,
\begin{equation}\label{eq:lspmin}
    \lambda_{\text{sp}, \, \text{min}}\, =\, \frac {1}{\sqrt{(d-1)(d-2)}} \, ,
\end{equation}
again independently of the values of $\lambda_{\text{sp}}$ and $p$. With these simple observations, one arrives to an interesting conclusion: the SSDC is the \emph{strongest} bound on the species scale exponential rate yet compatible with dimensional reduction on a circle. 

The fact that the KK tower of an $S^1$ yields this particular value was already pointed out in \cite{vandeHeisteeg:2023uxj} as an argument for this bound, in relation to the Emergent String Conjecture \cite{Lee:2019wij}. Their results for the 4d $\mathcal{N}=2$ setup moreover suggest that the bound could be stable beyond the strictly asymptotic regime, due to the positivity of the logarithmic corrections found therein.\footnote{We thank M. Wiesner for pointing this out to us.} Here we uncover that, if the $D$-dimensional theory has a moduli space, it is crucial to take into account the effective towers for the convex hull condition to be satisfied in the $d$-dimensional one. In particular, the saturation found in the direction of $\vec{\mathcal{Z}}_{\text{KK}}$ would present a problem for the condition to be satisfied along other (neighbouring) directions, unless other species scale vector were present to give rise to a boundary for the convex hull that were perpendicular to this $\vec{\mathcal{Z}}_{\text{KK}}$ direction. As we just saw, this role is played by $\vec{\mathcal{Z}}_{(\text{KK, t})}$. Interestingly, this happens independently of the details of the tower in $D$-dimensions, i.e., of $\lambda_{\text{sp}}$ and $p$. As discussed above, since we interpret $\hat \phi$ as parametrizing any geodesic trajectory in the moduli space of the $D$-dimensional theory, this conclusion is not limited to theories with a single modulus. In summary, we find a pretty robust mechanism protecting the SSDC convex hull condition whenever it is saturated by the KK tower of an extra dimension.

\begin{figure}[htb]
\begin{center}
	\subfigure[\label{sfig:4dp=0.2}$d=4$ and $p=0.2$]{\includegraphics[width=0.32\textwidth]{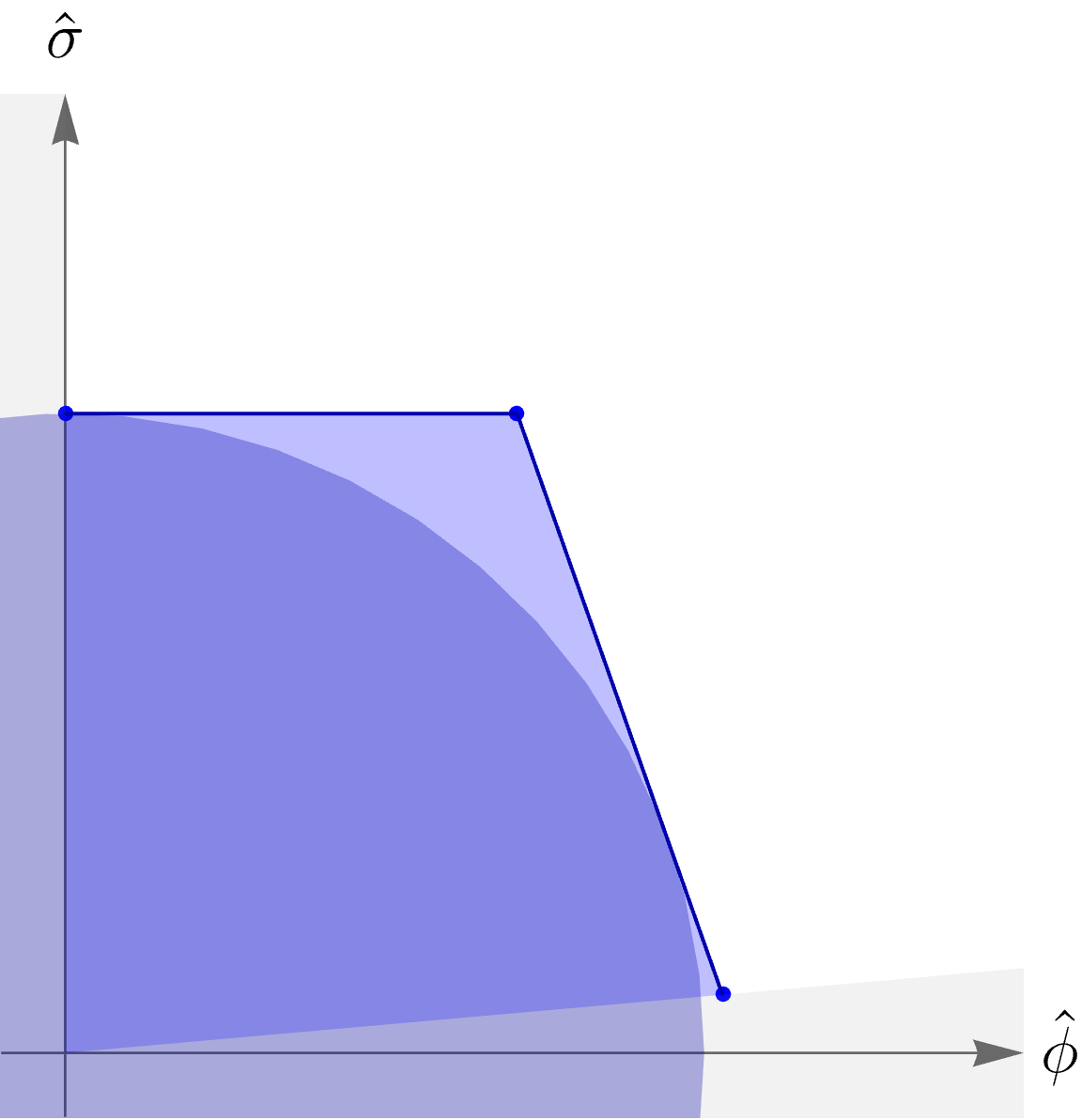}}
	\subfigure[\label{sfig:4dp=1}$d=4$ and $p=1$]{\includegraphics[width=0.32\textwidth]{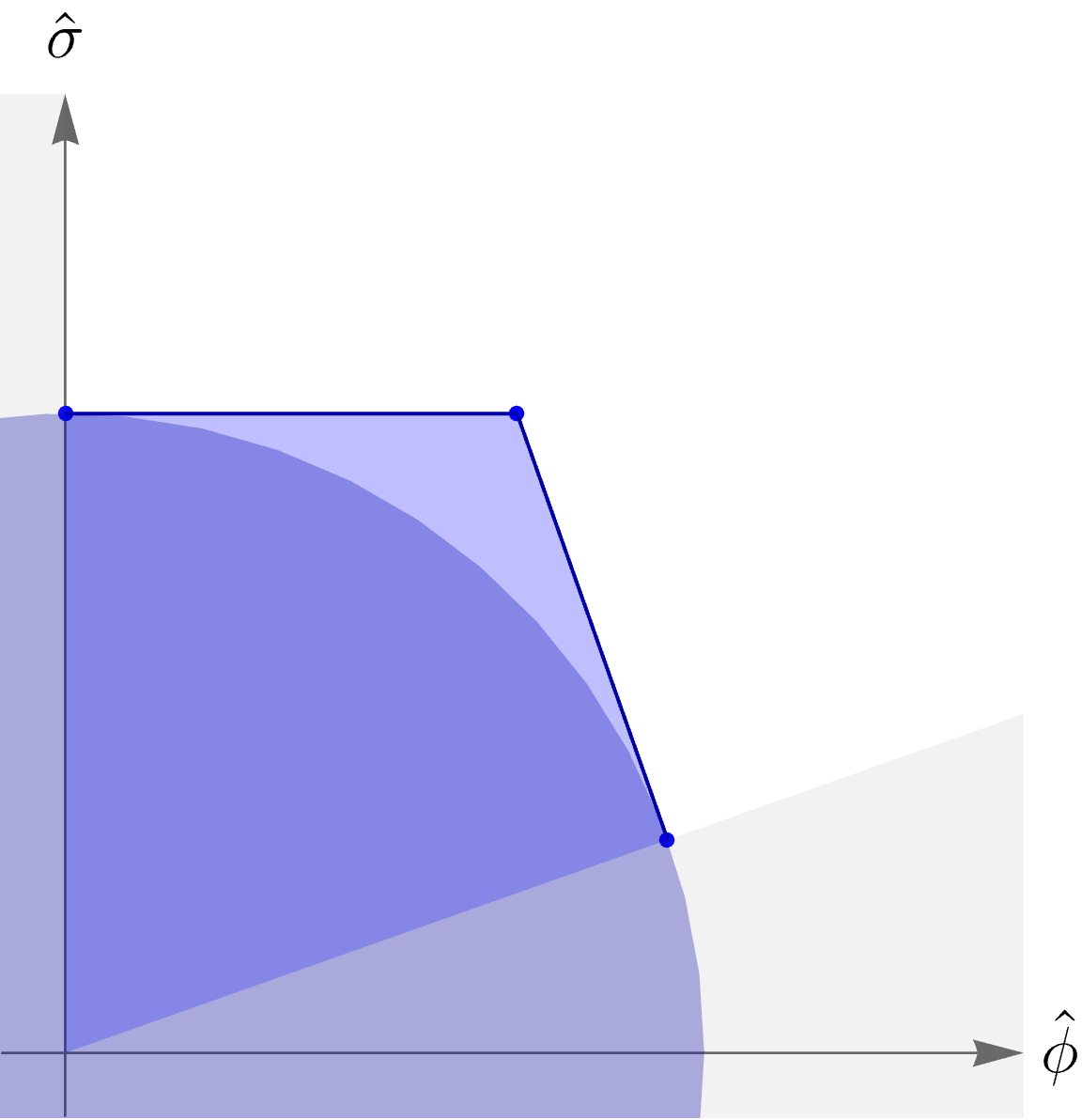}}
	\subfigure[\label{sfig:4dp=5}$d=4$ and $p=5$]{\includegraphics[width=0.32\textwidth]{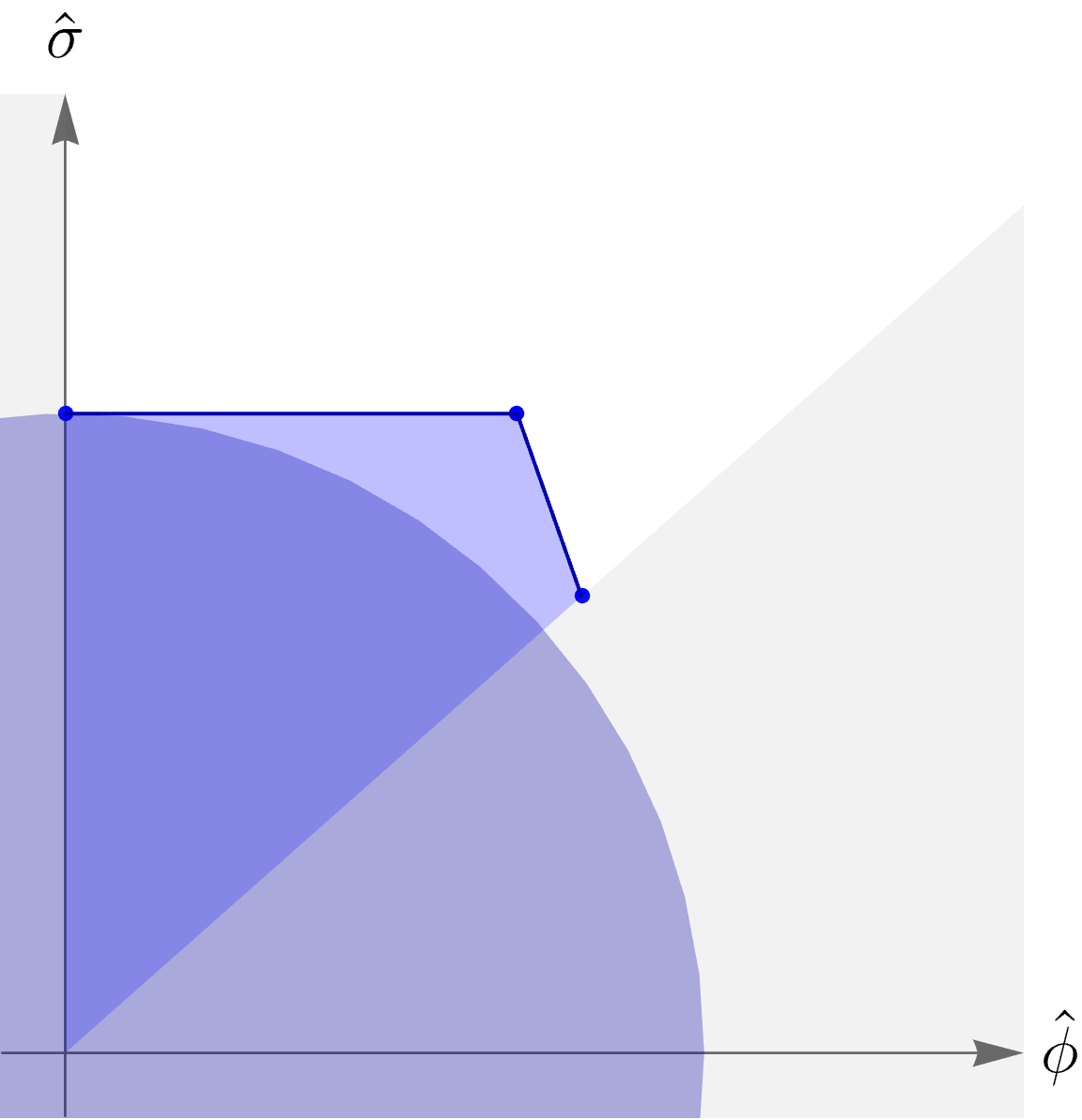}}
	\caption{Convex hull SSDC condition in $d=4$ dimensions for three different values of $p$ when $\lambda_{\text{sp}}$ saturates the SSDC in $D=5$ dimensions.}
	\label{fig:dim-red}	
\end{center}
\end{figure}

Let us now turn our attention to the line going from $\vec{\mathcal{Z}}_{(\text{KK, t})}$ to $\vec{\mathcal{Z}}_{\text{t}}$. First we notice that, for fixed $\lambda_{\text{sp}}$, varying $p$ only changes the length of this line but not its slope. Therefore, as we change $p$, the distance from the origin to this line is always bigger than the distance to its infinite length extension. Additionally, $\lambda_{\text{sp}}$ only appears in second components of the vectors in \eqref{vectors} in such a way that, for bigger $\lambda_{\text{sp}}$, this line will be further away from the origin. Physically, this is certainly meaningful, since it means that the closest the SSDC is to be violated in $D$-dimensions, the closest it gets to be violated after dimensional reduction, too. The most dangerous situation is then when the SSDC is saturated in $D$-dimensions. As shown in figure \ref{fig:dim-red}, even in this case the line under consideration does not lead to a violation of the SSDC, regardless of the value of $p$. Therefore, this line is compatible with our previous conclusion, that the SSDC is the strongest bound compatible with dimensional reduction on a circle. In fact, this line would also formally lead to saturation of the SSDC for $p\leq 1$. Let us however notice that, even though in this general approach we let $p$ take any value, having $p<1$ seems rather unphysical from our experience with string theory examples.

%%%%%%%%%%%%
%%%%%%%%%%%%

In conclusion, we see that the SSDC is preserved under dimensional reduction on a circle (within the directions in which field theory is enough to determine the convex hull). In fact, it turned out to be the strongest bound on the species scale exponential rate yet compatible with this procedure. Additionally, we saw that the effective tower is crucial to protect the convex hull condition when it is saturated by the KK tower of an extra dimension. All these conclusions are valid for any $p$ and any $\lambda_{\text{sp}}$ satisfying the SSDC in $D$-dimensions, thus are very robust.

\subsubsection*{Generalization for $D=d+n$ dimensions}

It is worth emphasizing that the previous conclusion also holds for the more general case of compactification on a $n$-dimensional internal manifold $\mathcal{X}_n$ (e.g. a $T^n$). We denote $\mathcal{V}_n$ the overall volume modulus associated to the internal compact manifold, $\mathcal{X}_n$, measured in $D$-dimensional Planck units. The relevant sector of the theory is thus described by the following action in $D$ dimensions
\begin{equation}\label{eq:Ddim}
	S_{D} = \int d^{D}x\, \sqrt{-g}\,  \left( \frac{1}{2\kappa_{D}^2} R - \frac{1}{2} \left(\partial \hat \phi \right)^2 \right)\, .
\end{equation}
Upon compactification on $\mathcal{X}_n$ one arrives at the following Einstein-frame action
\begin{equation}\label{eq:ddim}
	S_{d} = \int d^{d}x\, \sqrt{-g}\,  \left[ \frac{1}{2\kappa_{d}^2} \left(R - \frac{d+n-2}{n (d-2)} \left(\partial \log \mathcal{V}_n \right)^2 \right)- \frac{1}{2} \left(\partial \hat \phi \right)^2 \right]\, ,
\end{equation}
where we have retained only the massless scalar-tensor sector of the $d$-dimensional theory. The canonically normalized volume modulus is then given by
\begin{equation}\label{eq:canonicalvolume}
	\hat \sigma = \frac{1}{\kappa_d}\sqrt{\frac{d+n-2}{n(d-2)}} \log \mathcal{V}_n\, .
\end{equation}
One finds the following relevant scalar charge-to-mass vectors for the (isotropic) KK tower (with density parameter $p_{\text{KK}}=n$) and the original $D$-dimensional tower, respectively \cite{Etheredge:2022opl}:
\begin{equation}\label{eq:zvectornmfd}
	\vec{z}_{\text{KK},\, n} = \left( 0 \ ,\ \sqrt{\frac{d+n-2}{n(d-2)}} \right) \, , \quad \vec z_{\text{t}} = \left( \lambda_{\text{t}} \ ,\ \sqrt{\frac{n}{(d+n-2)(d-2)}} \right) \, .
\end{equation}
From these one obtains three species scale vectors (c.f. equation \eqref{eq:eff-vector})
\begin{equation}
\label{eq:Zspeciescompactification}
\begin{split} 
	&\vec{\mathcal{Z}}_{\text{KK},\, n} = \left( 0 \ ,\ \sqrt{\frac{n}{(d+n-2)(d-2)}} \right) \, ,\\
	&\vec{\mathcal{Z}}_{\text{t}} = \left( \frac{d-1+p}{d-2+p} \ \lambda_{\text{sp}} \ ,\ \frac{p}{d-2+p} \sqrt{\frac{n}{(d+n-2)(d-2)}} \right) \, ,\\
	&\vec{\mathcal{Z}}_{(\text{KK, t})}  = \left( \frac{d-1+p}{d-2+n+p} \ \lambda_{\text{sp}} \ ,\ \sqrt{\frac{n}{(d+n-2)(d-2)}} \right) \, .
\end{split}
\end{equation}
Notice that we again see that the line determined by $\vec{\mathcal{Z}}_{\text{KK},\, n}$ and $\vec{\mathcal{Z}}_{(\text{KK, t})}$ is horizontal. Contrary to the case of compactification on a circle, this is not actually needed to protect the SSDC bound since $\vec{\mathcal{Z}}_{\text{KK},\, n}$ does not saturate it for $n>1$. Despite this, the conclusion that this line preserves the SSDC bound holds for $n>1$. We just found that the most constraining compactification for the exponential rate of the species scale corresponds to $n=1$, i.e., a circle/interval compactification.

Similarly to the case of a circle compactification, the slope the line connecting $\vec{\mathcal{Z}}_{t}$ with $\vec{\mathcal{Z}}_{(\text{KK, t})}$ is independent of $p$. It also holds that this line gets further away from the origin for larger $\lambda_{\text{sp}}$. In the most dangerous situation, in which the SSDC is saturated in $D=d+n$ dimensions, this line again does not lead to violation of the SSDC in $d$-dimensions. In fact, this line again leads to saturation of the SSDC for $p\leq 1$. The convex hull in this case looks very much like the ones in figure \ref{fig:dim-red}, but with the horizontal line being shifted above according to the second component of $\vec{\mathcal{Z}}_{\text{KK},\, n}$ while keeping fixed that the infinite length extension of the other line touches the ball. All in all, the conclusion that the SSDC is preserved under dimensional reduction (within the directions in which field theory is enough to determine the convex hull), at least along the direction given by the overall volume modulus, gets extended to $n>1$.   

\subsection{Preservation and Saturation Under Dimensional Reduction}
\label{ss:preservation&saturation}

In the previous section we considered a situation where any combination of the parameters $\lambda_{\text{sp}}$ and $p$ for the tower satisfying the SSDC in $D$-dimensions was allowed, as long as they were compatible with the bound $\lambda_{\text{sp}}\geq 1/\sqrt{(D-1)(D-2})$. Let us remark, however, that if an EFT observer can probe the species scale (i.e. can know the parameter $\lambda_{\text{sp}}$), it should also have access to all the states below such energy, and therefore the corresponding $p$ should also be known. Hence, even though they can indeed be independent parameters, having access to one of them implies knowing about the value of the other. One of the goals of this section is then to explore this idea and, in particular, whether knowledge of $\lambda_{\text{sp}}$ with respect to the SSDC bound (namely whether it is saturated or not, and if not, by how much), combined with preservation under dimensional reduction, can restrict the possible values for $p$.

To do so, we study how the asymptotic decay rates for the mass scale of the towers and the species scale behave under dimensional reduction, both for Kaluza-Klein and emergent string limits. As in the previous subsection, we start by considering the compactification of some $D$-dimensional theory down to $d=D-1$ spacetime dimensions. The  mass scales of the relevant towers decay exponentially as determined by the scalar charge-to-mass ratios \eqref{eq:zvectornmfd} (i.e. in $d$-dimensional Planck units they scale as $m\, \sim\, e^{-\vec{\Phi}\cdot \vec{z}} \,$, with $\vec{\Phi}= ( \hat{\phi} \, ,\ \hat{\sigma})$, for each possible value of  $\vec{z}$). With this, one can then examine how $\lambda_D = |\Vec{z}|$ should depend on the spacetime dimension, $D$, in order to be preserved under the compactification process \cite{Etheredge:2022opl}. This amounts to requiring
\begin{equation}\label{eq:preservationS1}
	\lambda_{d+1}^2+ \frac{1}{(d-1)(d-2)}=\lambda_d^2\, ,
\end{equation}
and the general solution to this equation reads \cite{Etheredge:2022opl}\footnote{This solution can be obtained upon iteration of the recursion relation to give
\begin{equation}
 \lambda_{d}^2-\lambda_{d=3}^2  = \sum_{i=4}^d \lambda_{i}^2-\lambda_{i-1}^2=\sum_{i=4}^d-\dfrac{1}{(i-2)(i-3)}=\frac{1}{d-2}-1 \, ,
\end{equation}
from which we then obtain \eqref{eq:solpreservationtower} upon relabeling the $d$-independent coefficient $\beta=\lambda_{d=3}^2-1$ that arises as an initial condition.}
\begin{equation}
\label{eq:solpreservationtower}
    \lambda_{d}^2=\frac{1}{d-2} + \beta\, , 
\end{equation}
where $\beta$ denotes some arbitrary parameter that does not depend on  $d\,$. 

Notice that, upon iterating $(n-1)$ times equation \eqref{eq:preservationS1} in order to relate $\lambda_{d+n}$ with $\lambda_d^2$, we get
\begin{equation}\label{eq:preservation}
	\lambda_{d+n}^2+ \frac{n}{(d+n-2)(d-2)}=\lambda_d^2\, ,
\end{equation}
which is nothing but the condition for preservation of the charge-to-mass ratio associated to some already existing tower upon compactifying from $D=d+n$ to $d$ dimensions. This is, the recursion relation obtained from the compactification of just one dimension already contains all the necessary information about the $d$-dependence of $\lambda_d$ for it to be preserved (along the plane spanned by the original scalar modulus and the canonically normalized volume direction) upon compactifying on an $n$-dimensional manifold.

Let us now consider the condition for preservation of the species scale decay rate, $\lambda_{\text{sp}}$, upon compactification from $D=d+1$ to $d$ dimensions. Using equation \eqref{eq:Zspeciescompactification} for the $D$-dimensional tower with arbitrary $p$ and identifying $\lambda_{\text{sp},\, d}=|\vec{\mathcal{Z}}_t|$, when $n=1$, yields
\begin{equation}
  \left( \dfrac{d+p-1}{d+p-2}\right)^2 \, \lambda_{\text{sp},\, d+1}^2+ \dfrac{p^2}{(d+p-2)^2\, (d-1)\, (d-2)}\, =\, \lambda_{\text{sp},\, d}^2 \, .
\end{equation}
In fact, this is exactly the same relation that we obtain if we just rewrite equation \eqref{eq:preservationS1} in terms of its associated $\lambda_{\text{sp}}$ instead of $\lambda$. Therefore, it can be checked that the solution to this recursion relation coincides with the one we would obtain from computing the $\lambda_{\text{sp},\, d}$ associated to the $\lambda_d$ in equation \eqref{eq:solpreservationtower} above, that is\footnote{In analogy to the previous case, the $d$-independent parameter $\beta$ is related to an initial condition for the recurrence relation given by ${\beta= \left( \frac{p+1}{p}\right)^2 \lambda_{\text{sp}, \, d=3}^2 -1}$, and one can check that, as expected for $d=3$, ${\lambda_{d=3}^2=\left( \frac{p+1}{p}\right)^2 \lambda_{\text{sp},\, d=3}^2}\,$.}
\begin{equation}
\label{eq:solpreservationspecies}
    \lambda_{\text{sp},\, d}^2\, =\, \left(\dfrac{1}{d-2}+\beta \right)\, \left( \dfrac{p}{d+p-2}\right)^2 \, .
\end{equation}
To sum up, the asymptotic decay rates associated to the mass and species scales from higher-dimensional towers given by eqs. \eqref{eq:solpreservationtower} and \eqref{eq:solpreservationspecies} are preserved upon dimensional reduction for any $d$-independent (but possibly $p$-dependent) $\beta$. For the former, $\beta=0$ was proposed in \cite{Etheredge:2022opl} as the preferred value based on examples and on the fact that it yields the lowest possible $\lambda_{\text{min}}^2=\frac{1}{d-2}$. Moreover, this is the only non-negative $\beta$ that gives rise to preservation of the convex hull condition along the directions comprised between the higher-dimensional tower and the Kaluza-Klein tower associated to the compactification if we assume saturation in the higher dimensional theory. Notice that since this bound is not sensitive to the density parameter, this argument alone did not allow for an identification of the nature of tower that would saturate it, even though in all the examples the relevant towers were stringy.

For the species-scale charge to mass ratio \eqref{eq:solpreservationspecies}, however, since it depends on $p$, we can try to obtain some information about what towers would saturate our bound \eqref{eq:lambdaspeciesbound}. In particular, we can ask under which conditions saturation in $(d+1)$-dimensions leads to saturation in $d$-dimensions. Thus, we need to find the possible solutions to the equation
\begin{equation}
    \left(\dfrac{1}{d-2}+\beta \right)\, \left( \dfrac{p}{d+p-2}\right)^2\, = \, \dfrac{1}{(d-1)(d-2)}\, ,
\end{equation}
in terms of a $d$-independent $\beta$. The general solution for such equation is
\begin{equation}
\label{eq:betasolution}
    \beta=\dfrac{d-2-p(p-2)}{p^2 \, (d-1)} \, ,
\end{equation}
which is clearly not $d$-independent in general. The only universal, $d$-independent value for $p$ that gives rise to a $d$-independent $\beta$, and hence saturation of our bound, is $p=1$ (which also yields $\beta=1$).\footnote{The only two options to get a $d$-independent $\beta$ in equation \eqref{eq:betasolution} are either having a vanishing numerator, or getting it numerator to cancel the $(d-1)$ dependence in the denominator, possibly leaving some extra dependence on $p$. The first option yields $p=1\pm \sqrt{d-1}$, which is in general not an integer and also cannot be fulfilled for all $d$. The second one is in general more complicated, but if we allow for a possible monomial dependence of $\beta=p^\alpha$, we find the surprising result that the only $d$-independent solution is $p=1$, and thus $\beta=1$, apart from some particular $d$-dependent solutions that also depend on $\alpha$ and are in typically irrational. It would be interesting to investigate further these non-generic solutions that could in principle give rise to saturation only in certain dimensions and for particular density parameters, such as e.g. a $p=3$ tower in $d=5$, but we leave this for future work.} Thus we expect it to be saturated in general only by single Kaluza-Klein towers, as we observe in the different examples throughout this paper. Let us emphasize, however, that this argument is completely independent of examples and clearly distinguishes $p=1$ towers as those saturating our entropic bound on the species scale decay rate, $\lambda_{\text{sp, min}}=1/\sqrt{(d-1)(d-2)}$. Additionally, it is remarkable that a string tower could never saturate our bound and be preserved under dimensional reduction, since $p\to \infty$ would inevitably yield a $d$-dependent $\beta$, which is inconsistent. 

Equipped with these insights, let us explore the choice $\beta=1/p$, which yields 
\begin{equation}
\label{eq:solpreservationtower1/p}
    \lambda_{d}^2=\dfrac{d+p-2}{p(d-2)}\, , \qquad \lambda_{\text{sp},\, d}^2=\dfrac{p}{(d+p-2)(d-2)}\, ,
\end{equation}
and can be motivated from different angles. On the one hand, this way of including the dependence on $p$ interpolates between the case of a string tower ($p\to \infty$) and a single Kaluza-Klein tower ($p=1$), in such a way that the minimum value for $\lambda_d$ is recovered for the former (effectively recovering the results from \cite{Etheredge:2022opl} with the extra output that string towers are the ones that saturate their bound), and the minimum value for $\lambda_{\text{sp}, \, d}$ for the latter. Besides, these also recover the typical values exhibited by KK and string towers (for all the corresponding values of $p$) obtained in simple compactifications that are preserved upon dimensional reduction, as seen in the different examples that we consider (see section \ref{ss:MthyT3} below). Thus, we see that whereas for the single Kaluza-Klein tower the charge-to-mass ratio and the species scale vector are maximally separated, for the stringy case they converge to the same value, with all the values of $p$ in between giving rise to intermediate separations. This explains why the lower bounds for $\lambda_d$ are expected to be saturated by towers of strings, whereas the lower bounds for $\lambda_{\text{sp}, \, d}$ are expected to be saturated by single Kaluza-Klein towers, since each of them gives respectively the smallest possible vector. Moreover, notice that the string tower case is some kind of fixed point in the space of charge-to-mass and species scale vectors (see e.g. figure \ref{fig:ch2}), due to the fact that the species scale and the string mass coincide in such limits. In fact, all the other pairs of vectors are somehow dual with respect to the sphere with radius given by the string towers vectors, and one member of the pair is always in the interior of such  ball (but still far enough from the center so as to satisfy the convex hull condition for the species scale), whereas the other has to be in the exterior in order to  satisfy the convex hull condition for the charge-to-mass ratio associated to the mass scale of the tower (see \cite{Castellano:2023stg, Castellano:2023jjt} for a precise treatment of this correspondence). 

Finally, let us review how the actual convex hull condition for the species scale is satisfied whenever the norms of the corresponding species scale vectors are preserved upon dimensional reduction. So far, we have checked that if the species vector has the form \eqref{eq:solpreservationspecies}, then the Species Scale Distance Conjecture will be preserved along the direction of the dimensionally reduced species scale vector, but what about the remaining asymptotic directions, such as the one associated to the volume modulus considered in section \ref{ss:field-theory}? By sticking to the choice $\beta=1/p$ that we have just argued for, we can see that upon compactification from $D=d+n$ down to $d$ dimensions, the relevant species scale vectors given by \eqref{eq:Zspeciescompactification} become
\begin{equation}\label{eq:vectorsspeciesscalebox}
\begin{split} 
	&\vec{\mathcal{Z}}_{\text{KK}} = \left( 0 \, , \ \sqrt{\frac{n}{(d+n-2)(d-2)}} \right) \, ,\\
	&\vec{\mathcal{Z}}_{\text{t}} = \left( \sqrt{\dfrac{p\, (d+n+p-2)}{(d+p-2)^2 \, (d+n-2)}} \, , \ \sqrt{\dfrac{n \, p^2}{(d+p-2)^2 \, (d+n-2) \, (d-2)}} \right) \, ,\\
	&\vec{\mathcal{Z}}_{(\text{KK, t})} = \left( \sqrt{\frac{p}{(d+n-2)(d+n+p-2)}}\,  , \ \sqrt{\frac{n}{(d+n-2)(d-2)}} \right) \, .
\end{split}
\end{equation}
Actually, the three vectors (and not only $\vec{\mathcal{Z}}_{\text{t}}$) satisfy the condition \eqref{eq:solpreservationtower1/p} in $d$ dimensions for the suitable identifications of the density parameter of the corresponding tower, namely $p = \lbrace n, \ p,\  n+p \rbrace$, respectively. In fact, notice that this is effectively recovering the convex hull associated to a compactification on a product space of the form $ \mathcal{Y}_p \times \mathcal{X}_n$, where we keep track of their corresponding volume directions. 

Besides, there are two important facts associated to the triplet \eqref{eq:vectorsspeciesscalebox} regarding the preservation of the convex hull condition in the $d$-dimensional theory: first, the line determined by $\left( \vec{\mathcal{Z}}_{\text{KK}},  \vec{\mathcal{Z}}_{(\text{KK, t})} \right)$ is always horizontal; and second, the segment $\left( \vec{\mathcal{Z}}_{(\text{KK, t})},  \vec{\mathcal{Z}}_t \right)$ can be seen to be perpendicular to the vector $\vec{\mathcal{Z}}_{\text{t}}\, $.\footnote{To check this we just compute $\vec u = \vec{\mathcal{Z}}_{(\text{KK, t})} - \vec{\mathcal{Z}}_{\text{t}}$, which verifies $\vec{\mathcal{Z}}_t \cdot \vec u=0$ regardless of the particular values for the dimension of the compactification space, $n$, and the density parameter of the tower, $p$.} This means, in particular, that if both $\vec{\mathcal{Z}}_{\text{KK}}$ and $\vec{\mathcal{Z}}_{\text{t}}$ satisfy $\lambda_{\text{sp}} \geq \lambda_{\text{min}}$, the convex hull generated by the three vectors in \eqref{eq:vectorsspeciesscalebox} automatically contains the ball of radius $\lambda_{\text{min}}$. Therefore, since the minimum of $|\vec{\mathcal{Z}}_{\text{KK}}|$ (alternatively $|\vec{\mathcal{Z}}_{\text{t}}|$) happens precisely when $n$ (or $p$) is equal to one, one finds that our bound \eqref{eq:lambdaspeciesbound}
is indeed exactly preserved under dimensional reduction, in the sense that saturation in $D=d+n$ dimensions (i.e. when $p=1$) leads also to saturation in $d$ dimensions. Moreover, in the special case when $n=1$, both segments of the convex hull become tangent to the ball. Whenever $n, \ p >1$, the bound is satisfied, but not saturated. 

\subsection{Beyond Field Theory: Including Strings and Windings}
\label{ss:compactificationstring}

In this section we go beyond compactification of a field theory by considering a toy model with two strings in $D$-dimensions. This is, we will consider a setup (clearly inspired by type IIB string theory) with two different strings becoming tensionless along two infinite distance limits, namely as $\hat \phi \to \pm \infty$. The corresponding string excitation modes will become asymptotically massless with an exponential rate $\lambda_{\text{str}}$, which in this case coincides with the exponential rate for the species scale. 

The dimensional reduction procedure on a circle goes through exactly as in section \ref{ss:field-theory}, with the difference that we have extra towers in $d$-dimensions coming from the winding modes of these extended objects. The relevant scalar charge-to-mass vectors read \cite{Etheredge:2022opl}
\begin{equation}\label{eq:zvectorsp=1}
\begin{split} 
	\vec z_{\text{KK}} =& \left( 0 \ ,\ \sqrt{\frac{d-1}{d-2}} \right)\, , \qquad \vec z_{\text{wind}, \, 1} = \left( 2 \lambda_{\text{str}} \ ,\ - \frac{d-3}{\sqrt{(d-1)(d-2)}} \right) \, ,\\
	\vec z_{\text{wind}, \,2} =& \left( -2 \lambda_{\text{str}} \ ,\ - \frac{d-3}{\sqrt{(d-1)(d-2)}} \right) \, ,
\end{split}
\end{equation}
for the $p=1$ towers, whilst for the strings one finds
\begin{equation}\label{eq:zvectorsStrings}
  \vec{z}_{\text{str},\, 1} =\left( \lambda_{\text{str}} \ ,\ \frac{1}{\sqrt{(d-1)(d-2)}} \right) \, , \qquad 
  \vec{z}_{\text{str},\, 2} = \left( - \lambda_{\text{str}} \ ,\ \frac{1}{\sqrt{(d-1)(d-2)}} \right) \, .
\end{equation}
Moreover, upon using equation \eqref{eq:eff-vector} above, we can translate the set \eqref{eq:zvectorsp=1} into the following species scale vectors
\begin{equation}\label{vectors2}
\begin{split} 
	\vec{\mathcal{Z}}_{\text{KK}} =& \left( 0 , \frac{1}{\sqrt{(d-1)(d-2)}} \right)\, , \qquad \qquad \qquad \ \ \vec{\mathcal{Z}}_{\text{wind}, \, 1} = \left( \frac{2 \lambda_{\text{str}}}{d-1} , - \frac{d-3}{(d-1)^{3/2}\sqrt{(d-2)}} \right) \, ,\\
	\vec{\mathcal{Z}}_{\text{wind}, \,2} =& \left( -\frac{2 \lambda_{\text{str}}}{d-1} , - \frac{d-3}{(d-1)^{3/2}\sqrt{(d-2)}} \right) \, , \quad \vec{\mathcal{Z}}_{\text{wind}, \,(12)} = \left( 0 , - \frac{2(d-3)}{d\sqrt{(d-1)(d-2)}} \right) \, .
\end{split}
\end{equation}
Here we have taken into account several things. First, notice that strings do not form `effective towers', since they already saturate the species scale alone. This can be intuitively understood \cite{Castellano:2021mmx} upon taking $p\to\infty$ in equation \eqref{eq:eff-vector}. Additionally, we have assumed the two towers of winding modes to be multiplicative. In type IIB string theory this comes about from having not only the fundamental and the D1-string, but also the spectrum of $(p,q)$ bound states thereof \cite{Witten:1995im}. We will see later how this fits nicely with the results. Finally, even though a priori one could also consider winding and KK modes to be multiplicative, since they never become light simultaneously (indeed they are T-dual towers), they are to all effects irrelevant for testing the SSDC.

With this information, we are now ready to test the behavior of the SSDC under dimensional reduction within the present toy model. First, notice that as also happened in section \ref{ss:field-theory}, $\lambda_{\text{str}}$ only appears in the first components of (some of) the vectors in \eqref{eq:zvectorsStrings} and \eqref{vectors2}, in such a way that for larger values all these vectors go further away from the origin. This again fits the physical intuition that the smaller $\lambda_{\text{str}}$ is, the closer will be the convex hull in $d$-dimensions to violate the SSDC. One can then try to fix $\lambda_{\text{str}}$ so as to saturate the SSDC in the parent theory, i.e.,
\begin{equation}
    \lambda_{\text{str}} = \frac{1}{\sqrt{(D-1)(D-2)}} = \frac{1}{\sqrt{d(d-1)}} \, .
\end{equation}
Plugging this into the previous set of $\mathcal{Z}$-vectors and drawing the convex hull, one realizes that there is a violation of our bound for any spacetime dimension $d$. For illustrative purposes, this is depicted in the $d=9$ case in figure \ref{fig:ch3} below. Let us stress, however, that this by itself is \emph{not} a violation of the SSDC in Quantum Gravity, as we are just considering a toy model in which a stringy decay rate $\lambda_{\text{str}}$ is taken to saturate the SSDC in $D$-dimensions. In fact, in Quantum Gravity critical strings do not present this value for the exponential decay rate but instead a larger one. In this sense, consistency under dimensional reduction of the SSDC is telling us that, indeed, this toy model in which the strings saturate the SSDC belongs to the Swampland. Interestingly, under some circumstances the SSDC becomes stronger when imposing its consistency under dimensional reduction.

%%%%%%%%%%%%
\begin{figure}[htb]
\begin{center}
\includegraphics[width=0.45\textwidth]{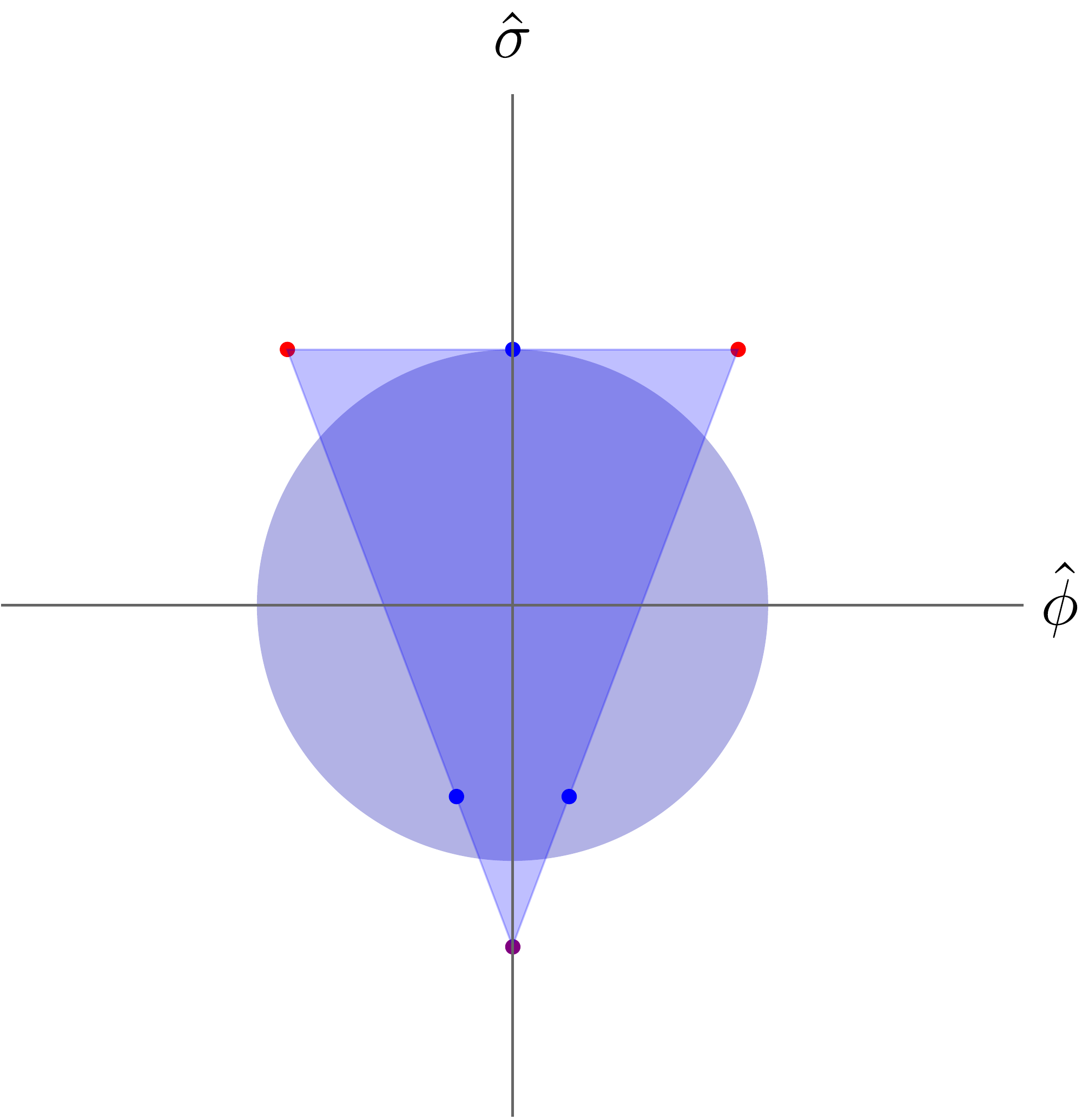}
\caption{\small Convex hull of a theory with strings saturating the SSDC in $D=10$ dimensions compactified on a circle.} 
\label{fig:ch3}
\end{center}
\end{figure}
%%%%%%%%%%%%

\begin{figure}[htb]
\begin{center}
	\subfigure[\label{sfig:5dto4d}$D=5 \, \rightarrow\, d=4$]{\includegraphics[width=0.45\textwidth]{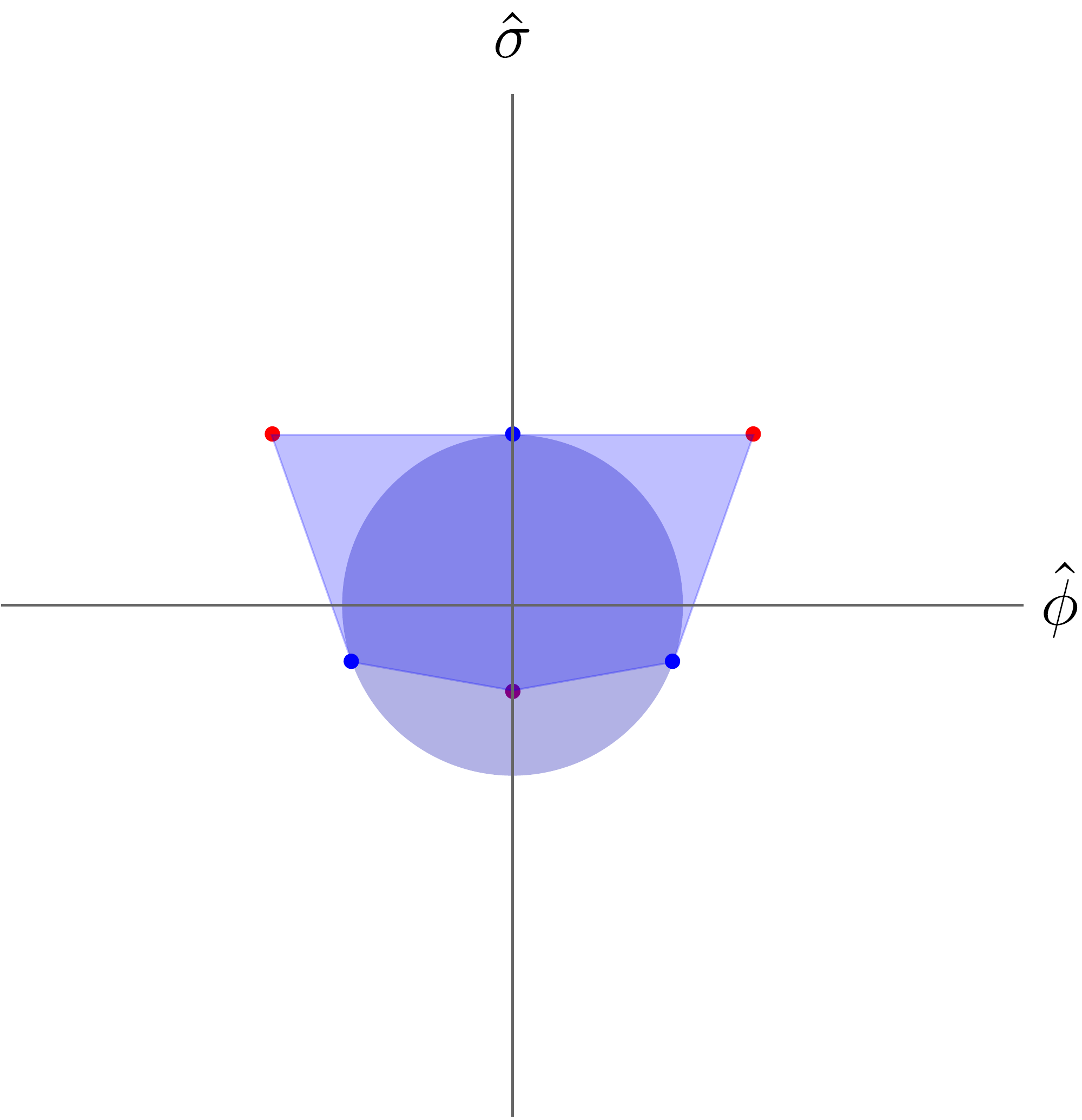}}\qquad 
	\subfigure[\label{sfig:9dto8d}$D=9 \, \rightarrow\, d=8$]{\includegraphics[width=0.45\textwidth]{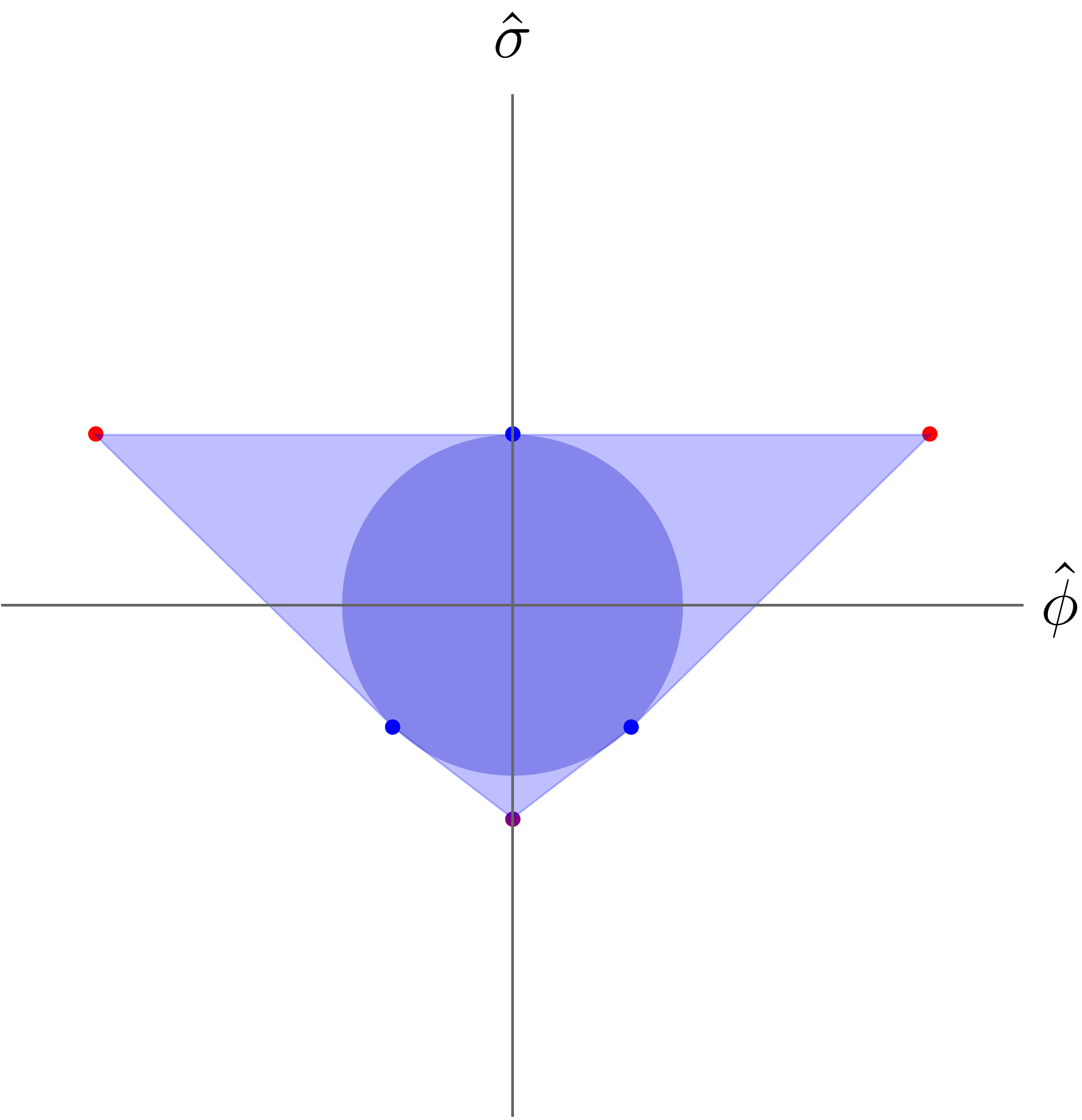}}
	\caption{Convex hull of a theory with strings with $\lambda_{s}=1/\sqrt{D-2}$ in $D=5$ and $D=9$ dimensions compactified on a circle. Even though it is difficult to see by eye, the $D=9 \rightarrow d=8$ case violates the SSDC convex hull condition by a tiny amount around the two lower blue dots.}
	\label{fig:ch4}	
\end{center}
\end{figure}

\begin{figure}[htb]
\begin{center}
\includegraphics[width=0.45\textwidth]{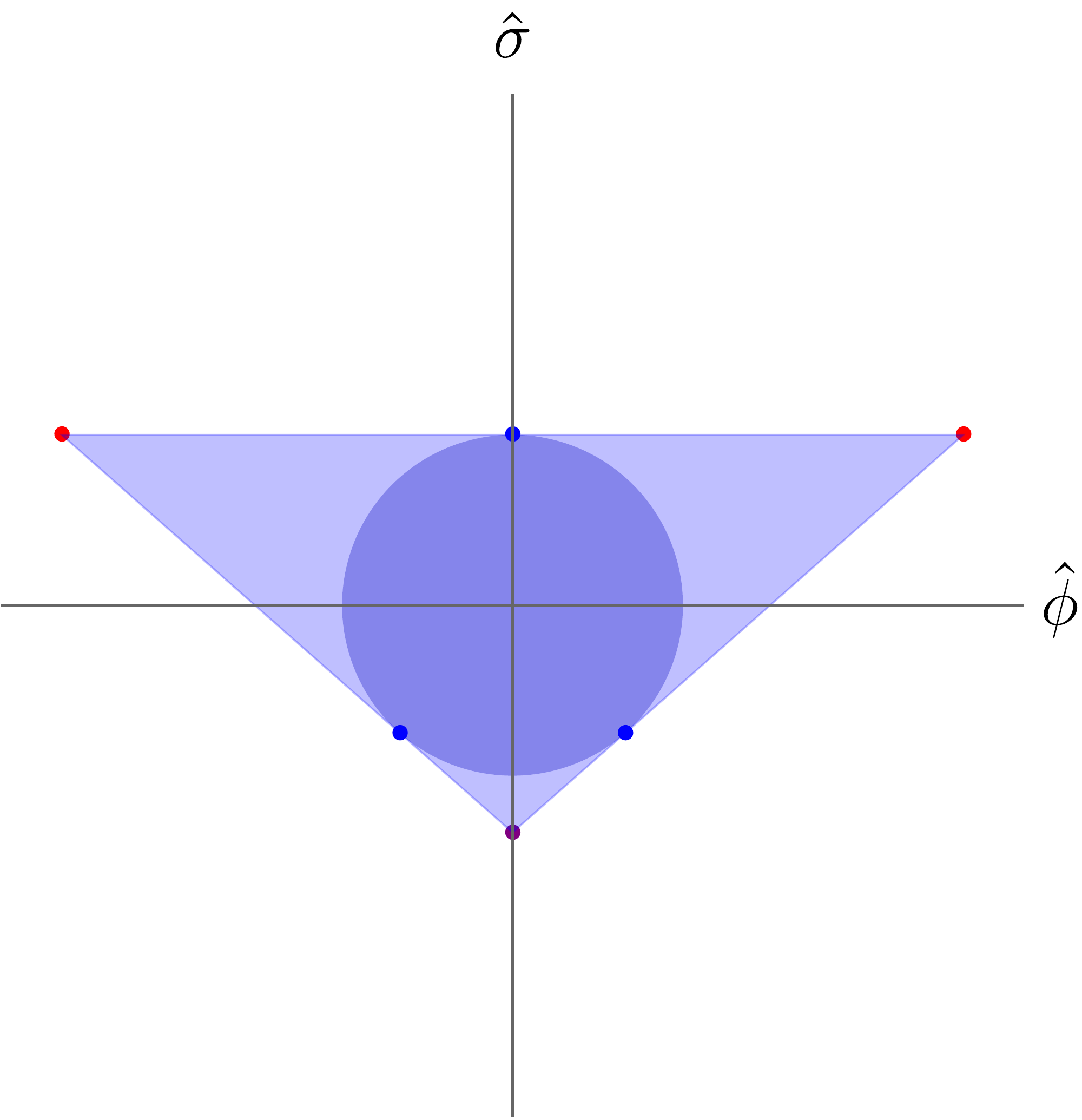}
\caption{\small Convex hull of a theory with strings with $\lambda_{s}=1/\sqrt{D-2}$ in $D=10$ dimensions compactified on a circle. It recovers a rotated version of figure \ref{fig:ch1}.} 
\label{fig:ch5}
\end{center}
\end{figure}

On the other hand, as found in \cite{Etheredge:2022opl} (see also the discussion around equation \eqref{eq:solpreservationspecies}), a better justified value for $\lambda_{\text{str}}$ would be
\begin{equation}
    \lambda_{\text{str}} = \frac{1}{\sqrt{D-2}} = \frac{1}{\sqrt{d-1}} \, .
\end{equation}
After plugging this into \eqref{eq:zvectorsStrings} and \eqref{vectors2} so as to draw the convex hull, we find that the SSDC is satisfied for $D\geq 10$ and violated for $D\leq9$ (see figure \ref{fig:ch4} for the particular cases of $D=5$ and $D=9$). Furthermore, we find saturation precisely for $D=10$. Indeed, notice that, as shown in figure \ref{fig:ch5}, this latter case precisely recovers a rotated version of the convex hull depicted in figure \ref{fig:ch1} above. Of course, this is not a coincidence, since in this case our toy model precisely reproduces type IIB string theory on $S^1$, which is dual to M-theory on $T^2$ \cite{Vafa:1996xn}.

%%%%%%%%%%%%
%%%%%%%%%%%%

%%%%%%%%%%%%
%%%%%%%%%%%%

Very remarkably, we conclude that this toy model, which presents only strings to start with, does not satisfy the SSDC in $D\leq 9$ after dimensionally reducing on a circle. This is moreover consistent with our experience from string theory, since upon compactification down to $d\leq8$ dimensions, not only strings and winding modes will arise. Higher-dimensional objects, such as D$p$-branes wrapping some of the internal cycles of the compact manifold, will give rise to new infinite towers of states. An explicit realization of this will be presented later on in section \ref{ss:MthyT3}, where we consider M-theory compactified on $T^3$. There, we will not only verify the SSDC but we will also be able to construct a picture very similar to the one shown in figure \ref{sfig:9dto8d}, with additional towers which indeed `protect' the SSDC convex hull condition.

Before closing up this subsection, let us compare our results with the ones obtained for the sharpened SDC discussed in \cite{Etheredge:2022opl}. For this same toy model, it was shown there that the winding modes of the critical strings were in fact sufficient to make the sharpened SDC hold in $d\geq5$. Additionally, it was argued that a tower of KK monopoles prevents this condition to be violated in $d=4$. We find that the SSDC seems to ask for more than this. As argued above, the presence of higher-dimensional objects in $D$-dimensions proves to be crucial for the SSDC to be satisfied in lower dimensional vacua. Again, it turns out that under some circumstances, the SSDC becomes more powerful after imposing its consistency under dimensional reduction.

\section{Further Evidence in M-theory Compactifications}
\label{s:Mthycompactifications}

After having discussed the behavior of the SSDC under dimensional reduction, in this section we move into verifying it within effective field theories which are known to be consistent with Quantum Gravity. In particular, we focus on maximally supersymmetric theories, as arising from M-theory toroidal compactifications. The first non-trivial example, namely M-theory on $T^2$, was already discussed in \ref{ss:MthyT2}. Thus we start discussing the case of M-theory on $T^3$ in section \ref{ss:MthyT3}. Such eight-dimensional model will be particularly useful to show how the appearance of new infinite towers of states change the naive picture in the RHS of figure \ref{fig:ch4}, such that the convex hull condition is satisfied along every asymptotic direction in moduli space. Additionally, the lessons we will learn in this setup, together with those from our previous analysis in section \ref{ss:field-theory}, will allow us to give a simple argument in support of the SSDC in M-theory on $T^n$ with $n>3$.

\subsection{M-theory on $T^3$}
\label{ss:MthyT3}

Let us consider M-theory compactified on a $T^3$, leading to a 8d $\mathcal N=2$ supergravity theory. The gravitational and scalar sectors present the following action after switching to the eight-dimensional Einstein frame \cite{Etheredge:2022opl}
\begin{equation}\label{eq:8d}
	S^{\text{8d}}_{\text{M}} \supset \frac{1}{2\kappa_8^2} \int d^{8}x\, \sqrt{-g}\,  \left[ R - \frac{1}{4} \left(g^{m m'} g^{n n'} + \frac{1}{6} g^{m n} g^{m' n'}\right) \partial g_{m n} \cdot \partial g_{m' n'} - \frac{1}{2 \mathcal{V}_3^2} \partial b \cdot \partial b \right]\, ,
\end{equation}
where $g_{mn}$ is the internal metric, $\mathcal{V}_3$ its overall volume in M-theory units and $b= C^{(3)}_{123}$ denotes the axionic field arising from the 11d 3-form with all its legs along the three-torus. Notice that, despite not being apparent from equation \eqref{eq:8d} above, the moduli space in this eight-dimensional theory is a coset space of the form $\mathcal{M}_{\text{8d}}=SL(2, \mathbb{Z})\backslash SL(2, \mathbb{R})/U(1) \times SL(3, \mathbb{Z})\backslash SL(3, \mathbb{R})/SO(3)$, where the discrete piece corresponds to the U-duality group of the theory\cite{Hull:1994ys}.

The idea here is to show how Quantum Gravity avoids the naive violation of the bound for the species scale in $D\leq9$, as discussed in our simple toy model from section \ref{ss:compactificationstring}. The mechanism by which this should happen is of course the appearance of new towers of light states along certain asymptotic directions. Therefore we want to check whether 
\begin{equation} \label{eq:bound8d}
  \lambda_{\text{sp}} \geq \frac{1}{\sqrt{(d-1)(d-2)}} \stackrel{\text{8d}}{=} \frac{1}{\sqrt{42}}\, ,
\end{equation}
is satisfied or not in the present setup. In principle, one could adopt a more general approach and take also into account the axionic dependence both in the mass scale and scalar charge-to-mass ratios of the infinite towers of states. However, for simplicity and in light of our analysis in nine dimensions from section \ref{sss:axions}, we will henceforth `freeze' all the axions in the theory, exploring geodesics in moduli space which move just along the non-compact `saxionic' directions. In any event, by making use of $SL(2, \mathbb{Z}) \times SL(3, \mathbb{Z})$ duality, one can relate such paths to any other exploring some analogous infinite distance singularity in moduli space.

To make contact with the logic followed in section \ref{ss:compactificationstring}, let us rewrite the scalar lagrangian in equation \eqref{eq:8d} as if it was obtained by circle-reduction from the 9d theory in section \ref{ss:MthyT2} instead of directly compactifying M-theory on $T^3$. Thus, we take the 9d metric in \eqref{eq:11dmetric} and propose the following ansatz
\begin{equation}\label{eq:9dmetric}
	ds^2_{9} = e^{-\sqrt{1/21}\,\rho} ds_8^2 + e^{\sqrt{12/7}\,\rho} \left(dz^3 \right)^2\, ,
\end{equation}
with the radion field $\rho$ being related to the extra radius by $R_3=e^{\sqrt{3/7}\,\rho}$. Next, we dimensionally reduce the 9d action \eqref{eq:9d} yielding 
\begin{equation}\label{eq:8dcanonical}
	S^{\text{8d}}_{\text{M}} \supset \frac{1}{2\kappa_8^2} \int d^{8}x\, \sqrt{-g}\,  \left[ R - \left( \partial \hat U \right)^2 - \left(\partial \hat \tau\right)^2 - \left( \partial \hat \rho \right)^2 \right] + \left( \text{axions} \right)\, ,
\end{equation}
where we have introduced the canonically normalized fields $\hat \rho=\frac{\rho}{\kappa_8 \sqrt{2}}$ as well as those in \eqref{eq:canonicalnormalization}. Note that the compactification process just described could be seen as a $T^2$-fibration over the $S^1$.

With this, we can now start computing the relevant scales of the infinite tower of states, their `charge-to-mass' ratio (c.f. \eqref{eq:z-vectors}) as well as their (combined) species scales, similarly as we did for the 9d setting. The relevant towers can be found in table \ref{tab:BPSstates} below. 

%%%%%%%%%%%%%%%%%%%%%%%%%%%%%%%
	\begin{table}[h!!]\begin{center}
			\renewcommand{\arraystretch}{1.00}
			\begin{tabular}{|c||c|c|c|}
				\hline
				$\frac{1}{2}$-BPS states &  construction  &  tension/mass/action  & degeneracy  \\
				\hline 
				strings& str, $i=$ M2 on $S^1_i$  &    $T_{\text{str},\, i} \propto (2\pi R_i)$  &   $3$ \\
				\hline
				non-perturbative particles & $\text{M}_{ij}=$ M2 on $S^1_i \times S^1_j$  &    $m_{ij} \propto (2\pi R_i)(2\pi R_j)$  &   $3$ \\
				\hline
				perturbative particles &  KK$_{i}=$ KK from $S^1_i$  &    $m_{\text{KK}, i}= 1/ R_i$  &   $3$\\
				% \hline
				% instantons & M2 on $T^3$  &  $S_{\text{M2}} \propto R_1 R_2 R_3$  
    %             &   1  \\
				\hline
			\end{tabular}
			\caption{Relevant towers of $\frac{1}{2}$-BPS states together with their mass scales (in Planck units) in the eight-dimensional setup.}
			\label{tab:BPSstates}
		\end{center}
	\end{table}
 %%%%%%%%%%%%%%%%%%%%%%%%%%%%%%%%%
 
%Notice that apart from the states discussed in that section, we get an additional\footnote{There are other BPS extended objects in the theory, such as M5-branes and descendants obtained from wrapping it along different cycles of the $T^2$ manifold. We refuse to discuss their multiplicity and kinematics since they play no role in our present discussion.} M2-brane with tension
%
%\begin{equation}\label{eq:9dM2}
%	T_{\text{M2}} \sim \frac{2\pi}{\ell_9^3} e^{-3U/7} \sim \frac{2\pi}{\ell_9^3} e^{-2\hat U/\sqrt{14}}\, ,
%\end{equation}
%
%where $\ell_9$ denotes the nine-dimensional Planck length.

Let us first study the case of critical strings. It is easy to see that apart from the set of $(p,q)$-strings which were already present in 9d, we get an additional one by wrapping the M2-brane along the extra circle $S^1_3$, whose tension is given by
\begin{equation}\label{eq:8dstring}
	T_{\text{str},\, 3} \sim \frac{2\pi}{\ell_8^2} e^{-2\hat U/\sqrt{14}}\, e^{2\sqrt{2}/\sqrt{21}\, \hat \rho} \, \Longrightarrow \, \frac{m^{(\text{str})}_{\text{str},\, 3}}{M_{\text{pl},\, 8}}  \sim e^{-\frac{1}{\sqrt{14}}\hat U}\, e^{\sqrt{\frac{2}{21}}\, \hat \rho}\, .
\end{equation}
Since the species scale associated to a critical string is the string scale itself, we arrive at the following species scale vectors
\begin{equation} \label{eq:stringvectors}
\begin{split} 
	\vec{\mathcal{Z}}_{\text{str},\, 1} &= \left( \frac{1}{2\sqrt{2}} , \frac{1}{\sqrt{42}}, -\frac{1}{2 \sqrt{14}} \right)\, , \qquad \vec{\mathcal{Z}}_{\text{str},\, 2} = \left( -\frac{1}{2\sqrt{2}} , \frac{1}{\sqrt{42}}, -\frac{1}{2 \sqrt{14}} \right) \, ,\\
	\vec{\mathcal{Z}}_{\text{str},\, 3} &= \left( 0 , -\sqrt{\frac{2}{21}}, \frac{1}{\sqrt{14}} \right) \, ,
\end{split}
\end{equation}
where the notation is $\vec{\mathcal{Z}} = \left(\mathcal{Z}_{\hat \tau}, \mathcal{Z}_{\hat \rho}, \mathcal{Z}_{\hat U} \right)$. To obtain the first two we made use of equation \eqref{string-species} together with the second relation in \eqref{eq:zvectorafterdimreduction}. Notice that they satisfy $|\vec{\mathcal{Z}}_{\text{str},\, i}|^2=1/(d-2)=1/6$, as expected from the analysis in section \ref{ss:preservation&saturation}. They can be seen as type IIA strings and 8d S-duals upon choosing one cycle as the reference M-theory circle.

For the Kaluza-Klein towers we proceed similarly to what we did for the strings above, namely we borrow the results from the 9d setup and then compute the additional KK scale associated to the extra circle. One finds:
\begin{equation} \label{eq:KKvectors}
\begin{split} 
	\vec{\mathcal{Z}}_{\text{KK},\, 1} &= \left( \frac{1}{7\sqrt{2}} , \frac{1}{7 \sqrt{42}}, \frac{3}{7 \sqrt{14}} \right) \, , \qquad \vec{\mathcal{Z}}_{\text{KK},\, 2} = \left( -\frac{1}{7\sqrt{2}} , \frac{1}{7 \sqrt{42}}, \frac{3}{7 \sqrt{14}} \right) \, ,\\
	\vec{\mathcal{Z}}_{\text{KK},\, 3} &= \left( 0 , \frac{1}{\sqrt{42}}, 0 \right) \, .
\end{split}
\end{equation}
These vectors all verify $|\vec{\mathcal{Z}}_{\text{KK},\, i}|^2=1/(d-1)(d-2)=1/42$, thus saturating our proposed bound. In string theory language, they become two distinct KK towers and D0-branes upon choosing without loss of generality some cycle as the M-theory circle. Notice that the first two, which were already present in nine dimensions, have a $\lambda_{\text{sp}}$-parameter whose functional form is preserved upon dimensional reduction, see section \ref{ss:preservation&saturation}.

Lastly, the other relevant set of $\frac{1}{2}$-BPS particles arises from M2-branes wrapping any 2-cycle of the internal manifold, as shown in table \ref{tab:BPSstates}. These can be either seen as winding modes associated to each one of the three critical strings in \eqref{eq:stringvectors}, or alternatively as the 8d analogues of the F-theory tower discussed around equation \eqref{eq:Ftheorytower}. In any event, it is clear from the above considerations that they are nothing but KK towers associated to decompactification limits in a dual type IIB frame. Moreover, their species scale vectors read
\begin{equation} \label{eq:M2vectors}
\begin{split} 
	\vec{\mathcal{Z}}_{\text{M}_{23}} &= \left( \frac{1}{7\sqrt{2}}, -\frac{5}{7 \sqrt{42}}, -\frac{1}{7 \sqrt{14}} \right) \, , \qquad \vec{\mathcal{Z}}_{\text{M}_{13}} = \left( -\frac{1}{7\sqrt{2}}, -\frac{5}{7 \sqrt{42}}, -\frac{1}{7 \sqrt{14}} \right) \, ,\\
	\vec{\mathcal{Z}}_{\text{M}_{12}} &= \left( 0, \frac{1}{7 \sqrt{42}}, -\frac{\sqrt{8}}{7 \sqrt{7}} \right)\, .
\end{split}
\end{equation}
where one can see that again $|\vec{\mathcal{Z}}_{\text{M}_{ij}}|^2=1/(d-1)(d-2)=1/42$. 

We are not done yet, though. As we learned from the general analysis of section \ref{s:dimensionalreduction} and the 9d example above one has to take into account combinations of towers of states, whose effective species scale may be lower than naively expected. For the case at hand, the ones that will be relevant are those formed by bound states involving only Kaluza-Klein states, those comprised by M2-particles alone and a mixture of these two sectors\cite{Obers:1998fb}. The KK bound states lead to the following set of vectors
\begin{equation} \label{eq:KKvectorscombined}
\begin{split} 
	\vec{\mathcal{Z}}_{\text{KK},\, (12)} &=  \left( 0, \frac{1}{4 \sqrt{42}}, \frac{3}{4 \sqrt{14}} \right) \, , \qquad \qquad \, \vec{\mathcal{Z}}_{\text{KK},\, (13)} =  \left( \frac{1}{8 \sqrt{2}}, \frac{1}{ \sqrt{42}}, \frac{3}{8 \sqrt{14}} \right) \, ,\\
	\vec{\mathcal{Z}}_{\text{KK},\, (23)} &=  \left( -\frac{1}{8 \sqrt{2}}, \frac{1}{ \sqrt{42}}, \frac{3}{8 \sqrt{14}} \right) \, , \qquad \vec{\mathcal{Z}}_{\text{KK},\, (123)} =  \left( 0, \frac{1}{ \sqrt{42}}, \frac{2}{3 \sqrt{14}} \right) \, .
\end{split}
\end{equation}
whilst for the M2-particles one finds instead\footnote{One can alternatively denote the $\text{M}_{ij}$ tower by $\tilde{\text{M}}_k$ where $k$ is such that $\epsilon_{ijk}=1$. Thus, the combined tower of $\text{M}_{ij}$ and $\text{M}_{lm}$ particles may be denoted as $\tilde{\text{M}}_{(kp)}$, where again $k, p$ are such that $\epsilon_{ijk}=\epsilon_{lmp}=1$.}
\begin{equation} \label{eq:M2vectorscombined}
\begin{split} 
	\vec{\mathcal{Z}}_{\tilde{\text{M}},\, (21)} &=  \left( 0, -\frac{5}{4 \sqrt{42}}, -\frac{1}{4 \sqrt{14}} \right) \, , \qquad \qquad \ \ \vec{\mathcal{Z}}_{\tilde{\text{M}},\, (13)} =  \left( \frac{1}{8 \sqrt{2}}, -\frac{1}{ 2 \sqrt{42}}, -\frac{5}{8 \sqrt{14}} \right) \, ,\\
	\vec{\mathcal{Z}}_{\tilde{\text{M}},\, (23)} &=  \left( -\frac{1}{8 \sqrt{2}}, -\frac{1}{ 2 \sqrt{42}}, -\frac{5}{8 \sqrt{14}} \right) \, , \qquad \vec{\mathcal{Z}}_{\tilde{\text{M}},\, (123)} =  \left( 0, -\frac{1}{ \sqrt{42}}, -\frac{2}{3 \sqrt{14}} \right) \, .
\end{split}
\end{equation}
Finally, one can find effective towers of M2-particles with non-trivial KK momentum along the 1-cycle they do not wrap. These will be denoted as
\begin{equation} \label{eq:KK&M2vectorscombined}
\begin{split} 
	\vec{\mathcal{Z}}^{\,3, 12}_{\text{KK},\, \text{M}} &=  \left( 0, \frac{1}{\sqrt{42}}, -\frac{1}{2 \sqrt{14}} \right) \, , \qquad \vec{\mathcal{Z}}^{\,2, 13}_{\text{KK},\, \text{M}} =  \left( -\frac{1}{4 \sqrt{2}}, -\frac{1}{2 \sqrt{42}}, \frac{1}{4 \sqrt{14}} \right) \, ,\\
	\vec{\mathcal{Z}}^{\,1, 23}_{\text{KK},\, \text{M}} &=  \left( \frac{1}{4 \sqrt{2}}, -\frac{1}{2 \sqrt{42}}, \frac{1}{4 \sqrt{14}} \right) \, .
\end{split}
\end{equation}
The interpretation of the above set of species scales is straightforward. The first three of each set can be seen to be effective towers with $p=2$ implementing some decompactification to 10d and they moreover satisfy $|\vec{\mathcal{Z}}_{ij}|^2=1/24$, whilst the last vector of both \eqref{eq:KKvectorscombined} and \eqref{eq:M2vectorscombined} take us back to 11d M-theory (one can check that $|\vec{\mathcal{Z}}_{123}|^2=1/18$). The triplet \eqref{eq:KK&M2vectorscombined} also implement some double decompactification to 10d.

\subsubsection*{Plotting the Convex Hull}

\begin{figure}[htb]
		\begin{center}
			\subfigure{
				\includegraphics[width=0.45\textwidth]{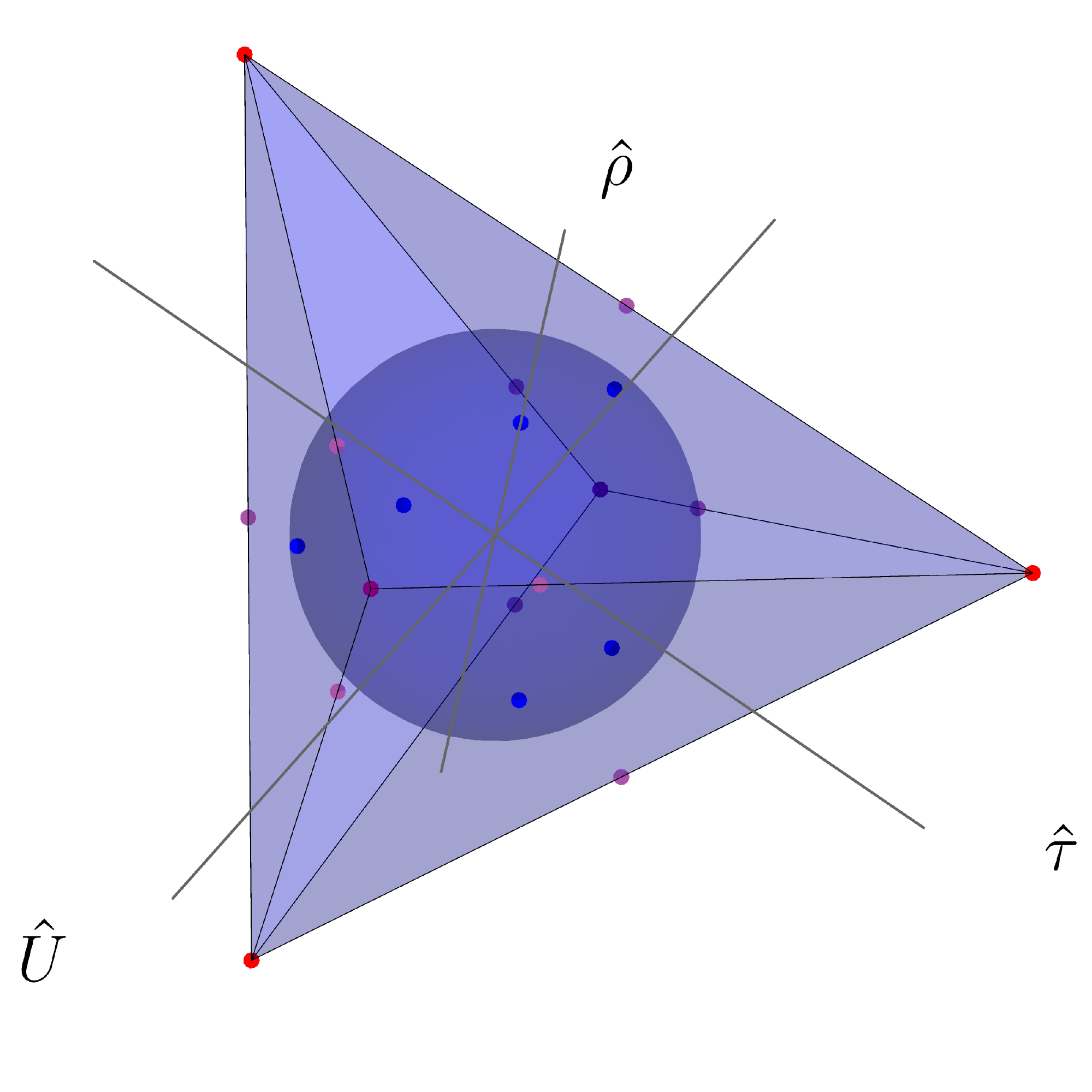}\label{sfig:slice3dCH}
			}
			\subfigure{
				\includegraphics[width=0.45\textwidth]{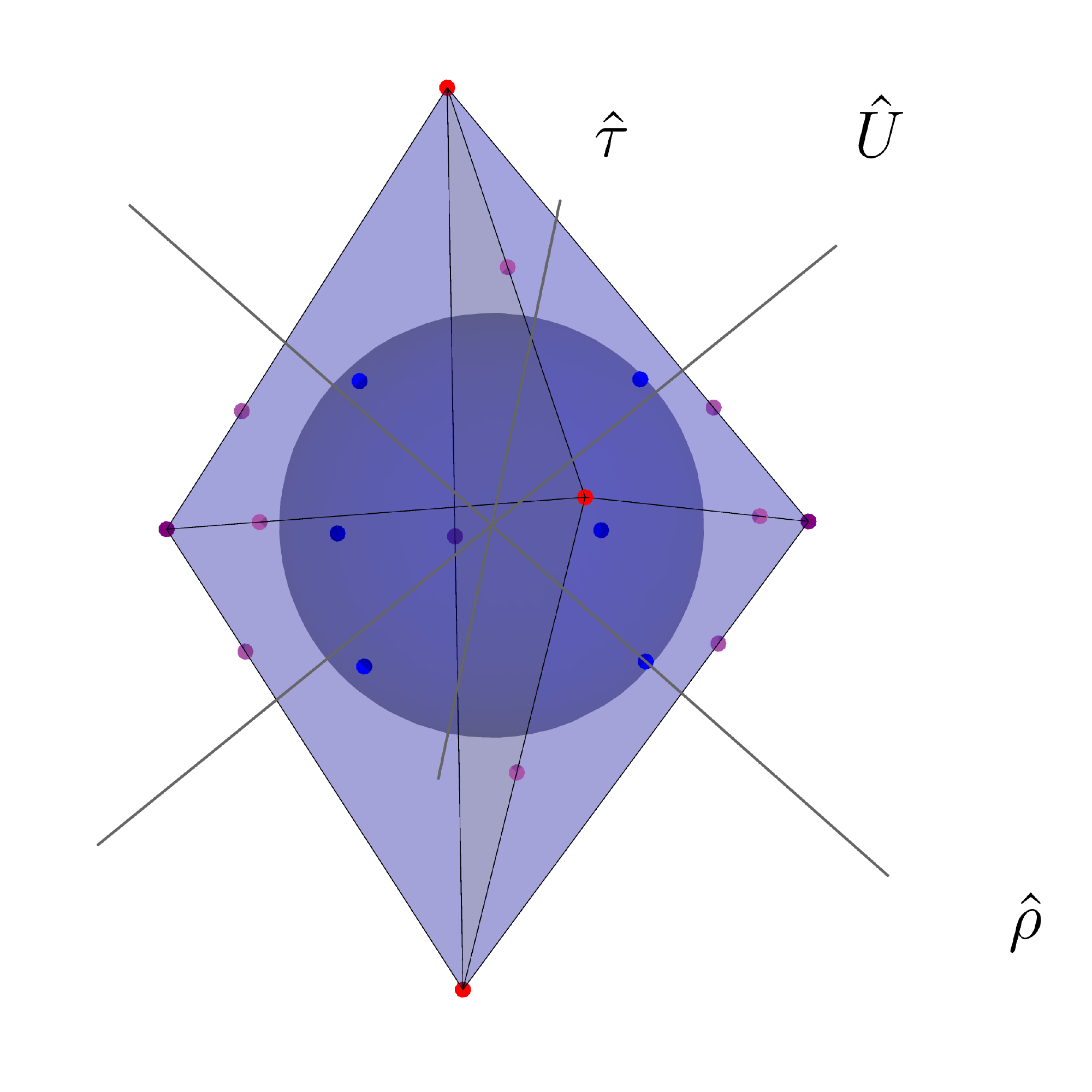}\label{sfig:2ndslice3dCH}
			}
			\caption{\small Convex hull for the species scale in M-theory on $T^3$ with the axions set to a constant value from two different angles. The blue dots in the faces of the convex hull correspond to single KK towers ($p=1$), the light purple dots in the edges indicate double KK towers ($p=2$) and the purple and red dots in the vertices correspond to triple KK towers ($p=3$) and string towers, respectively.}
			\label{fig:ch8dgeneric}
		\end{center}
\end{figure}

Once we have all the $\mathcal{Z}$-vectors associated to the individual species scales, we can plot them in a 3d graph to check whether the SSDC is satisfied. As shown in figure \ref{fig:ch8dgeneric} from two different perspectives, the convex hull for the present 8d example contains the ball of radius $\lambda_{\text{sp, min}}$ (see equation \eqref{eq:lspmin}), thus fulfilling the conjecture. 

% %%%%%%%%%%%%

Interestingly, and similarly to what happened with the 9d setup discussed in section \ref{ss:MthyT2}, the convex hull for the species scale vectors presents some nice structure capturing both the symmetries of the quantum theory as well as the relevant physics associated to the infinite distance boundaries in $\mathcal{M}_{\text{8d}}$. Indeed, from figure \ref{sfig:2ndslice3dCH} one can see at least a $\mathbb{Z}_3 \times \mathbb{Z}_2$ symmetry\footnote{\label{fnote:symmgroup}In fact, the complete group can be seen to correspond to $\mathsf{D}_3 \times \mathsf{D}_1$\cite{Castellano:2023jjt}, where $\mathsf{D}_n$ denotes the dihedral group of order $2n$.} of the polyhedron, which may be thought of as a discrete remnant of the U-duality group in the eight-dimensional theory. This means, in particular, that the convex hull for the species scale is completely encoded within some \emph{fundamental domain}, which is replicated upon acting with the discrete symmetry group. This simplification will be important later on in section \ref{ss:MthyTn} so as to extend the present analysis to setups with maximal supersymmetry in $d<8$.

Regarding the structure exhibited by the diagram, let us first notice that the convex hull is generated again by the species scale vectors associated to either emergent string limits (the red dots in the diagram) or full decompactification to 11d M-theory (the purple dots). Moreover, the blue dots, corresponding to $\mathcal{Z}$-vectors for $p=1$ towers saturate the SSDC bound and appear precisely at the faces of the convex hull, the latter being perpendicular to the former. In fact, the faces of the polyhedron turn out to be nothing but the corresponding convex hull diagram of the 9d theory (see figure \ref{sfig:9dCH-2}), which in turn include at its edges the convex hull for the different 10d theories. Therefore, we find an inductive sequence (resembling that of a matryoshka) which informs us about all possible infinite distance limits that can be explored either directly from the lower dimensional perspective, or rather by passing first through an intermediate higher dimensional frame.

Finally, and in order to appreciate the crucial role that the effective towers of states play in order for the convex hull condition to be verified, let us study one particular 2d slice of figure \ref{fig:ch8dgeneric}, namely that spanned by the $(\hat \rho, \hat \tau)$ directions. Therefore, upon projecting the $\mathcal{Z}$-vectors down to a plane characterized by its normal $\hat n = \partial_{\hat{U}}$ as
\begin{equation}\label{eq:2dprojection}
	\vec{\mathcal{Z}}_{\hat n} = \vec{\mathcal{Z}} - \left( \hat n \cdot \vec{\mathcal{Z}}\right) \hat n\, ,
\end{equation}
one obtains precisely what is shown in figure \ref{fig:ch-comparison-toymodel} below. The reason for choosing such slice is because it can be easily connected to the situation discussed previously in section \ref{ss:compactificationstring}. There we showed that, upon compactifying a $D$-dimensional theory in which we start with two S-dual strings on a circle, the SSDC seemed to be naively put in danger whenever $D\leq9$. The idea was that the $D-1$ avatars of the pair of strings together with their winding modes --- and effective combination thereof --- which are crucial to saturate $\lambda_{\text{sp, min}}$ in $D=10$, were in fact not sufficient to preserve the bound when starting from the non-critical dimension (see figure \ref{sfig:toymodel8d}). As we have already seen, when considering M-theory on $T^3$, i.e. an eight-dimensional EFT really coming from Quantum Gravity instead of a toy model, there is no violation of the SSDC convex hull condition. In fact, the projection in figure \ref{fig:ch-comparison-toymodel} turns out to have a similar structure to that exhibited by the toy model, which is also shown in the RHS for comparison. In both cases there are two strings that correspond to the red dots appearing on the upper half of the image. They are indeed connected by a line which is tangent to the `critical ball' of radius $\lambda_{\text{sp, min}}$, touching it at the point where the KK tower associated to the extra $S^1$ becomes relevant. On the opposite direction, namely along the negative $\hat \rho$ axis, where the (effective) tower of windings do not give rise to a convex hull containing the critical ball, a new dual string appears in the LHS. This new tower, absent in the toy model, ensures that the bound is preserved along every direction in the two-dimensional graph.

%%%%%%%%%%%%
\begin{figure}[htb]
\begin{center}
	\subfigure[M-theory on $T^3$]{\includegraphics[width=0.45\textwidth]{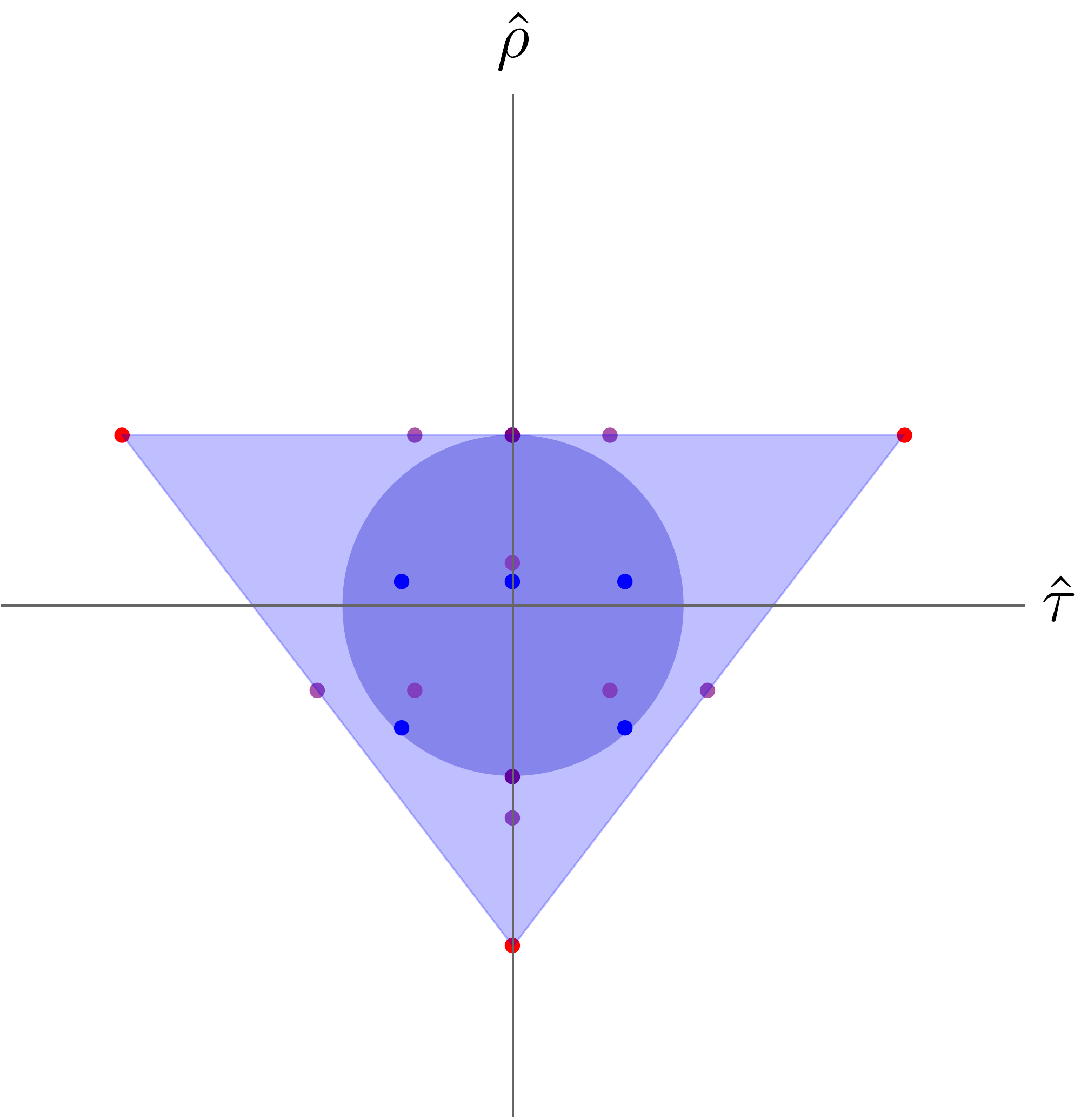}}\label{sfig:2dslice}
    \quad
	\subfigure[Toy model]{\includegraphics[width=0.45\textwidth]{CH-5.pdf}}\label{sfig:toymodel8d}
	\caption{\small a) Two-dimensional projection of the convex hull for the species scale in M-theory on $T^3$. b) Convex hull diagram for the 9d to 8d compactification of the toy model discussed in section \ref{ss:compactificationstring}.}
	\label{fig:ch-comparison-toymodel}	
\end{center}
\end{figure}
%%%%%%%%%%%%

\subsubsection*{On the Quest for a Global Species Scale}

From the previous discussion one can draw a couple of nice lessons. First, as expected based on general considerations in section \ref{ss:compactificationstring}, upon compactifying to lower dimensions one starts to need from genuine QG ingredients in order to satisfy the SSDC, namely new infinite towers of states. In a sense, from a bottom-up perspective, one could say that the naive imposition of the convex hull condition to certain directions in this 8d setup \emph{postdicts} the existence of new infinite distance limits, which turn out to have a nice UV embedding from the 11d perspective. 

Second, by inspection of the 3d plot in figure \ref{sfig:2ndslice3dCH}, one finds a highly symmetric arrangement of the points. This becomes manifest upon projecting on a plane whose unit normal vector lies along the direction associated to the $p=3$ limits, see figure \ref{fig:ch8dSL3} below. What is behind is precisely the U-duality group $G= SL (2, \mathbb{Z}) \times SL (3, \mathbb{Z})$ of the eight-dimensional theory. The species scale vectors arrange into representations of the group $G$, which is a reflection of the fact that the infinite distance limits are oftentimes related to each other by some duality transformation. In table \ref{tab:Udualityirreps} we show the different irreducible representations that one can find within the set of $\mathcal{Z}$-vectors.

%%%%%%%%%%%%%%%%%%%%%%%%%%%%%%%
\begin{table}[h!!]\begin{center}
		\renewcommand{\arraystretch}{1.00}
 			\begin{tabular}{|c||c|c|c|}
 				\hline
 				Density parameter &  $G$-representation &  QG Resolution  \\
 				\hline 
 				$p\to\infty$ (strings) & $(\textbf{1}, \textbf{3})$  &    Emergent String Limit \\
 				\hline
 				$p=1$ (KK$_i$ and $\text{M}_{ij}$ single towers) & $(\textbf{2}, \textbf{3})$  &    Decompactification to 9d \\
 				\hline
 				$p=2$ (KK$_{ij}$ and $\tilde{\text{M}}_{ij}$ double towers) & $(\textbf{2}, \textbf{3})$  &    Decompactification to 10d \\
 				\hline
                 $p=2$ (bound states of KK$_i$ and $\text{M}_{ij}$) & $(\textbf{1}, \textbf{3})$  &    Decompactification to 10d \\
                 \hline
 				$p=3$ (KK$_{123}$ and $\tilde{\text{M}}_{123}$ triple towers)  &  $(\textbf{2}, \textbf{1})$  &    Decompactification to 11d \\
 				\hline
 			\end{tabular}
 			\caption{Infinite distance limits and their representations in terms of the U-duality group of maximal supergravity in 8d.}
 			\label{tab:Udualityirreps}
 		\end{center}
 	\end{table}
%%%%%%%%%%%%%%%%%%%%%%%%%%%%%%%%%

Finally, let us comment about the redundancy of the convex hull diagram. Indeed, U-duality allows us to focus just on a `fundamental region' comprised by (half of) a single face of the polyhedron, see footnote \ref{fnote:symmgroup}. This will be important for the discussion in the next subsection, and moreover takes into account the quantum symmetries of the theory. Incidentally, one may think of the 3d graph under discussion as some sort of (asymptotic) `boundary condition' for the (logarithmic) derivative of a would-be \emph{globally} defined species scale function, $\Lambda_{\text{sp}} (\vec{\Phi})$, where $\Vec{\Phi}$ denotes collectively the moduli of the 8d theory. Therefore, if one were able to define such global function (similarly to recent 4d attempts \cite{vandeHeisteeg:2022btw, Cribiori:2023sch}), one would then need to make sure that it respects the symmetries of the theory and matches the asymptotic behavior here described. Moreover, such a `Master Species Scale Function' would be expected to behave as some generalized automorphic form of $SL (2, \mathbb{Z}) \times SL (3, \mathbb{Z})$ \cite{Castellano:2023aum,vandeHeisteeg:2023dlw}.

%%%%%%%%%%%%
\begin{figure}[htb]
\begin{center}
\includegraphics[scale=.4]{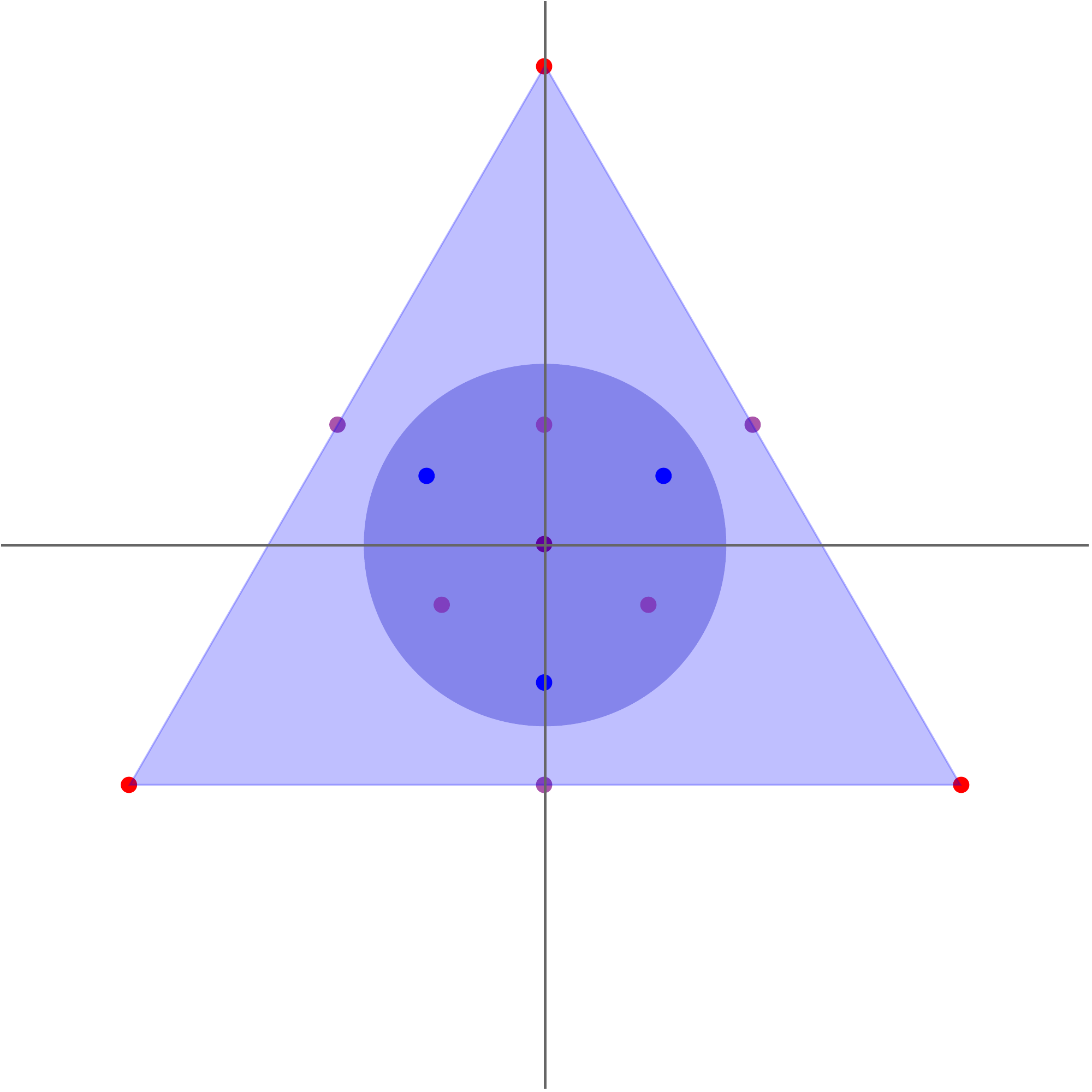}
\caption{\small Two-dimensional projection of the convex hull for the species scale in M-theory on $T^3$ with the axions set to a constant. The plane is chosen to be such that its normal $\hat n \propto \vec{\mathcal{Z}}_{\text{KK},\, (123)}$, making manifest the discrete remnant of the $SL(3, \mathbb{Z})$ symmetry (sub-)group.} 
 \label{fig:ch8dSL3}
 \end{center}
 \end{figure}
 %%%%%%%%%%%%

\subsection{M-theory on $T^n$}
\label{ss:MthyTn}

The aim of the present section is to extend the results from the previous examples in nine and eight-dimensional maximal supergravity to M-theory compactifications on $T^n$ with $n>3$. This will provide more evidence in favour of the SSDC and the idea that there exists a minimum rate at which the species scale can fall off asymptotically.

Our argument will proceed inductively, relying heavily both in U-duality and the uniqueness of the supergravity theory for $d<9$ spacetime dimensions. Indeed, U-duality tells us that the particle states comprising infinite towers with $p=1$ arrange themselves into a \emph{single} irreducible representation of the symmetry group (see e.g. \cite{Obers:1998fb}).\footnote{Actually, this is not true for maximal supergravity in 9d, where the KK-like towers form a $\mathbf{2} \oplus \mathbf{1}$ representation of the $SL(2, \mathbb{Z})$ duality group. This hinges on the fact that there exist two theories in ten dimensions with 32 supercharges one can go to from 9d, differing only on their chirality\cite{Hull:1994ys}.} Such orbit includes both perturbative (i.e. KK, winding modes, etc.) and non-perturbative states (wrapped $p$-branes, KK-monopoles, etc.), as seen from the original frame. For example, consider M-theory compactified on $T^4$ down to 7d. The U-duality group is identified with $SL(5, \mathbb{Z})$ in this case \cite{Hull:1994ys}, and the particle multiplet transforms as the $\mathbf{10}$ of $SL(5, \mathbb{Z})$. These states may be understood microscopically as four Kaluza-Klein towers associated to the compact directions, as well as six additional infinite sets of wrapped M2-particles.

Thus we start with a couple of insights provided by the examples from sections \ref{ss:MthyT2} and \ref{ss:MthyT3} above. There we saw that the states with $p=1$ saturated the bound and appeared precisely at the faces of the polyhedron. Additionally, these faces turn out to be nothing but the convex hull of the theory in one dimension more. This is easy to understand, since upon dimensionally reducing the supergravity theory, all $\mathcal{Z}$-vectors give rise in particular to analogous ones combined with the KK tower associated to the extra circle. The latter generate the same polyhedron already existing in the higher dimensional theory, which is moreover orthogonal to the species scale vector of the new compact dimension, $\vec{\mathcal{Z}}_{\text{KK}}$ (see section \ref{ss:field-theory}). Finally, from the analysis of section \ref{s:dimensionalreduction} we learned that if the theory we start with verifies the SSDC then, upon compactification on a circle, we end up with a bunch of species vectors which still satisfy the conjecture along all intermediate directions.

Let us now put everything together to argue that the SSDC is indeed satisfied in $d$-dimensional maximal supergravity for $d\geq4$. We work by induction, such that we first assume the SSDC to hold for M-theory compactified on $T^n$, with species vectors denoted by $\vec{\mathcal{Z}}_{\text{t}}$. In a next step, we dimensionally reduce the theory on a circle, leading to M-theory on $T^n \times S^1 \cong T^{n+1}$.\footnote{We freeze the axions to zero vev, since as we learned from the previous examples they play no important role in our analysis (see section \ref{sss:axions}).} Based on the general considerations from the previous paragraph we conclude that the (sub-)polyhedron spanned by the set of vectors $\lbrace \vec{\mathcal{Z}}_{\text{KK}}, \vec{\mathcal{Z}}_{(\text{KK, t})}\rbrace$ satisfy the SSDC, saturating the bound precisely along the direction determined by $\vec{\mathcal{Z}}_{\text{KK}}$ (c.f. equation \eqref{vectors}). However, from U-duality we know that this is already enough to ensure that the SSDC holds in every asymptotic direction, since upon acting with the discrete remnant of the symmetry group one can indeed reconstruct the rest of the convex hull diagram. The latter presents as many `faces' (which are hence identical to the one just described) as the dimension of the representation into which the $p=1$ towers fit into. The faces being identical follows from the uniqueness of the supergravity action, at least for $d<9$. Therefore, upon noticing that the Species Scale Distance Conjecture was satisfied for M-theory compactified on $T^k$ for $k=1,2,3$, we thus demonstrate that the conclusion extends to maximal supergravity in lower dimensions as well.

\section{Conclusions \& Outlook}
\label{s:conclusions}

In this work we have tackled the Distance Conjecture  by combining three different perspectives:  entropy bounds in Quantum Gravity, scalar running solutions in EFTs inspired by Dynamical Cobordisms, and the geometric properties of convex hulls. In particular, by combining these with the key idea of the species scale as the natural UV cut-off in an EFT of Quantum Gravity, we have provided the first general bottom-up argument (to our knowledge) reproducing all the main features of the Distance Conjecture (see \cite{Hamada:2021yxy,Stout:2021ubb,Stout:2022phm} for previous arguments recovering some of the features).\footnote{One could also argue that the Emergence Proposal (see \cite{Grimm:2018ohb,Corvilain:2018lgw,Heidenreich:2018kpg} for the original references and \cite{Castellano:2022bvr, Hamada:2021yxy,Marchesano:2022axe,Castellano:2023qhp,Blumenhagen:2023yws,Kawamura:2023cbd,Seo:2023xsb,Blumenhagen:2023tev,Calderon-Infante:2023uhz} for subsequent follow-ups), if regarded truly as a bottom-up condition, can be employed to explain the essential features captured by the Distance conjecture. However, we believe that such a proposal should be actually regarded as an UV microscopic explanation, rather than a bottom-up criterion.} That is, we have shown that the Quantum Gravity cut-off must decay at least \emph{exponentially} with the \emph{geodesic distance} in moduli space as we approach \emph{any} infinite distance limit. Thus, upon translating this exponential bound on the species scale to a bound on the tower of states responsible for its descent, we obtained the behavior predicted by the Distance Conjecture, and also reformulated this as a convex hull condition on the species scale itself.

The bottom-up argument was obtained by confronting the Covariant Entropy Bound\cite{Bousso:1999xy}, which restricts the entropy of a region of spacetime to scale at most as the area of its boundary (in Planck units), with the characteristic extensive scaling of the entropy with the volume in local field theory. This logic yields the upper bound $\Lambda_{\text{UV}}^{d-1}\leq\,  A/ \text{vol} \sim L^{-1}$, which was used in \cite{Castellano:2021mmx} to provide a bottom-up argument for the ADC upon identifying the relevant length scale with the AdS one. In order to apply a similar reasoning, a way to relate the spacetime distance, $L$, with the moduli space distance, $\Delta_\phi$, was missing. Inspired by the ideas from Dynamical Cobordism \cite{Buratti:2021yia,Buratti:2021fiv,Angius:2022aeq,Blumenhagen:2022mqw,Angius:2022mgh,Blumenhagen:2023abk,Angius:2023xtu,Huertas:2023syg}, we have found such a relation by considering the gravitational and scalar sectors of a general EFT (without a potential), and solving their equations of motion for a codimension-one running solution that effectively implements an end-of-the-world (ETW) brane. The Einstein equations, together with the equations of motion for the scalars, suffice to produce a field profile that explores \emph{geodesics} in moduli space and for which the relevant length scale goes as $L\, \sim\, e^{\sqrt{\frac{d-1}{d-2}}\,\Delta_{\phi}}$ arbitrarily far away from the ETW brane. By applying then the CEB to such configurations, we obtain the exponential upper bound for the UV cut-off, which we identify as the species scale, given by
\begin{equation}
 \Lambda_{\text{sp}} \lesssim e^{-\lambda_{\text{sp,} \text{min}}\, \Delta_\phi}\, , \qquad  \lambda_{\text{sp,}\, \text{min}} \, = \,  \dfrac{1}{\sqrt{(d-1)(d-2)}}\, .
  \label{eq:lambdaminconclusions}
\end{equation}
Let us emphasize two interesting properties of our argument. First, apart from the assumption of the existence of a moduli space, it is completely independent of supersymmetry, and  we expect it to apply equally well to any configuration with a hierarchically suppressed potential. Second, it works for any trajectory in moduli space that locally fulfills the geodesic equations and approaches the infinite distance limit, as opposed to the idea that only the shortest geodesic towards the singularity should be subject to the Distance Conjecture. It is also worth stressing at this point that the entropy to which we refer in our discussion is associated to the \emph{local}  EFT description, thus being extensive by construction and incorporating potential contributions from the phase space available for finite temperature configurations of the system. This is in contrast to the approach usually taken in zero temperature configurations, such as BPS black hole or species counting, where one only accounts for the total number of microstates compatible with a given zero temperature macrostate \cite{Cribiori:2022nke, vandeHeisteeg:2023ubh,vandeHeisteeg:2023uxj,Cribiori:2023ffn}.

Furthermore, let us remark that this exponential \emph{upper} bound on the UV cut-off is precisely the crucial property required to recover the Distance Conjecture. This is the case since it really captures the idea that the cut-off in Quantum Gravity must be arbitrarily low when arbitrarily large distances are probed, and upon identifying it with the species scale it enforces the existence of an infinite tower becoming light and also bounded above by the corresponding exponential behavior. In fact, when combined with the exponential lower bounds coming from consistency of the EFT description, it completely fixes the exponential behavior of the tower and the species scale (see \cite{vandeHeisteeg:2023ubh} for an alternative argument in favor of such an exponential lower bound, also based in consistency of the EFT description). From a different point of view, our exponential upper bound, together with duality properties of the species scale and tower convex-hulls (see \cite{Castellano:2023stg, Castellano:2023jjt} for a detailed discussion on this duality) could be enough to also give an exponential lower bound.

Regarding the application of the CEB to our running solutions, the strategy to obtain these exponential bounds has been to focus on an arbitrarily large region far away from the ETW brane. Still, let us recall that it is necessary to assume that the solution is somehow UV complete, namely that the ETW brane has some UV resolution. The existence of such objects is precisely the claim of the Cobordism Distance Conjecture \cite{Buratti:2021yia,Buratti:2021fiv,Angius:2022aeq,Angius:2022mgh,Blumenhagen:2023abk,Angius:2023xtu}, but they may still seem mysterious, particularly in the case at hand where the absence of a potential automatically implies that they must be non-BPS. For this reason, we presented an explicit example of how such running backgrounds could be uplifted to a complete UV solution, which in the present case was nothing but a non-compact orbifold in one dimension more\cite{Bagger:1986wa, Kachru:1995sj}. This was merely a proof of principle for the existence of these objects, but a more general analysis of their properties and realisation would be of great interest. Nevertheless, let us remark that once the UV completion exists, our bottom-up argument is completely independent of its details, as expected in order to obtain the kind of \emph{universal} behavior predicted by the Distance Conjecture. 

Upon applying the same logic to running solutions in the presence of exponentially suppressed scalar potentials, we have also been able to constrain the region of allowed values for the decay rates of the potential and the species scale (see figure \ref{fig:CEB-potential-bounds}). In particular, we have recovered the ADC bounds in the relevant regime, and assuming the previous lower bound for the species scale held in the present case, we have also been able to recover the TCC bound on the scalar potential decay rate, namely $  c\, \geq\, 2/\sqrt{(d-1)(d-2)}\,$ \cite{Bedroya:2019snp}.

Moreover, considering the lower bound on the species scale decay rate in multi-dimensional moduli spaces led us to formulate it as a convex hull condition. Hence, we have presented the \emph{Species Scale Distance Conjecture} (SSDC),
which states that the convex hull of the species scale vectors must contain the ball of radius $\lambda_{\text{sp,}\, \text{min}} \, = \,  1/\sqrt{(d-1)(d-2)}\, $. We would like to stress that our derivation comes from first principles and does not need the extra input of the scalar WGC to support the existence of a lower bound $\lambda_{\text{min}}$ to begin with.

By exploring different M-/String theory examples we observed that this bound is saturated by single KK towers (i.e. towers with $p=1$ that correspond to the decompactification of a single dimension). Saturation can be dangerous in the context of dimensional reduction, since it can lead to violation along directions infinitesimally close to the original one in the lower dimensional theory. For that reason, we studied the consistency of our bounds under dimensional reduction and observed several interesting facts:
\begin{itemize}
    \item[$i$)]{Preservation of the bound in higher dimension ensures preservation in lower dimensions within the region of field space defined by the cone generated by the species scale vectors associated to the (dimensionally reduced) higher dimensional tower and the KK towers of the compact dimensions. To see this, it is crucial to consider the \emph{combined} species scale tower that account for the cases when the species scale is not saturated  by a single tower (see e.g. figure \ref{fig:dim-red}).}
     \item[$ii$)]{By studying the general functional form (with the spacetime dimension $d$) of the bounds that are preserved under dimensional reduction, the single KK tower ($p=1$) is selected as the only one that can saturate the bound both before and after the compactification in any number of dimensions. This gives a general bottom-up argument for the observation in examples that these single KK towers saturate our bound. Special solutions that saturate the bound only in certain dimensions also seem to be allowed and it would be interesting to study them further.}
     \item[$iii$)]{Preservation of the convex hull condition along the directions beyond the cone defined by the field theory states previously mentioned  generally requires extended objects, which give rise to additional effective species scale vectors.}
\end{itemize}

In addition to these general lessons from consistency under dimensional reduction,  explicit compactification of M-theory on $T^n$ also provide several interesting insights about our convex-hull condition. We have studied them in detail to check our conjecture and the main lessons are the following:
\begin{itemize}
    \item[$i$)]{The species scale vectors associated to single KK towers ($p=1$) saturate the species scale bound, and in fact the full convex hull can be generated by the facets perpendicular to them. Moreover, combined KK towers with higher $p$ live in higher-codimension edges defined by the different intersections of such faces. }
     \item[$ii$)]{The vertices of the convex hull always correspond to either string towers or KK towers associated to full decompactification ($p=n$). This seems to capture the idea that, at the end of the day, we always expect the species scale to encode either the ten-dimensional string scale or the eleven-dimensional Planck mass. On top of that, the fact that the vertices required to build the convex hull only correspond to these two cases signals a certain robustness of the convex hull under elimination of towers with lower $p$ (associated to partial decompactifications), as could be the case in the presence of reduced supersymmetry. }
     \item[$iii$)]{For general $T^n$ compactifications of M-theory, the convex hull condition can be argued to be always satisfied by making use of maximal supersymmetry and the fact that it is fulfilled for $T^2$ compactifications (that we checked explicitly in section \ref{ss:MthyT2}). The argument can be summarized in the observation that the facets of the convex hull for some $T^n$ are comprised by different copies of the convex hull obtained from the $T^{n-1}$ compactification.}
\end{itemize}
Several other properties related to these convex hull remain to be explored, but we expect this to be a first step along that direction, emphasizing the importance of the species scale as the intrinsically quantum gravitational cut-off. A natural arena to test the SSDC further would be to consider 9d minimal supersymmetric models arising from e.g. heterotic string theory compactified on a circle, whose particular tower structure and scalar charge-to-mass vectors are analyzed in \cite{
Etheredge:2023odp} (based on duality results from \cite{Aharony:2007du, Polchinski:1995df}). There, a `jumping' phenomenon associated to the convex hull diagram for the $z$-vectors is found, which indeed change depending on the particular asymptotic direction that is explored. Interestingly, it is also shown in \cite{Etheredge:2023odp} that one can essentially recover an analogous diagram --- both for the charge-to-mass and species vectors --- to the one arising in 9d maximal supergravity (see figure \ref{fig:ch2}), except for those asymptotic geodesics which are parallel to the self-dual line (pointing towards strong coupling) where the theory decompactifies to some running solution in type I' string theory.\footnote{We acknowledge correspondence with I. Valenzuela and I. Ruiz regarding this point.} This means, in particular, that the SSDC is verified at least for any asymptotic direction not parallel to the latter, which indeed deserves further study.

All in all, the combination of ideas and insights from holographic quantum gravitational entropy bounds, dynamical cobordisms, and convex hulls seems to be a very fruitful approach in order to study the general properties of infinite distance limits and towers of states, and we hope that this work will encourage interesting findings along related directions.

\vspace{1.5cm}
\centerline{\bf Acknowledgments}
\vspace{0cm}
		
We would like to thank G. Aldazabal, F. Marchesano, M. Montero, D. Prieto, I. Ruiz, A. Uranga,  I. Valenzuela and M. Wiesner for useful discussions and correspondence. J.C. would like to thank the Instituto de F\'isica Te\'orica in Madrid for hospitality during early stages of this work. A.C. would like to thank the Theoretical Physics Department at CERN for hospitality and support during the last stages of this work. This work is supported through the grants CEX2020-001007-S and PID2021-123017NB-I00, funded by MCIN/AEI/10.13039/5011\\
00011033 and by ERDF A way of making Europe. The work by J.C. is partially supported by the FPU grant no. FPU17/04181 from the Spanish Ministry of Education. The work of A.C. is supported by the Spanish FPI grant No. PRE2019-089790 and by the Spanish Science and Innovation Ministry through a grant for postgraduate students in the Residencia de Estudiantes del CSIC. The work of A.H. is supported by the ERC Consolidator Grant 772408-Stringlandscape.

%\newpage

\appendix

\section{Running Solutions in Kaluza-Klein Theory}
\label{ap:runningKK}

In this appendix we discuss an explicit example of the general running solution studied in section \ref{sss:runningsol}. We will particularize to the familiar case of KK reduction on a circle, to see whether there may be some sensible interpretation of the solution displayed in equation \eqref{eq:metricprofile} in terms of the higher dimensional theory.

Let us start by writing down again the running solution discussed in section \ref{sss:runningsol}:
\begin{equation} \label{eq:running-solution-app}
\begin{split}
    ds_{d}^{2} &= \left( \frac{r}{r_0} \right)^{\frac{2}{d-1}} \eta_{\mu\nu} dx^\mu dx^\nu + dr^2 \, , \\
    \Delta_{\phi} &= \pm \sqrt{\frac{d-2}{d-1}} \log \left( \frac{r}{r_0} \right) \, .
\end{split}
\end{equation}
Notice that we have reinserted an integration constant $r_0$ with respect to the expressions in section \ref{sss:runningsol}. This will allow us to understand better how this a priori meaningless integration constant from the EFT perspective can actually parametrize different UV completions. We will come back to this point at the end of the analysis. In fact, in general one would have a different integration constant in the denominator of the logarithm in both lines of the equation. However, they can be set to be the same without loss of generality by a rescaling of the longitudinal coordinates $x^\mu$. Unlike for the case of $r_0$, this will be innocuous also for the UV completion of this solution.

As pointed out in \cite{Buratti:2021fiv,Angius:2022aeq}, the fact that this solution explores infinite field space distance as the ETW singularity is approached suggest that its resolution should be related to the physics in these asymptotic regimes. In particular, one would expect it to be related to the nature of the SDC tower in such a limit, i.e., to the QG resolution of the infinite distance limit. Let us then take the easiest example, in which the infinite distance limit is related to the presence of an extra dimension (along with its KK tower). This is, from now on we consider the $d$-dimensional effective action as coming from dimensionally reducing a $(d+1)$-dimensional theory on a circle.

Considering the effective action in $(d+1)$-dimensions
\begin{equation}
	S_{d+1} = \frac{1}{2\kappa_{d+1}^{2}} \int d^{d+1}x\, \sqrt{-g}\,  R_{d+1} \, ,
\end{equation}
as well as the compactification ansatz
\begin{equation} \label{eq:dim-red}
    ds_{d+1}^{2} = e^{-\frac{2}{\sqrt{(d-1)(d-2)}} \hat{\rho} } ds_{d}^2 + e^{2 \sqrt{\frac{d-2}{d-1}} \hat{\rho}} dy^2 \, , \qquad y \sim y + 2 \pi y_0 \, ,
\end{equation}
one arrives at the $d$-dimensional effective action
\begin{equation}
    S_{d} = \frac{1}{\kappa_d^2} \int d^{d}x\, \sqrt{-g}\,  \left[ \frac{1}{2} R_d - \frac{1}{2} \left( \partial \hat{\rho} \right)^2 \right] \, .
\end{equation}
This indeed takes the form of the more general effective action considered in section \ref{sss:runningsol}, with the moduli space being parametrized by the radion, $\hat\rho$. In fact, the ansatz was precisely designed to get a this lower dimensional effective action already in the Einstein frame and with the radion being canonically normalized. Notice that we are considering the minimal ingredients that are relevant for the analysis. For instance, we are ignoring the KK photon. Furthermore, we focus in the Einstein-Hilbert term in $(d+1)$-dimensions. Additional fields, such as massless scalars, $p$-forms, etc., can be added without changing the conclusions. In particular, adding extra scalars would amount to including other directions in the $d$-dimensional moduli space, and our analysis would be thus recovered upon focusing on the (geodesic) direction parametrized by the radion $\hat{\rho}$.

Having this setup, one can now build the running solution in equation \eqref{eq:running-solution-app} with the moduli space distance being identified with the canonically normalized radion, i.e.,
\begin{equation} \label{eq:rho-profile}
    \hat \rho = \sqrt{\frac{d-2}{d-1}} \log \left( \frac{r}{r_0} \right) \, .
\end{equation}
From this lower-dimensional perspective, the solution resembles some sort of codimension-one singularity located at $r=0$ (see section \ref{sss:runningsol}), with a non-trivial profile for the scalar field $\hat \rho(r)$ as displayed in \eqref{eq:rho-profile}. What we want to do now is to lift such a solution, which we have chosen to interpolate between the small radius regime at $r = 0$ and the decompactification limit upon approaching $r\to \infty$. In this sense, this solution is reminiscent of Witten's bubble of nothing (BoN) \cite{Witten:1981gj}.

Plugging the $d$-dimensional metric \eqref{eq:running-solution-app} and the scalar profile \eqref{eq:rho-profile} into the dimensional reduction ansatz \eqref{eq:dim-red} we get the following $(d+1)$-dimensional metric:
\begin{equation}
    ds_{d+1}^{2} = \eta_{\mu\nu} dx^\mu dx^\nu + \left( \frac{r}{r_0} \right)^{- \frac{2}{d-1}} \left( dr^2 + \left( \frac{r}{r_0} \right)^{2} dy^2 \right) \, .
\end{equation}
Remarkably, this metric is of the form of a Cartesian product of $(d-1)$-dimensional flat space with a two-dimensional cone. Taking into account the periodicity of the coordinate $y$, the deficit angle of the two-dimensional cone reads:
\begin{equation}
    \theta_{\text{def}} = 2 \pi \left( 1 - \frac{d-2}{d-1} \frac{y_0}{r_0} \right) \, .
\end{equation}
Whenever the combination
\begin{equation}
    k = \frac{d-1}{d-2} \frac{r_0}{y_0} \, 
\end{equation}
takes non-negative integer values, this solution is just $\mathbb{R}^{d-2} \times \mathbb{R}^2/\mathbb{Z}_k$, i.e., an orbifold of flat space. For these values, the running solution is then uplifted to a well-behaved background in string theory (see e.g. \cite{Bagger:1986wa, Kachru:1995sj}). We thus find particular values for the integration constant $r_0$ for which the running solution is resolved to a consistent background in Quantum Gravity. In particular, there is a value for which the uplift corresponds to just $(d+1)$-dimensional flat space. In this case one does not even need the string theory input of orbifold singularities being well-behaved in Quantum Gravity.

Let us remark that the different uplifts of the running solutions are solely controlled by the ratio of $y_0$ and $r_0$. This is to be expected, since a change in $y_0$ can be reabsorbed by a shift in the canonically normalized modulus $\hat \rho$, which in turns can be reabsorbed by a change in $r_0$ in the running solution. Interestingly, a naive effective field theorist could have disregarded $r_0$ as a meaningless integration constant since it can be reabsorbed by a shift in the modulus. However, this relies on the shift symmetry for this scalar, which is known to be absent in Quantum Gravity due to the presence of the towers whose mass depend on its vev. Somehow, the absence of global symmetries in Quantum Gravity is telling us that this integration constant is relevant for the UV completion of this solution.

\bibliography{references}
\bibliographystyle{JHEP}

\end{document}